\documentclass[10pt]{book}

\input{epsf}

\begin{document}


\newcommand{\bb}{\begin{equation}}
\newcommand{\ee}{\end{equation}}
\newcommand{\bbb}{\begin{eqnarray}}
\newcommand{\eee}{\end{eqnarray}}
\newcommand{\vc}[1]{\mbox{$\vec{{\bf #1}}$}}
\newcommand{\mc}[1]{\mathcal{#1}}
\newcommand{\del}{\partial}
\newcommand{\lk}{\left}
\newcommand{\ave}[1]{\mbox{$\langle{#1}\rangle$}}
\newcommand{\re}{\right}
\newcommand{\pd}[1]{\frac{\del}{\del #1}}
\newcommand{\pdd}[2]{\frac{\del^2}{\del #1 \del #2}}
\newcommand{\Dd}[1]{\frac{d}{d #1}}
\newcommand{\sech}{\mbox{sech}}
\newcommand{\pref}[1]{(\ref{#1})}

\newcommand
{\sect}[1]{\vspace{20pt}{\LARGE}\noindent
{\bf #1:}}
\newcommand
{\subsect}[1]{\vspace{20pt}\hspace*{10pt}{\Large{$\bullet$}}\mbox{ }
{\bf #1}}
\newcommand
{\subsubsect}[1]{\hspace*{20pt}{\large{$\bullet$}}\mbox{ }
{\bf #1}}

\def\ie{{\it i.e.}}
\def\eg{{\it e.g.}}
\def\cf{{\it c.f.}}
\def\etal{{\it et.al.}}
\def\etc{{\it etc.}}

\def\AA{{\cal A}}
\def\BB{{\cal B}}
\def\CC{{\cal C}}
\def\DD{{\cal D}}
\def\EE{{\cal E}}
\def\FF{{\cal F}}
\def\GG{{\cal G}}
\def\HH{{\cal H}}
\def\II{{\cal I}}
\def\JJ{{\cal J}}
\def\KK{{\cal K}}
\def\LL{{\cal L}}
\def\MM{{\cal M}}
\def\NN{{\cal N}}
\def\OO{{\cal O}}
\def\PP{{\cal P}}
\def\QQ{{\cal Q}}
\def\RR{{\cal R}}
\def\SS{{\cal S}}
\def\TT{{\cal T}}
\def\UU{{\cal U}}
\def\VV{{\cal V}}
\def\WW{{\cal W}}
\def\XX{{\cal X}}
\def\YY{{\cal Y}}
\def\ZZ{{\cal Z}}

\def\sinh{{\rm sinh}}
\def\cosh{{\rm cosh}}
\def\tanh{{\rm tanh}}
\def\sgn{{\rm sgn}}
\def\det{{\rm det}}
\def\exp{{\rm exp}}
\def\sh{{\rm sh}}
\def\ch{{\rm ch}}

\def\ell{{\it l}}
\def\str{{\it str}}
\def\lp{\ell_{{\rm pl}}}
\def\blp{\overline{\ell}_{{\rm pl}}}
\def\ls{\ell_{{\str}}}
\def\bls{{\bar\ell}_{{\str}}}
\def\bM{{\overline{\rm M}}}
\def\gs{g_\str}
\def\gym{g_{Y}}
\def\geff{g_{\rm eff}}
\def\eff{{\rm eff}}
\def\r11{R_{11}}
\def\kel{{2\kappa_{11}^2}}
\def\kten{{2\kappa_{10}^2}}
\def\lpten{{\lp^{(10)}}}
\def\alp{{\alpha'}}
\def\tgs{{\tilde{g}_s}}
\def\talp{{{\tilde{\alpha}}'}}
\def\tlp{{\tilde{\ell}_{{\rm pl}}}}
\def\tr11{{\tilde{R}_{11}}}
\def\wtilde{\widetilde}
\def\what{\widehat}
\def\hlp{{\hat{\ell}_{{\rm pl}}}}
\def\hr11{{\hat{R}_{11}}}
\def\hf{{\textstyle\frac12}}
\def\coeff#1#2{{\textstyle{#1\over#2}}}
\def\CY{Calabi-Yau}
\def\lessapprox{\;{\buildrel{<}\over{\scriptstyle\sim}}\;}
\def\greaterapprox{\;{\buildrel{>}\over{\scriptstyle\sim}}\;}
\def\inbar{\,\vrule height1.5ex width.4pt depth0pt}
\def\IC{\relax\hbox{$\inbar\kern-.3em{\rm C}$}}
\def\IR{\relax{\rm I\kern-.18em R}}
\def\IP{\relax{\rm I\kern-.18em P}}
\def\Z{{\bf Z}}
\def\R{{\bf R}}
\def\One{{1\hskip -3pt {\rm l}}}
\def\sst{\scriptscriptstyle}
\def\osc{{\rm\sst osc}}
\def\lam{\lambda}
\def\lc{{\sst LC}}
\def\pr{{\sst \rm pr}}
\def\cl{{\sst \rm cl}}
\def\D{{\sst D}}
\def\bh{{\sst BH}}
\def\vev#1{\langle#1\rangle}

\begin{titlepage}
\rightline{EFI-99-26}

\rightline{hep-th/9906044}

\vskip 3cm
\begin{center}
{\LARGE{Black Holes and Thermodynamics\\
of~Non-Gravitational~Theories}}

\vskip 3cm 
Vatche Sahakian\footnote{\texttt{isaak@theory.uchicago.edu}}
\footnote{Address
after August 1, 1999: Laboratory of Nuclear Studies, Cornell University,
Ithaca, NY 14853}
\vskip 12pt
\centerline{\sl Enrico Fermi Inst. and Dept. of Physics}
\centerline{\sl University of Chicago}
\centerline{\sl 5640 S. Ellis Ave., Chicago, IL 60637, USA}

\vskip 2cm 

{\Large{Abstract}}
\end{center}

\vspace{12pt}
This is a thesis/review article that combines some of the results
of~\cite{LMS,MSSYM123,MSFIVE}
with a short discussion of introductory background material;
an attempt has been made to present the work in a self-contained manner.
The first chapter mostly targets readers who
are vaguely familiar with traditional and contemporary string theory.
Chapter two discusses in detail the thermodynamics of the
$0+1$ dimensional Super Yang-Mills (SYM) theory as an illustrative 
example of the main ideas of the work.
The third chapter outlines the phase structures of
$p+1$ dimensional SYM theories on tori for $1\le p \le 5$,
and that of the D1D5 system; we avoid presenting the technical details of
the construction of these phase diagrams, focusing instead on
the physics of the final results. The last chapter discusses the 
dynamics of the formation
of boosted black holes in strongly coupled SYM theory.

\end{titlepage}

\newpage

\begin{center}
\Large{Acknowledgments}
\end{center}

\vspace{12pt}
The work presented in this thesis is a compilation of the
papers~\cite{LMS,MSSYM123,MSFIVE}; it was submitted to the
division of Physical Sciences at the University of Chicago
as a PhD thesis.
I thank my collaborators Emil Martinec and Miao Li
for many fruitful discussions
and a pleasant atmosphere of collaboration. In particular,
I am grateful to my advisor Emil Martinec for teaching me physics,
for encouragement, for guidance, and for support; all these despite 
having spilled on him hot coffee.
 
I am grateful to Peter Freund and Robert Geroch for raising 
my appreciation for teaching physics. 
I thank my office-mates
Bruno Carneiro da Cunha,
Cristi\'{a}n Garc\'{\i}a,
Ajay Gopinathan,
Julie and Scott Slezak and
Li-Sheng Tseng
for company and for many pleasant, as well as 
sometimes very strange, discussions.
As for people from the world outside the office, I thank
Aleksey Nudelman for many interesting conversations, and
Emil Yuzbashyan for genetic company.

Last but not least, I thank my family for support and encouragement.

\vspace{24pt}
April 19, 1999

\newpage
\vspace{2in}
\begin{center}
In memory of April 24, 1915

\vspace{0.25cm}
\epsfxsize=5cm \centerline{\leavevmode \epsfbox{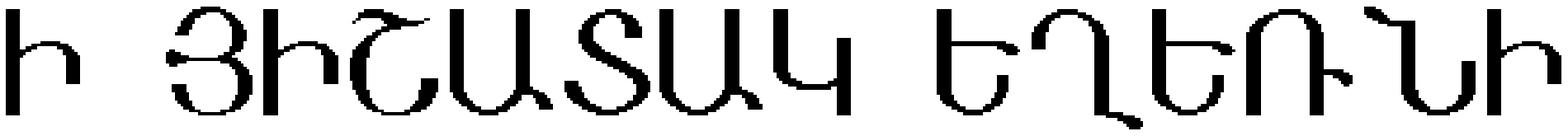}}
\end{center}

\tableofcontents

\setcounter{page}{1}

\chapter{Introduction}

\section{Motivation}

There are two main frameworks through which laws of physics can be studied.
The first is a microscopic setting; one arranges a few asymptotic states in 
a given theory, throws them at each other, and observes the outcome. 
The dynamics of the theory can in principle be decoded out
of such experiments. Another platform of exploration is thermodynamics; 
one takes a large number of degrees of freedom, 
prepares a thermodynamic phase, and traces the system as a function of
the thermodynamic parameters. In this
latter setting, the important attributes of
the dynamics manifest themselves via critical phenomena. 
For example, 
the transition between the normal and superconductive phases in a metal
is the thermodynamic signature of the bound state formation 
phenomena between pairs of electrons. 
Phase transitions are typically reflections of some of the
most interesting characteristics of the underlying microscopic physics. 
Furthermore, the concepts of
critical phenomena and thermodynamic phase structure are fundamentally
related to our modern understanding of the hierarchy and
connections between physical theories.

It is now believed 
that the proper framework to correctly formulate a quantum theory
of gravity has been identified. As the most significant recent
evidence in support of this view has been the accounting for the 
degrees of freedom responsible for the entropy of black holes
~\cite{STROMVAFA,MALDATHESIS}.
These ideas have emerged by embedding general relativity into the low
energy regime of a new theory of string theoretical origin. 
While a great deal remains to be understood about this theory, 
significant progress in unraveling its intricacies has been achieved
in the past few years~\cite{MWITTEN,HULLTOWN,MAT1,MALDA1}.
The focus of this thesis is 
to study thermodynamics and critical phenomena
in this fundamental theory.

Let us shift the discussion
away from gravitational physics and consider what appears
to be the unrelated topic of the
thermodynamics of non-gravitational theories.
Consider a gas of non-gravitating but 
otherwise weakly interacting particles 
in a square $p$ dimensional 
box at fixed and high temperature. Accord $g$ degrees 
of freedom for each cell of the phase space of each particle.
The interactions being very weak, the equation of
state can be sketched easily. The entropy must be extensive, so it
is proportional to the volume of the box $\Sigma^p$, where $\Sigma$
is the length of a side of the box. Assuming Boltzmann
statistics at high enough temperatures, 
the entropy is proportional
to $g$. Finally, the power of the temperature is determined by 
dimensional analysis
\bb
S\sim g \Sigma^p T^p\ ,
\ee
while the energy scales as $E\sim T S$.
Putting these together, we write the energy as a function of the entropy as
\bb\label{Gaseos}
E=\frac{S^{\frac{p+1}{p}}}{N^{\frac{2}{p}} \Sigma}\ ,
\ee
where, for future reference, we have set $g=N^2$.
This equation of state may get subleading corrections
due to the interactions between the constituents
of the gas; for weak coupling, a perturbative expansion
in the coupling constant can in principle be written.
As we cool the system,
the interactions may become strong, correlations between
various parts of the gas may grow stiffer, and a new phase
may emerge after the crossing of a point of phase transition. All of
this may happen in a non-perturbative regime of the theory;
questions regarding the state of
the system then generically become intractable by conventional physics.
The focus of this thesis is to study such phenomena in a certain
class of non-gravitational theories.

The intended
implication of our last comment is that the two separate issues
that we raised, thermodynamics of a gravitational theory and that of
certain non-gravitational ones, are related. 
Recent progress in string theory
indicates that {\em gravity can be encoded in non-perturbative
regimes of certain non-gravitational theories}~\cite{MAT1,MAT2,
MALDA1,WITHOLO,GUBSER}; in particular,
supersymmetric Yang-Mills theories, at strong coupling 
and for large ranks of the gauge group, appear to describe elaborate
quantum theories of gravity~\cite{MALDA1,MALDA2}.
This revelation can be qualified nothing less than remarkable.
It is leading to a fundamental reassessment of our understanding
of gravity, space-time and quantum field theories.
In the forthcoming sections of this chapter, we intend to
systematically review these ideas.

An attempt has been made to present most of the necessary background
physics in a self-contained manner. 
The casual reader may focus on reading 
Chapters 1, 2, and Sections 3.1, 3.2, 4.1, and 4.2.
Further background material can be found in Appendices A and B. 
Beyond this, the discussion may become considerably less entertaining
for the non-specialist.

\section{Basics}

Generically, a theory of particle physics identifies a set of
degrees of freedom, and proposes a prescription for their 
dynamics and interactions. In practice, this setting is
often a low energy approximation of 
more fundamental physics. Beyond the domain of relevance associated
with the theory, new degrees of freedom may enter the game, modify
the dynamics, and the emerging picture may be endowed with a 
fundamentally different character. It is proposed that string theory
is a description of 
physics at the most fundamental level. The degrees of freedom and their
dynamics form a correct account of ``reality'' at the smallest possible
length scales. Simpler, less fundamental but not necessarily uninteresting
physics is to emerge
from string theory at progressively lower energies.

Consider an eleven dimensional
supersymmetric theory, which we will refer to as {\em M theory},
entailing the dynamics of certain flavors of extended objects.
The dimensionful parameters of M theory consist of
$\hbar$, $c$, and the gravitational coupling in eleven dimensions
$\kel$. We choose units such that
$\hbar=c=1$, and all dimensionful
observables are henceforth measured in units of length set
by the Planck scale $\kel=(2\pi)^8 \lp^9$. 
The regime of low energy (with respect to the Planck scale) of this
theory is $\NN=1$, $11$d supergravity, a well known and relatively
simple supersymmetric theory of gravity.
The high energy dynamics of M theory
is considerably better understood when it is compactified to lower
dimensions. Particularly, compactifying
on a circle of sub-Planckian size leads to a ten dimensional
theory known as
the {\em type IIA string theory}.
The latter is parameterized by the string length scale
$\ls$, and a dimensionless coupling constant $g_s<1$.
These two variables
are related to the parameters of the M theory from which
the IIA theory descends by
\bb\label{MIIA}
\ls^2\equiv \alp = \frac{\lp^3}{R_{11}}\ , \ \ \ 
g_s^2=\lk(\frac{\r11}{\lp}\re)^3\ ,
\ee
where $2\pi R_{11}$ is the circumference of the compactified dimension.
The gravitational constant in ten dimensions is then given by
\bb\label{g10}
\kten = \frac{\kel}{2\pi R_{11}}= (2\pi)^7 \gs^2 \ls^8\ .
\ee

The degrees of freedom of the IIA string theory consist of:
\begin{itemize}
\item A one dimensional extended object, the closed
string (F1); its tension 
is defined by
\bb
T_{F1}=\frac{1}{2\pi \alp}\ .
\ee
\item The magnetic dual of this string; this is a five dimensional extended
object referred to as the Neveu-Schwartz five brane (NS5). Its tension
is given by
\bb
T_{NS5}=\frac{1}{(2\pi)^5 g_s^2 {\alp}^3}\ .
\ee
\item Various $p$ dimensional extended objects referred to as $Dp$ branes
(Dp). Their tension is
\bb\label{TDp}
T_{Dp}=\frac{1}{(2\pi)^p g_s \alp^{(p+1)/2}}\ ,
\ee
with $p$ an even integer for the type IIA theory.
\end{itemize}

All of these
objects of the IIA string theory originate from two objects in M theory;
a membrane (M2), and its magnetic dual, a five brane (M5). Their tension
is set by the eleven dimensional Planck scale. These are the only objects
allowed in eleven dimensional M theory by symmetry considerations.

F1 strings, NS5 branes and Dp branes 
interact with each other in various well understood, as well as
sometimes ill-understood, ways. The F1 string mediates some of
the interactions between the other types of objects, in addition to 
interacting with other F1 strings. The strength of all these processes is tuned
by $g_s\ll 1$. In this setting, the NS5 and Dp branes are heavy
compared to the fundamental string as can be seen from the tension
formulae above.
The dynamics is therefore dominated by that of the
string, and the physics is described by a perturbative expansion in the
string coupling $g_s$.

The closed strings of the IIA theory can break upon collision with the surface
of a Dp brane. The result of such a process is an open string with its
endpoints confined to the surface; this is ``the ripple''
created on the surface of the brane as a result of the collision.
More generally, the fluctuations of
the surface of excited Dp branes
are described through the dynamics of a gas of such open strings.
The time reversed version of the collision process just depicted represents
``the evaporation'' from an excited Dp brane.

A fundamental characteristic of string theory, non-local dynamics,
is due to the fact that closed and open strings are extended objects that 
can vibrate. At $g_s=0$, the spectrum of a vibrating string is like that of
the quantized harmonic oscillator,
with each level representing a quantum degree of freedom
propagating in ten dimensional space-time.
The spacing of the energy levels is
set by the string tension $T_{F1}$; the ground state
has zero energy, \ie\ it corresponds to massless quanta. 
The polarizations of the vibrational modes encode the 
spin of the quanta. String theory thus involves an infinite number 
of flavors of particles with arbitrarily large mass and spin. At low
energies with respect to the string scale, the degrees of freedom
consist of massless particles, the ground state modes of the
string spectrum, with the maximum bosonic spin being two.

String theory is endowed with a myriad of symmetries. 
In addition to the ten dimensional Lorentz symmetry, supersymmetry
assures that the quanta associated with the spectrum of string excitations
fall into supersymmetry multiplets, with a pairing between bosonic and 
fermionic degrees of freedom. Furthermore, there exists strong
evidence for another class of symmetries underlying the theory. These are
the so called {\em dualities}~\cite{HULLTOWN,SENDUAL,SENREV}.
These symmetries are of fundamental 
importance since they have lead to an understanding of the theory 
beyond the perturbative expansion in $g_s$. Furthermore, the five
known flavors of string theories, of which we have only described the IIA
theory above, transform into each other under the action of these duality
transformations. These connections between the various theories become
richer in structure as we compactify the string theories to lower dimensions. 
The global picture that emerges suggests the existence of
a single theory underlying M theory and
all five flavors of string theory.
A slightly more detailed account of this
subject is given in Appendix~\ref{dualapp} which the reader is encouraged
to consult.
For the casual reader, we briefly state the duality nomenclature
we will make use of.
{\em S duality} relates a weakly coupled regime of
a theory to a strongly coupled regime by inverting the string coupling
$\gs\rightarrow 1/\gs$. {\em T duality} relates a theory compactified
on a circle of size $R$ to one compactified on the dual circle of
size $\alp/R$. {\em M-IIA duality} identifies the strongly coupled regime
of IIA theory with eleven dimensional M theory as described above.

The previous discussion, presented as such, 
may appear as a particularly 
imaginative excerpt from a fairy tale. It is important to emphasize
that the structure of the theory we outlined is severely restricted by
various mathematical and physical considerations. A handful of fundamental
physical principles and a few intuitively
motivated postulates can be identified as the foundations of this
elaborate theory.

\section{Closed strings and gravity}

The spectrum of a free closed string
consists of an infinite tower of string vibrational modes,
the levels separated by $1/\ls$. Focusing on dynamics at large distances
with respect to the string scale simplifies the theory dramatically.
A low energy effective description of a closed string theory is given
by the dynamics of the quanta corresponding to the zero energy ground states
of the free string spectrum. Hence we have a quantum field theory of
massless quanta propagating in ten (or lower) dimensions and
arranged into representations of the Lorentz and supersymmetry
algebras. The symmetries typically 
uniquely determine the field content of these quantum field theories.

It is found that the low energy
regime of closed string theories always yields a gravitational theory.
It is then implied that the diseases of quantum gravity at small 
length scales are cured by the emergence of the additional degrees of freedom
that one carelessly truncates away in the low energy approximation.
A simplified and generic
form for the bosonic part of the
low energy effective action that one encounters 
in a closed string theory is~\cite{GLIOZZI1,CFMP}
\bb\label{Sgrav}
S=\frac{1}{\kten} 
\int\ d^{10} x\ \sqrt{-g} 
\lk\{ 
e^{-2\phi} 
\lk[ 
R+4\nabla_\mu \phi \nabla^\mu \phi - \frac{1}{12}
{H_{(3)}}^2
\re]
-\frac{1}{2 (p+2)!} {F_{(p+2)}}^2 
\re\} 
+ \cdots\ .
\ee
The massless fields in this action are the graviton $g_{\mu\nu}$, the
dilaton $\phi$,
a $3$-form 
field strength $H_{(3)}$ of a $2$-form gauge field coupling to the
fundamental string, and 
a $p+2$-form field strength $F_{(p+2)}$ of a $p+1$-form gauge field
coupling to Dp brane charge.
$R$ is the Ricci scalar constructed from the metric $g_{\mu\nu}$.
Shifting the dilaton in this action
corresponds to rescaling $g_s$ in $\kten$. 
The string coupling can hence be thought of as a dynamical
variable of the theory, $e^\phi\rightarrow \gs$. 
In this thesis,
we will adopt the convention of absorbing the asymptotic value of the
dilaton field in the definition of the string coupling; \ie\ the
dilaton $\phi$ will always go to zero at large distances away from a source.

The action~\pref{Sgrav} admits classical solutions describing the
geometry and gauge fields cast about various objects that can arise in
string theory. Solutions describing black holes, fundamental strings, Dp
branes and NS5 branes are known and have been extensively studied.
In writing~\pref{Sgrav}, it is assumed that scales of curvatures under
consideration are much smaller than the string scale so that the low energy
truncation is consistent; and that the
string coupling, \ie\ the dilaton, is small enough for perturbative
string theory to make sense. Classical solutions are then required
to respect these conditions.

\begin{figure}
\epsfxsize=10cm \centerline{\leavevmode \epsfbox{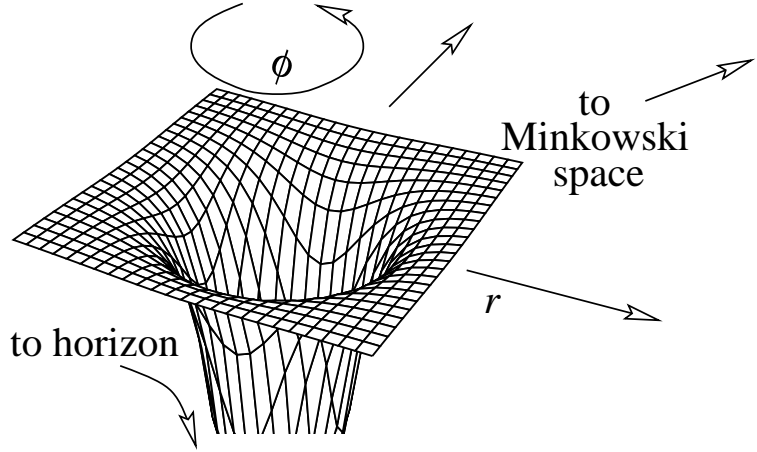}}
\caption{\sl A cross section of the metric~\pref{Dpmetric} in the
$r-\phi$ plane, where $\phi$ is the lateral polar angle in the
transverse sphere. The local extrinsic curvature is an approximate
representation of the curvature scale appearing in the original
metric at the corresponding point in space. 
Far way from the center, the geometry becomes flat.}
\label{nhfig}
\end{figure}

Consider the solution to the equations of motion of~\pref{Sgrav}
describing the fields about $N$ excited Dp branes
~\cite{HORSTROM,STRINGSOLITONS}
\bb\label{Dpmetric}
ds^2=H^{-1/2} \lk( -h dt^2+d{\vec y}_p^2\re) + H^{1/2} \lk( h^{-1} dr^2
+r^2 d\Omega_{8-p}^2\re)\ ,
\ee
\bb\label{Dpdilaton}
e^\phi=H^{(3-p)/4}\ ,
\ee
\bb\label{DpRRfield}
F_{r t y_1\ldots y_p}=
\del_r H^{-1}\ \ \ \mbox{with the other components equal to 0}\ .
\ee
where
\bb\label{harmH}
h\equiv 1-\lk(\frac{r_0}{r}\re)^{7-p}\ ,\ \ \ \ 
H\equiv 1+\lk(\frac{q}{r}\re)^{7-p}\ .
\ee
$r=r_0$ denotes the location of the horizon; the Dp branes
are siting at $r=0$. 
This horizon has finite area, and hence it is associated
with finite entropy; the corresponding
geometry is referred to as ``black'' to remind us of the thermodynamic 
nature of the state being described, reminiscent of black hole geometries.
Setting $r_0\rightarrow 0$ yields the geometry of {\em extremal}
Dp branes, \ie\  it is the zero temperature limit.
It can be shown that half of the supersymmetry generators leave this state
unchanged; the $r_0\neq 0$ solution however breaks all supersymmetries.
The variable
$q^{7-p}\propto N$ measures the amount of charge $N$ carried by 
the solution under the corresponding field strength. We will often compactify
the $p$ space coordinates parallel to the surface of the Dp branes,
denoted in~\pref{Dpmetric} by $y_{(p)}$,
on a torus $T^p$ of equal cycle sizes $\Sigma$. 
The space transverse to the Dp branes is mapped by the radial coordinate
$r$ and the $8-p$ angular coordinates parameterizing an $8-p$ dimensional
sphere; the metric on this sphere
is denoted by $d\Omega_{8-p}^2$.
A pictorial representation of this Dp brane geometry is shown in 
Figure~\ref{nhfig}.

By standard techniques
of general relativity, one can calculate the entropy
of~\pref{Dpmetric}, as well as its energy content
as measured by an observer at infinity. The reader is referred
to Appendix~\ref{grapp} for some of the technical details regarding such
computations. We find that
the equation of state for the thermodynamic state 
comprised of $N$ non-extremal Dp branes is given by~\cite{MALDA2,KLEBTSEYT,DUFF}
\bb\label{DpEOS}
E\sim \lk( \frac{S^2}{N} \re)^{\frac{7-p}{9-p}}
\lk(
\frac{\Sigma^{p(p-5)} \gs^{3-p}}{\alp^{\frac{(p-3)^2}{2}}}
\re)^{\frac{1}{9-p}}
\ .
\ee
Here $E$ is the energy measured above the mass of the extremal branes,
and $S$ is the entropy. The specific heat for this geometry is
positive for $p<6$. The system is metastable in the sense that it
evaporates slowly through Hawking radiation.

The solution given in~\pref{Dpmetric}-\pref{DpRRfield}
is a faithful
description of the Dp branes for regions of space where the 
local curvature scale
is small relative to the string scale, and the string coupling
$g_s e^\phi< 1$. Elsewhere, the original Lagrangian is corrected by new
physics of string theoretical origin. Particularly, the size of these
two observables, curvature and coupling,
as measured at the horizon $r=r_0$ determine the window of
validity of the equation of state~\pref{DpEOS}, since the latter
is a statement extracted from geometrical
information at the horizon. These two restrictions are found to be
\bb\label{couplingp}
\lk(
\gs^{-1} \lk( \frac{\Sigma^2}\alp\re)^{\frac{p-3}{2}}
\re)^{p(p-7)}
< \lk( N^{8-p} S^{p-7}\re)^{p-3} \ ,
\ee
for the condition $g_s e^\phi|_{hor}<1$, and
\bb\label{curvaturep}
\gs^p \lk(\frac{\Sigma^2}\alp\re)^{p\frac{3-p}{2}}
> S^{3-p} N^{p-6}\ ,
\ee
for the curvature scale at the horizon to be small with respect
to $1/\alp$; we will refer to this latter criterion as the Horowitz-Polchinski
correspondence point. 
We thus have a gravitational
description of the thermodynamic state of $N$ excited Dp branes for certain
entropies $S$, asymptotic string coupling $\gs$,
and torus size $\Sigma$. 

\section{Open strings and Yang-Mills theories}

We concluded the previous section by describing 
Dp branes as seen by the low energy dynamics of closed string theory.
In this section, we focus on Dp brane
physics as emerging from the dynamics of
the open strings propagating on their surfaces.

As in the closed
string sector, open string dynamics is simplified by studying the low energy 
regime, truncating away the degrees of freedom 
associated with the vibrational modes of the open strings. This
is again 
a quantum field theory of massless fields, 
quanta corresponding to the ground states of the open
string spectrum. It is not however a gravitational theory.
We further restrict to the regime where the interactions
of these quanta with the gravitational closed string sector in
the geometry projected by the Dp branes is negligible; this
is achieved by requiring that the gravitational coupling is small. 
For $N$ Dp branes, the
resulting theory is the dimensional reduction of $\NN=1$ $U(N)$ ten dimensional
Super Yang Mills (SYM) theory to the 
$p+1$ dimensional world-volume of the $N$ branes~\cite{GLIOZZI1,SYMORIG,
POLCHTASI}
\bbb\label{SYMaction}
S=-\frac{1}{4\gym^2} \int\ d^{p+1}y\ 
\mbox{Tr}\lk( F_{\mu\nu}^2 + 2 D_\mu X_i D^\mu X^i -[X_i,X_j]^2\re) \nonumber \\
+2 i \mbox{Tr} \lk( \overline{\lambda} \Gamma^\mu D_\mu \lambda
+ i \overline{\lambda} \Gamma^i [X_i,\lambda] \re)\ .
\eee
All fields are $N\times N$ hermitian matrices in the adjoint of $U(N)$. 
Throughout this thesis, we will assume that $N\gg 1$
for reasons that will become evident in the next Chapter. 
$D_\mu\equiv \del_\mu - i [A_\mu,\ ]$ is the standard covariant 
derivative.
We have the gauge field strength $F_{\mu\nu}$, where $\mu,\nu$ run over
the $p+1$ coordinates parameterizing the world-volume of the Dp branes;
the scalars $X^i$, where $i$ runs over the space dimensions transverse to
the branes; and the sixteen ``gaugino'' field $\lambda^\alpha$ as
Majorana Weyl spinors of $SO(9,1)$. The Yang-Mills coupling $\gym$ is
given by
\bb\label{gym}
\gym^2= (2\pi)^{p-2} \gs \alp^{\frac{p-3}{2}}\ .
\ee
We will also take the coordinates $y_\mu$ that
parameterize the world-volume of the branes to live on a torus with cycle
circumferences equal to $\Sigma$, \ie\ the Dp branes are
wrapped on a torus of size $\Sigma$. 

An intuitive insight can be obtained as to the meaning of the scalar
fields in the SYM action as follows. Expanding around a background consisting
of diagonal scalar matrices $X^i$,
$\gym^2$ is found to tune the masses of the off-diagonal modes of the 
scalar matrices.
For the branes separated from
each other by super-Planckian distances, 
the diagonal elements of these matrices
can be thought of as representing the positions
of the $N$ branes in the transverse space to their worldvolumes. The
massive off-diagonal elements may then be integrated out in an 
adiabatic approximation scheme
to yield an effective description of the dynamics of the diagonals. 
Remarkably, the resulting potential between the diagonal modes is found
to be of gravitational character.
This is suggesting that closed string physics and
space-time are encoded in the low energy open string dynamics.
For sub-Planckian brane separations, this natural division of roles
between diagonal and off-diagonal matrix elements cannot be made.
The physics is more complex, and our intuitive
picture of a smooth fabric for space
falters. 
Suggestions have been made that this is
indicative of a non-commutative structure of space-time at the 
eleven dimensional Planck scale.

The action~\pref{SYMaction} receives 
corrections
as we probe physics at string scale energies due
to the effect of the stringy vibrational modes of the open strings. 
Interactions
with the closed string sector further alter the dynamics. Processes
involving closed strings from the surrounding bulk 
space breaking on the surface of the branes,
and open strings evaporating to closed strings, are tuned by the
strength of the gravitational coupling. The action~\pref{SYMaction}
however becomes progressively better reflection of the correct Dp brane physics
in the limit
\bb\label{Maldalimit}
\alp\rightarrow 0\ \ \mbox{with}\ \ \gym^2,\ \Sigma,\ E\ \mbox{held fixed,}
\ee
henceforth referred to as {\em the Maldacena limit}. $E$
here is the energy scale of an excitation in the  field theory.
The gravitational coupling is given by
\bb\label{decoupling}
\kten \sim \gym^4 \alp^{7-p} \rightarrow 0\ \ \mbox{for $p<7$}\ .
\ee
However, we need to be careful before we conclude that all is
well for $p<7$. 
This is because equation~\pref{gym} implies that in the Maldacena limit
$\gs\rightarrow \infty$ for $p>3$. This means that, 
for $p$ odd, we need to look at the S dual picture;
for $p$ even, we need 
to look at the gravitational coupling in eleven dimensions.
As discussed in Appendix~\ref{dualapp},
equation~\pref{decoupling}
is invariant under S and T duality transformations. Hence, we only have
to look at the $p=4$ and $p=6$ cases more carefully.
For $p$ even and $p>3$, the eleven dimensional M theory
gravitational coupling is
\bb
\kel\sim \kten \r11\sim \lk(\gym^2 \alp^{\frac{6-p}{2}}\re)^3
\rightarrow 0\ \ \mbox{for $p<6$.}
\ee
Combining these observations,
we conclude that, for $p<6$ and in the Maldacena limit~\pref{Maldalimit},
both gravitational and
stringy effects are scaled out, while the parameters of the field theory
are held fixed. This statement defines a certain regime of energy
and string coupling of interest from the perspective of our field theory.

We define the effective dimensionless coupling at large
$N$, with 't Hooft-like scaling, by
\bb\label{geffeq}
\geff^2 \equiv \gym^2 N T^{p-3},
\ee
where $T$ is the substringy 
energy scale at which we study a given process with the
action~\pref{SYMaction}. We see three different scenarios arising
as a function of $p$ that need to be distinguished:

\begin{itemize}
\item For $p<3$, the theory is super-renormalizable and well
defined in the UV
\footnote{Note that the UV cutoff must always be smaller than the string
scale, which is taken to infinity in the Maldacena limit.}. The
effective Yang-Mills coupling increases at low energies.

\item For $p=3$, we have $3+1$d $\NN=4$ SYM theory. This is a superconformal
theory; its beta function vanishes.

\item For $3<p<6$, the Yang-Mills coupling is irrelevant; it blows up at
high energies, indicating an ill-defined theory at small length scales.
A non-renormalizable theory arising at low energies may be thought of
as a descendent from an otherwise more elaborate but
well-defined theory in the UV; at higher energies, the degrees of
freedom of the UV theory set in and regularize the dynamics. It is
believed that the SYM theories for $3<p<6$ are connected in the UV
to theories describing the dynamics of the NS5 branes of string theory.
These connections will be tackled at slightly
greater length in Section~\ref{5branesec}.

\item For $6\le p\le 9$, the situation is more complicated. As we saw above,
the gravitational coupling does
not vanish indicating that the closed string sector does not decouple
from the open string dynamics. 
Except for $p=6$,
these cases will be omitted from the discussion in this thesis.
\end{itemize}

Consider exciting a gas of quanta in these SYM theories. The variable $T$
in equation~\pref{geffeq} denotes then the temperature.
For $\geff\ll 1$, we have a weakly interacting gas of bosons and fermions.
As discussed in the Introduction, the
equation of state of the finite temperature vacuum in this regime is
given by
equation~\pref{Gaseos}.
Perturbation theory can be applied to refine this free gas picture.
As the temperature is tuned past $\geff\sim 1$,
our understanding of the dynamics via standard 
field theoretical means breaks down.
The physics of the theory beyond this point is then a mystery.
A mystery, that is, until the advent of Maldacena's conjecture.

\section{Comments on five brane theories}\label{5branesec}

We observed in the previous section that $p+1$
dimensional SYM theories for $3<p<6$ are non-renormalizable; \ie\ they are low
energy effective descriptions
of physics of possibly different character that is
well defined at short length scales. 
This UV physics can be unraveled by making use of some of the special
symmetries arising in string theories.
Particularly, it is known that certain duality transformations of string
theories shuffle D4 branes with M5 branes, D5 branes with NS5 branes,
and NS5 branes with M5 branes (see Appendix~\ref{dualapp}).
This suggests that $4+1$d and $5+1$d SYM theories may connect in the UV
with theories describing the dynamics of five-branes. These
theories
are not very well understood; we will try to
outline in this section a few facts known about them
~\cite{STROMOPEN,DVV5D,DVVM5,GANORMOTL,SEIBLITTLE}.

We remind the reader that the NS5 brane is a
five dimensional extended object that arises in string theories as the
magnetic dual of the F1 string. Its dynamics appears to be
fundamentally different from that of Dp branes. 
In particular, one does not have a picture of open strings propagating on the
NS5 brane surface describing ripples of excitations. There exists however
a less understood picture of closed strings living in the $5+1$d
world-volume of the NS5 brane; their dynamics is presumably describing
the fluctuations of the brane.
This ``little closed string theory''
is {\em not} gravitational and is necessarily a non-local theory.

There are two flavors of NS5 branes; the one in
IIA theory and the one in IIB theory, related to each other by T duality 
transformations. The NS5 brane of the IIA theory (NS5A) descends from
the M theory five brane, the magnetic dual of the M theory membrane.
The 6d string theory on its surface has $(2,0)$ supersymmetry
with sixteen supersymmetry generators. Compactifying
this theory on a substringy circle (\ie\ wrapping the NS5A on the
circle) yields $4+1$d SYM theory. The uncompactified theory
appears at low energies and for well-separated branes as
a $5+1$d field theory of free massless fields forming a tensor
multiplet of supersymmetry.
The bosonic field content of this theory is given by five scalars and an
anti-selfdual two-form. 
The tension of the little strings is set by the tension of a membrane
wrapping the M theory circle and ending on the five brane.
We will see that it is held fixed in the Maldacena limit, 
signaling that the infinite tower of little string vibrational modes
does not decouple in the energy regime of interest; 
this signals that we will be dealing with
non-local theories
\footnote{
The non-locality is of spatial extent set by the length scale
determined by the little string tension.
}
arising in the UV of certain local SYM theories.
The NS5 brane of the IIB theory (NS5B) is a 6d $(1,1)$ conformal field theory
related to the $(2,0)$ theory by an odd number of T dualities. At low
energies, it is described by $5+1$d SYM theory.

\begin{figure}
\epsfxsize=10cm \centerline{\leavevmode \epsfbox{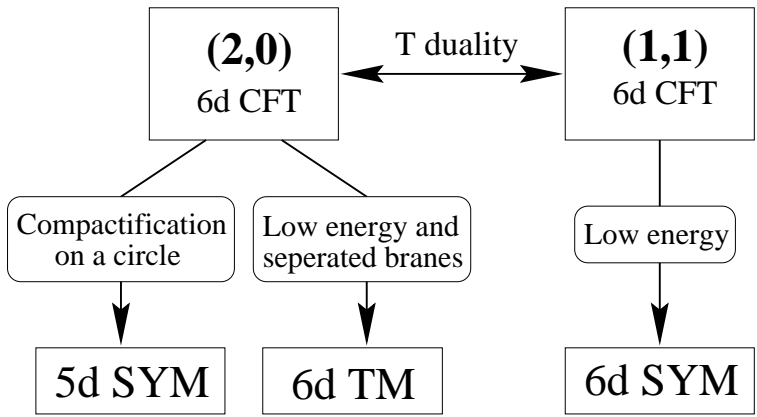}}
\caption{\sl Five brane theories and their connections to various
quantum field theories; TM stands for Tensor Multiplet
.}
\label{5branefig}
\end{figure}

Figure~\ref{5branefig} is a schematic diagram summarizing our comments
in this section.
When we will be studying 
the thermodynamics of $4+1$d and $5+1$d SYM theories, it will be 
appropriate
to label the discussion as an analysis
of the thermodynamics of five brane theories.

\section{Maldacena's conjecture}\label{Maldaconjsec}

In the previous sections, we outlined 
two different descriptions for a configuration of excited Dp branes.
One is through gravity, a picture consisting of a space-time
curved by the energy
content of the branes; the other is through the finite
temperature vacuum of a non-gravitational
SYM or little string theory. In the Maldacena limit, which defines 
a certain low energy regime,
the latter picture is 
in focus. There are two questions that arise after these observations.
First, what does Maldacena's energy regime correspond to in the
geometrical picture? And second, how does this energy regime relate to the
one encountered on the geometry side in equations~\pref{DpEOS},
~\pref{couplingp} and ~\pref{curvaturep}?
 
Consider perturbations propagating in the {\em extremal} background geometry
given by the metric~\pref{Dpmetric} with $r_0=0$. We may for 
example consider excitations in the supergravity fields, or stringy 
modes beyond the supergravity multiplet. The background 
geometry typically enforces a dispersion relation relating
the energy of the perturbation as measured by an observer at infinity
and the region of the background geometry where the field of the perturbation
is predominantly supported.
Consider for instance exciting the extremal state
by knocking out one of the $N$ Dp branes of the bound
system away from $r=0$. A measure of the energy of this excitation 
can be obtained by taking a snapshot of projected Dp brane at its 
maximum radial extent $r$ from the rest of the system. In the
SYM, this corresponds to giving a pronounced vev to a diagonal entry 
in the scalar matrices. On the geometry side,
we have a string radially stretched from the extremal zero area horizon at
$r=0$ to a coordinate distance $r$ away. Using the
metric~\pref{Dpmetric}, one finds that the energy of this stretched string
as measured at infinity is proportional to $r/\alp$. If we identify this
with the energy of the excitation in the field theory, we see that
that lower energy excitations tend to sit closer to the center 
of the geometry. Crudely put, the higher the energy of a perturbation of
the extremal state, the farther is the extent in the transverse space of the
``jet'' or ripple produced on the Dp branes. 
The Maldacena energy regime~\pref{Maldalimit} then
corresponds to focusing on physics near the core of the Dp brane geometry.
Equation~\pref{Maldalimit} essentially establishes
a UV cutoff on the open string theory, and corresponds to
restricting dynamics in the closed string sector 
to the background geometry of {\em the near horizon
region} of the extremal Dp brane solution. This close-up geometry of
the horizon that corresponds to the Maldacena energy regime~\pref{Maldalimit}
is easily obtained from the metric~\pref{Dpmetric} by dropping the
$1$ in the harmonic function $H$ of~\pref{harmH}.
Having made these observations,
Maldacena proposed the following~\cite{MALDA1,ADSLECT}:

\vspace{0.5cm}
{\bf Maldacena's conjecture\footnote{More general versions of this
statement have been proposed and studied; in this thesis, we confine our
discussion to the Dp brane case exclusively.}:}
{\em String theory in the background geometry of the near horizon region of
the extremal Dp branes is encoded in the corresponding SYM theory.}
\vspace{0.5cm}

By this it is meant that the partition functions given by the two pictures are
equal. 
The conjecture suggests a one to one correspondence between excitations of
the background geometry and excitations in the SYM quantum field theory.
This is essentially an equivalence between closed string theory
(the gravitational side)
and open string theory (the SYM side). Note that
one is to truncate
the open string dynamics to the ground states, while one is to keep
the infinite tower of closed string excitations
propagating in the Dp brane background. This is because,
in the Maldacena limit, the ratio of string scale
proper energy of a perturbation
in the bulk space to its energy measured at infinity (\ie\ the
energy in the open string sector) is finite, \ie\ independent of $\ls$.

We focus on perturbations of the extremal background that form 
metastable configurations corresponding
to the black solutions (\ie\ $r_0\neq 0$) of equation~\pref{Dpmetric}
~\cite{MALDA2}.
This amounts to exciting
the extremal Dp branes to finite temperature, and therefore
corresponds to studying a certain thermodynamic phase of the SYM theory.
Maldacena's limit zooms onto physics in the
near horizon region $r=r_0$. Our previous picture of
a Dp brane projected from the core as describing an excitation of the
extremal state suggests the following:
in this finite temperature scenario, the horizon $r=r_0$ may denote,
roughly, the extent in the transverse space
of a halo due to a gas of ripples on the surface
of the Dp branes. 
This observation may indicate 
that the standard classical procedure to analytically continue
the black geometry smoothly past the horizon 
is not necessarily well justified.
The equation of state of this phase of the
SYM theory is given by equation~\pref{DpEOS}
\bb\label{DpEOS2}
E\sim \lk( \frac{S^2}{N} \re)^{\frac{7-p}{9-p}}
\lk( 
\Sigma^{p (p-5)} (\gym^2)^{3-p}
\re)^{\frac{1}{9-p}}
\ ,
\ee
where we have written it in terms of the field theory
parameters. We see that, in the Maldacena energy regime~\pref{Maldalimit}, 
this energy scale is kept in focus. Equation~\pref{DpEOS2} can be trusted 
for a window of entropies such
that the curvature scale and the dilaton measured at the horizon are
small, as we saw in equations~\pref{couplingp} and~\pref{curvaturep}.
In the field theory variables, these become respectively
\bb\label{couplingp2}
\lk( N^{8-p} S^{p-7}\re)^{3-p} 
\lk(
\frac{\Sigma^{p-3}}{\gym^2}
\re)^{p(p-7)}
< 1\ ,
\ee
and 
\bb\label{curvaturep2}
\gym^{2p} \Sigma^{p(3-p)}
N^{6-p} > S^{3-p}\Rightarrow \geff>1\ .
\ee
We see that this entropy/energy window
is a subset of Maldacena's energy regime, and that the geometrical 
phase arises in a non-perturbative regime of the SYM theory.
Within this window, the equation of state of the
black geometry then describes a certain phase of the SYM theory.
The conjecture implies that the
analysis of the stability of such black geometries can be
translated to an analysis of critical phenomema in the SYM theory.

We thus have a powerful geometrical tool to explore the thermodynamics
of SYM or little string theories beyond the confines of
pertubration theory. We will
make use of it to systematically
map out the thermodynamic phase diagrams of these theories.

\section{The Matrix conjecture}\label{matsec}

Prior to Maldacena's conjecture, another proposal 
suggested a correspondence between gravitational and SYM theories.
The Matrix conjecture~\cite{MAT1,MAT2} proposed that 
Discrete Light Cone Quantized
\footnote{The DLCQ of a theory is obtained by compactifying the theory
on a light-like circle. We single out the combination 
$x^\pm = (x\pm t)/\sqrt{2}$ and treat $x^+$ as the new time variable.
Its canonical variable $p^-=p_+\equiv E_{LC}$ is called 
{\em the Light-Cone Hamiltonian}. The coordinate $x^-=x_+$ is compactified
on a circle of size $2\pi R_+$, and {\em the longitudinal momentum} $p_-$
is quantized in units of $1/R_+$. The relativistic dispersion
relation $E^2-p^2 -\vec{{\bf p}}^2=M^2$ then takes the non-relativistic form
\bb\label{ELC}
E_{LC} = \frac{M^2+\vec{{\bf p}}^2}{2p_-}\ .
\ee
Quanta propagating in this frame carry only positive longitudinal 
momentum $p_-$. This simplifies the physics by restricting
the number of degrees of freedom in a given sector of total
longitudinal momentum to a finite number.

It was argued in~\cite{WHYSEIB} that the DLCQ frame can be reached by a
combination of an infinite boost and compactification on a
vanishingly small cycle. The longitudinal momentum 
$p_{11}=N/\r11 \rightarrow \infty$ (with $N$ fixed)
is made of order the energy $E$ in
the boosted frame yielding the dispersion relation
\bb
(E+p_{11})(E-p_{11})-\vec{{\bf p}}^2=M^2
\sim 2p_{11} (E-p_{11})-\vec{{\bf p}}^2\equiv 2p_{11} E_{IMF}-
\vec{{\bf p}}^2\ ,
\ee
where $E_{IMF}\rightarrow 0$ with $E_{IMF}\ p_{11}$ held fixed.
One then identifies $p_{11}$ with $p_-$, and $E_{IMF}$ with $E_{LC}$
yielding equation~\pref{ELC}. Given this mapping, the reader is
warned that, in the upcoming chapters,
we freely mix the use of the infinitely
boosted picture with that of the DLCQ.
}
(DLCQ) M theory
on a $p$ dimensional torus is encoded in $p+1$d SYM theory.
Our thermodynamic analysis will lead to a better understanding
of this statement; we will conclude that the content of
this conjecture is a subset of Maldacena's proposal. 
We here briefly review 
the Matrix conjecture for future reference. The casual reader may
quickely browse through this section only to get accustomed to some of the
nomenclature we will make use of later.

A convenient way to summarize the Matrix theory conjecture
is to say that DLCQ 
M-theory on $T^p$ with $N$ units of
longitudinal momentum is a particular regime
of an auxiliary `$\bM$-theory' which freezes
the dynamics onto a subsector of that theory.
Consider such an $\bM$-theory, with eleven-dimensional
Planck scale $\blp$ (which we denote ($\bM$,$\blp$))
on a $p+1$d dimensional torus of radii ${\bar{R}}_i$, $i=1\ldots p$, and
$\bar{R}$ the `M-theory circle' of reduction to type $\overline{\rm IIA}$
string theory, in the limiting regime
\bb\label{limitlp}
\blp\rightarrow 0,\ \ \ 
\mbox{with } x\equiv \frac{\blp^2}{\bar{R}}\ \ \ \mbox{and }
y_i\equiv \frac{\blp}{\bar{R}_i}\ \ \ \mbox{fixed} ,
\ee
and $N$ units of momentum along $\bar{R}$. 
It is proposed that~\cite{MAT1,MAT2}
\begin{itemize}
\item 
This theory is equivalent to an ($M$,$\lp$) 
theory on the DLCQ background
we denote by $D^{1,1}\times T^p \times \R^{9-p}$, 
where $D^{1,1}$ is a $1+1$ dimensional subspace compactified
on a lightlike circle of radius $R_+$, and the torus $T^p$ has
radii $R_i$ ($i=1\ldots p$).
The map between the two theories is given by
\bb \label{map}
x=\frac{\lp^2}{R_+},\ \ \ 
y_i=\frac{\lp}{R_i} ,
\ee
with $N$ units of momentum along $R_+$.
\item
The dynamics of
the $\bM$ theory in the above limit can be described
by a subset of its degrees of freedom, that of
$N$ D0 branes of the $\overline{\rm IIA}$ theory.

\end{itemize}

The two propositions above, in conjunction, 
are referred to as the Matrix conjecture
\footnote{
Note that the statement of the conjecture, phrased as we have 
in the text above,
must restrict $p$ to $p\le 3$. 
For $3<p<6$, the Matrix conjecture proposes that DLCQ M theory is described
by the theory of the little strings living on wrapped five branes.
All these issues are best understood in a unified framework
through Maldacena's
conjecture; this is the approach we adopt in this thesis. 
}~\cite{MAT1,MAT2}.

T-dualizing on the $\bar{R}_i$'s, we describe the D0 brane physics 
by the $p+1$d SYM of N D$p$ branes
wrapped on the dualized torus.
The dictionary needed in this process is
\bbb 
\bar{R}=&~\bar{g}_s' \ls\ \ \ 
\blp^3=&\bar{g}_s' \ls^3\ ,\nonumber\\
\bar{g}_s=&~\bar{g}_s' \frac{\bls^p}{\Pi \bar{R}_i}\ \ \ 
\Sigma_i=&\frac{\bls^2}{\bar{R}_i}\ .\label{dictionary1}
\eee
The first line is the $\bM-\overline{\rm IIA}$ 
relation, the second that of T-duality.
The limit~\pref{limitlp} then translates in the new variables to
\bb \label{limitym}
  \bar{\alpha}'\rightarrow 0\quad, \quad \quad 
  \mbox{with }\ \  g_Y^2=(2 \pi)^{p-2} 
	\bar{g}_s {{\bar\alpha}}^{\prime \frac{p-3}{2}}\ \ \ 
  \mbox{and } \Sigma_i\ \ \ \mbox{fixed}\ ,
\ee
where the nomenclature 
$g_Y^2$ and $\Sigma_i$ refers to the coupling and
radii of the corresponding $p+1$d
$U(N)$ SYM theory. This is simply the Maldacena limit~\pref{Maldalimit}.

\chapter{A simple phase diagram}

We concluded in the previous chapter that the thermodynamic
state described by the Dp brane geometry given by~\pref{Dpmetric} with
equation of state~\pref{DpEOS} corresponds to a certain phase
in the thermodynamic phase diagram of $p+1$d SYM. This is obviously
a different phase than the one accesible by perturbation theory, whose
equation of state scales as~\pref{Gaseos}. In this chapter,
we focus on the case of D0 branes ($p=0$), and we investigate
the transition between the perturbative and geometrical phases. In the process, 
we will map the thermodynamic phase diagram of the $0+1$d SYM theory
well into non-perturbative regimes. We will present a relatively
detailed analysis since this simple case can be used to illustrate
some of the basic ideas involved in the more elaborate thermodynamic phase
diagrams we will encounter in the next chapter.

\section{Thermodynamics of D0 branes}

We are considering $0+1$d SYM theory which describes the dynamics of
D0 branes in IIA string theory. We take the number of D0 branes $N$
to be much greater than one.
The effective Yang-Mills coupling is given by equation~\pref{geffeq};
using equation~\pref{DpEOS2} and $E\sim T S$ with $p$ set to zero, this
yields
\bb\label{D0geff}
\geff^2\sim \lk(\frac{N^2}{S}\re)^{5/3}\ ,
\ee
At entropies much greater than $N^2$, we see that the system
is in a weakly coupled phase.
We have a quantum mechanical system with
$N^2$ weakly interacting particles propagating in an {\em infinite}
volume; this is because we have not restricted the SYM scalars,
the coordinates of the particles, to live within a confined region.
As such, the stability of such a phase is at issue; 
the physics of this phase will be discussed in greater detail below.
As the interactions become strong at $S\sim N^2$,
the dynamics becomes more interesting; we expect to form a ball of
strongly interacting particles that are
confined to a region of space by virtue of these strong
interactions. Putting the system in a box is not necessary anymore if we
are content to describe metastable phases that evaporate slowly.

We lower the entropy of our system past $S\sim N^2$, 
venturing into non-perturbative dynamics $\geff^2\gg 1$. 
Consider the near horizon geometry of $N$ excited D0 branes given
by~\pref{Dpmetric} with $p=0$
\bb\label{D0metric}
ds_{10}^2= -H^{-1/2} h dt^2+H^{1/2} \lk( h^{-1} dr^2 + r^2 d\Omega_8^2\re)\ ,
\ee
\bb
H\sim \gs \ls^7\frac{N}{r^7}\ .
\ee
$r_0$ is related to the entropy by the Hawking area law
\bb
r_0^9\sim \gs^3 \ls^9 \frac{S^2}{N}\ .
\ee
The equation of state is given by~\pref{DpEOS} with $p=0$
\bb\label{D0EOS}
E \sim \frac{\gs^{1/3}}{\ls} \lk(\frac{S^2}{N}\re)^{7/9}\ .
\ee
Maldacena's conjecture states that this is a phase in $0+1$d SYM quantum
mechanics.
It is a long-lived, metastable thermodynamic phase
that evaporates slowly by Hawking radiation.
Let us next look closer to the criteria making this geometrical
picture a physically consistent one.
Its curvature scale is set by
the angular part of the metric. We require that this scale, as measured
at the horizon $r_0$, be less than $1/\alp$. 
This yields the condition
\bb
\lk. H^{1/2} r^2 \re|_{r=r_0} > \alp \Rightarrow S<N^2\ .
\ee
We see that the geometrical phase complements the perturbative regime;
for $S>N^2$, we have the phase described by perturbation theory
and for $S<N^2$, strongly coupled
dynamics forces the system into another phase with equation of state
~\pref{D0EOS}. 
The point $S\sim N^2$ is most likely a phase
transition line, but our analysis is crude enough not to allow us to 
study the details of this transition. Our entire discussion
in this thesis will suffer from this deficiency. We will be able to track
the bulk phase structure of the phase diagrams and the rough scaling
of the critical curves. We will make no attempt at going into a more
detailed analysis of these critical phenomena.

The next task is to determine 
what happens to the strongly coupled phase as we decrease the entropy
further. To answer this, we need to look at the dilaton 
\bb
e^\phi = H^{3/4}\ .
\ee
As discussed in the
previous chapter, consistency of the geometrical description requires
us to assure that $\lk. \gs e^\phi\re|_{r=r_0} \ll 1$; 
\ie\ perturbative string theory
was assumed in deriving the low energy supergravity action.
This yields the condition
\bb
S> N^{8/7}\ .
\ee
For lower entropies, the size of the eleventh dimension of M theory 
is of super-Planckian size 
(\cf\ equation~\pref{MIIA}). The full eleven dimensional
nature of the underlying dynamics must be taken into account.
The connection between M and IIA theories 
at the low energy supergravity level is simply through the prescription
of dimensional reduction that is commonly used
in the compactification of supergravity theories. A few technical details
can be found in Appendix~\ref{dualapp}.
The eleven dimensional metric
from which our IIA metric~\pref{D0metric} descends is found to be
\bb
ds_{11}^2=H (dx_{11}-dt)^2+dx_{11}^2-dt^2
+H^{-1} (1-h) dt^2+h^{-1} dr^2 + r^2 d\Omega_8^2\ .
\ee
Here $x_{11}$ lives on a circle of size $2\pi R_{11}$, and the
M theory Planck scale is denoted by $\lp$.
This is simply a black wave propagating along $x_{11}$ (we
label it W11 for future reference). D0 branes of
IIA theory map onto gravitational waves in M theory. The equation
of state is unchanged; we are describing the same physics using
a new setting whose low energy dynamics we can trust beyond
the perturbative string theory regime.

\begin{figure}[t]
\epsfxsize=10cm \centerline{\leavevmode \epsfbox{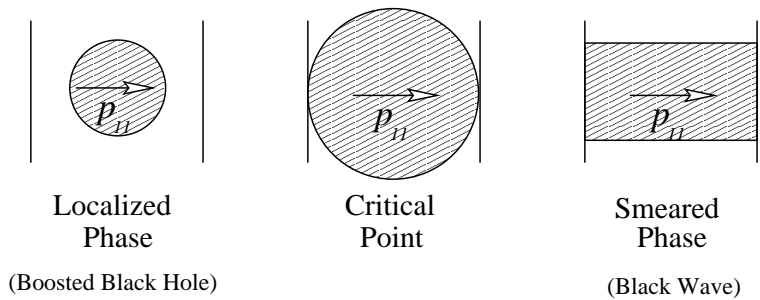}}
\caption{\sl The transition between a localized boosted black hole
and a black wave. The horizontal axis in these pictures
is the M theory cycle of size $\r11$
.}
\label{locfig}
\end{figure}

The question now becomes what happens to this eleven
dimensional geometry as we lower the entropy further.
This is answered by studying the stability of the black
wave geometry.
It is known
that a black wave in a box is unstable toward collapse into 
a boosted black hole localized in the box; 
the transition occurs at the point where
the boosted black hole has less free energy. As we lower the entropy, we cool
the black wave enough that the 
preferred configuration is one which is localized in the box
along the direction of wave propagation. Alternatively,
a boosted black hole in a box can be said to smear itself into a black 
wave geometry as the entropy is raised past the point where the size
of the horizon is the size of the box. This process is sketched
in Figure~\ref{locfig}. 

\begin{figure}[t]
\epsfxsize=10cm \centerline{\leavevmode \epsfbox{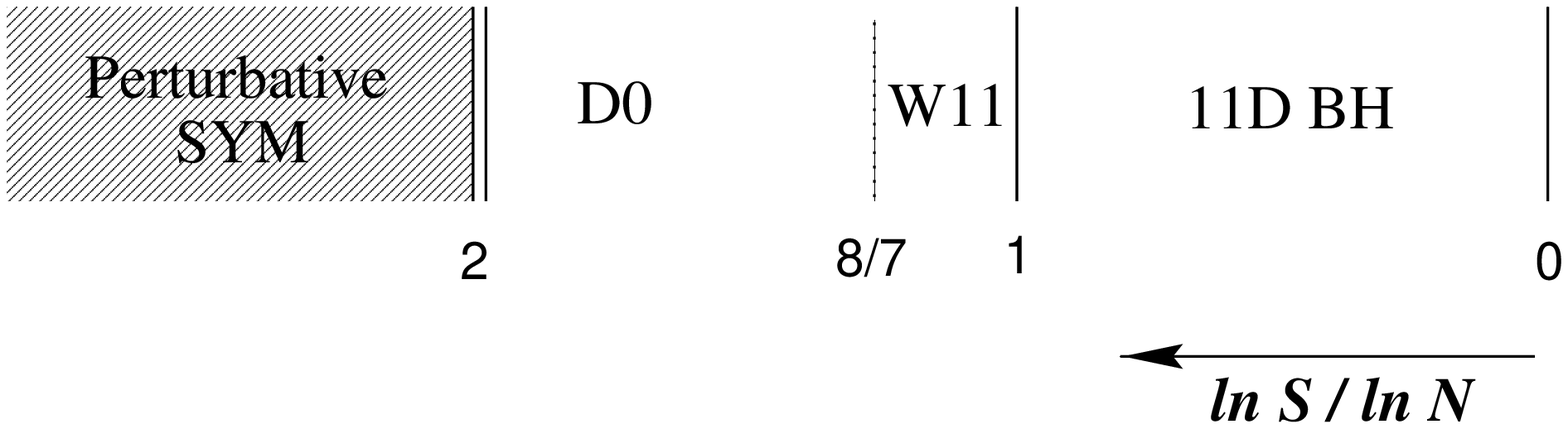}}
\caption{\sl The thermodynamic phase diagram of $0+1$d SYM. 
The dotted line represents the lift to M theory. We have a total
of three phases: a perturbative quantum mechanics phase, a phase
of black D0 branes (comprised of two geometrical patches),
and a phase of a boosted eleven dimensional black hole
.}
\label{D0phasefig}
\end{figure}

The boosted black hole geometry is obtained by
applying the appropriate Lorentz 
boost transformation on a Schwarzchild black hole (see Appendix~\ref{grapp}).
The equation of state of a black hole of mass $M$ boosted to a large
momentum $p_{11}=N/\r11=N/(\gs \ls)$ is given by
\bb\label{lbheos}
E\sim \frac{\gs^{1/3}}{\ls} \frac{S^{16/9}}{N}\ ,
\ee
following the comments in the footnote of Section~\ref{matsec}.
Note that we are identifying the SYM Hamiltonian with the Light Cone
energy in M theory.
We can find the transition point between this phase and the black wave phase
by either minimizing the Gibbs energies between~\pref{lbheos}
and~\pref{D0EOS}\footnote{We are only tracking the scaling of the transition
curves; distinction between free, Gibbs or internal energy is not
necessary for such purposes.}, 
or by equating the size of the box $x_{11}$ as measured
at the horizon to the size of the horizon.
This yields the point of transition at
\bb
S\sim N\ .
\ee
The conclusion is that, at $S\sim N$, there is a transition between
the phase described by the equation of state~\pref{D0EOS}, and the
phase of a boosted black hole described by the equation of state~\pref{lbheos}.
For $S<N$, the $0+1$d SYM theory is in a thermodynamic state which is
a boosted black hole, with the boost momentum related to the rank of
the gauge group. This phase, unlike that of a stationary black hole,
has positive specific heat.
Figure~\ref{D0phasefig} depicts graphically the phase 
diagram we have been constructing.

This simple example illustrates the basic idea we will make use of
in the next chapter in
mapping the thermodynamic phase diagrams of more elaborate cases.
We use Maldacena's conjecture to identify phases in SYM theories
which have strongly coupled dynamics
with Dp brane geometries; we analyze the stability and validity
of these geometrical phases, using the duality symmetries
if needed, and import the conclusions to SYM
thermodynamics. We will be able to systematically
chart in this manner the phase diagrams of SYM and little string theories
on tori. 

A few more comments regarding the role of the vevs of the scalars
in the dynamics 
are in order. We saw that the perturbative $0+1$d SYM phase consists of
a gas of $N^2$ weakly coupled particles in infinite volume. As such,
its thermodynamics is well defined provided that the 
kinetic energy content of the zero modes of
this gas is comparable to the potential
energy content; we then can expect that a long lived metastable phase
exists, a ball of a gas confined to a region of space by virtue
of the weak interactions between the constituents. The existence 
of a potential strong enough to bind the zero modes is crucial
in the $0+1$d case since the only dynamics available 
is one that can probe the infinite extent of the space.
Whether a stable or metastable phase sits on the left of Figure~\ref{D0phasefig}
is then in principle left as an open issue. We focus instead on the
$p+1$d SYM theories with $p\neq 0$, where this problem
will be considerably milder. 
These cases involve the physics of Dp 
branes {\em wrapped on a torus}. 
The difference between the two scenarios
is analoguous to the following more mundane situations: on the
one hand, consider a gas of a certain number of particles propagating on the
infinite two dimensional plane; on the other hand, consider a 
vibrating rubber band wrapping a cylinder. The thermodynamics of the
former case is similar to the $0+1$d SYM case we discussed.
In the rubber band scenario however, the physics involves in addition
the thermodynamics of the vibrational modes, quanta propagating along
the band (modding out the dynamics of the
center of mass of the rubber band, which readily explores the longitudinal
infinite extent of the cylinder). The contribution to the free 
energy of this sector of the dynamics dominates the thermodynamics; 
by increasing the temperature in the $p+1$d SYM
case with $p\neq 0$, we will see that one
transfers dynamics to the fluctuations on the surface
of the wrapped Dp branes; this phenomena will happen at $S\sim N^2$.
The perturbative $p+1$d SYM phases
for $p\neq 0$ will be better defined in this sense; for scaling purposes,
they can be readily approximated by
systems of free particles living on a torus.

Typically,
phases involving weakly interacting dynamics, such as the perturbative
SYM gas or, more interestingly, the
Matrix string, abut
black geometrical phases on our phase diagrams. We expect
that the vevs of the scalars in the weakly coupled phases
explore the
extent of the non-compact space for arbitrarily high temperatures. It is
not obvious that, starting from such a phase and moving toward
strongly coupled dynamics, we are guaranteed to collapse the
system into one of the black geometries we encounter along the
traced path. The system may have spread itself into large volumes of
the transverse space (or an infinite volume in the pathological 
$0+1$d SYM case) such that the interactions, which generically
fall in strength with distance, are effectively too weak to lead to a 
collapse at the point identified on the phase diagram.
This phenomena will be explored in detail in Chapter~\ref{bhformchap}. 
Our phase diagrams are faithful representation of the
physics if we trace paths starting from black objects
and moving 
toward weakly coupled phases; such paths are not necessary reversible. 
In Chapter~\ref{bhformchap}, we map out a thermodynamic phase diagram which can
be navigated with reversible processes. 
Note that this physics is again intimately tied with our choice
not to restrict the scalar vevs to a finite region of space.

Before we conclude this chapter, let us reflect on the boosted M theory
phases (the wave W11 and the black hole) that arose at low entropies.
Using the M-IIA relations~\pref{MIIA},
we cast the Maldacena limit in the parameters of this M theory;
this yields
\bb
\lp\rightarrow 0\ \ \mbox{with}\ \ \frac{\lp^2}{\r11}\ \ \mbox{held fixed.}
\ee
We see that this is identical to the energy regime of the Matrix
conjecture~\pref{limitlp}; \ie\ the Matrix regime is a statement dual to the
Maldacena limit, where the duality at play here is the M theory-IIA 
connection. Another map between the two limits
(involving a T duality transformation) was given
in the case of $p+1$d SYM with
$p\neq 0$ in Section~\ref{matsec}.
We will see how this relation gets embedded in the phase diagrams of the next
Chapter. The conclusion is the same:
the Maldacena and Matrix energy regimes are dual to each other.
The low entropy phases of our diagrams live in DLCQ M theory.
The charge $N$ carried by the D0 branes
is mapped onto the momentum in the Light-Cone direction. In this sense,
Maldacena's conjecture leads to the Matrix proposal.

Given the elaborate web of dualities between the various string theories that
connect vacua of the underlying theory, a conjecture
relating a field theory to string theory in a given background vacuum
is naturally
extended to a statement relating this field theory to string/M theory on any
dual background. Given that one typically needs to accord 
a window of energy or of
the thermodynamic parameter space to a given background vacuum, 
the field theory encodes naturally string/M theory in different backgrounds
for different regimes of energy or temperature.
A thermodynamic phase diagram constructed in this manner
paints a global picture of the thermodynamic vacuum of a single
theory; the role of the web of dualities in patching various
string theories into one big theory metamorphoses into the role
of combining various thermodynamical vacua of a single SYM theory into
a single phase diagram. We see in the $0+1$d SYM phase diagram 
a very simple realization of these ideas.

\section{The Matrix black hole}\label{mbsec}

We saw in the previous sections that a boosted black hole, or
a black hole in DLCQ M theory, is a thermodynamic phase of $0+1$d SYM
theory. We will find it a generic phase arising at low entropies in
$p+1$d SYM theories with $p\le 6$. The question then arises as to how
this phase is realized as excitations of the SYM fields at fixed temperature
or entropy.
A model for this was proposed in~\cite{HORMART,BFKS3,LIMART} 
which we now summarize.

At low entropies $S<N$, consider the scenario where
the SYM dynamics is dominated by that of the zero modes
of the fields with
the effective Yang-Mills coupling being large. Assuming that the
diagonal elements of the scalars $X^i$ do not
take degenerate values, the off-diagonal modes are massive with the
mass scale tuned by the Yang-Mills coupling. For large coupling, these
degrees of freedom can be integrated out to yield an effective low
energy potential between the diagonal elements. The latter represent,
as discussed earlier, the coordinates of $N$ D0 branes, each with
mass $1/\r11$ (\cf\ equations~\pref{MIIA} and~\pref{TDp}). So, we
have a gas of $N$ D0 branes interacting with some effective potential.
It is proposed that, for $S<N$,
there is a stable thermodynamic state in the SYM that can be modelled
as a gas of $K$ clusters of D0 branes, each with $N/K$ D0 branes.
It is assumed that the physics of this model
is dominated by the center of mass dynamics of these clusters of D0 branes.
The clustering phenomena of the D0 branes into nuclei with $N/K$ partons is
in principle an unknown, an assumption that will be justified
in the analysis of Chapter~\ref{bhformchap}.
The scaling of the kinetic energy content of this gas 
can be read off the Langangian~\ref{SYMaction}
\bb\label{kinterm}
\KK\sim K M_{clst} v^2\sim K \lk(\frac{N}{K\r11} \re) v^2\ ,
\ee
where $v$ is the velocity of a cluster. 
The potential energy content of the gas due to the interactions
between two clusters is found to be
\bb\label{potterm}
\VV\sim \frac{\kappa_{11-p}^2}{\r11} 
\frac{N}{K} \frac{\lk(M_{clst} v^2\re)^2}{r^{7-p}}
\sim \frac{\lp^9}{\r11 R^p} \lk(\frac{N}{K}\re) \lk( \frac{N}{K\r11} v^2\re)^2
\frac{1}{r^{7-p}}\ ,
\ee
where $r$ is the typical seperation between the clusters,
or the size of the system. We are considering
a boosted black hole in M theory on a
$p$ dimensional torus $T^p$ of equal cycle sizes $R$; this changes the
gravitational coupling and the power of the cluster seperation distance $r$
as shown.
We have also enhanced the interaction
by a factor $N/K$. At this stage, this is
an assumption; it will be justified in Chapter~\ref{bhformchap} 
and Appendix~\ref{stramplapp}. 

We now make use of three additional ingredients. 
We assume that the clusters are distinguishable 
from each other. We then have the entropy of the gas scaling as $S\sim K$.
We assume that the wavefunctions of the clusters 
saturate the quantum mechanical uncertainty bound
\bb\label{uncertainty}
p\ r \sim 1 \Rightarrow \frac{N}{K\r11} v\  r\sim 1
\ee
Furthermore, the virial theorem dictates
\bb\label{virial}
\KK\sim S^2 \VV\Rightarrow r^{9-p}\sim \frac{\lp^9}{R^p} S\ .
\ee
Puting things together in equation~\pref{kinterm} or~\pref{potterm}, 
we get that the total energy for this gas scales as
\bb
E\sim \frac{\r11}{N} \lk(S^{\frac{8-p}{9-p}} \lk( 
\kappa_{11-p}^2 \re)^{\frac{1}{p-9}}\re)^2 \sim \frac{M^2}{p_{11}}\ ,
\ee
where we have assumed in the last step that the SYM Hamiltonian is
the Light-Cone energy as in equation~\pref{ELC}.
The mass $M$ obtained in this equation,
up to numerical coefficients, is that
of a Schwarzchild black hole in an eleven dimensional space-time 
compactified on a torus $T^p$ of size $R$.
The proposed model then reproduces black hole physics from SYM theory.
Various assumptions made in constructing this model will be
justified in the analysis of Chapter~\ref{bhformchap}.

\section{The Matrix string}\label{matstsec}

Another phase that we will encounter in the various
phase diagrams is that of the Matrix string. This is a highly
excited string in DLCQ string theory with entropy $S\gg 1$. 
Its equation of state has
the Light-Cone structure~\pref{ELC}
\bb\label{ELCstring}
E=\frac{M^2}{2 p_+}\ ,
\ee
where $M^2$ is the mass of a free string at a level of degeneracy $e^S$
\bb
M^2\sim \frac{S^2}{\alp}\ .
\ee
This phase arises, for example, in the IR of $1+1$d SYM theory where
a free conformal
field theory is believed to sit. An explicit realization of the free
string as excitations in this theory was given~\cite{MOTL,DVV}. We
summarize briefly this proposal.

In the IR, the relevant Yang-Mills coupling is large; this makes the
commutator terms in the action~\pref{SYMaction} costly in energy. Configurations
where all matrix fields are diagonal are energetically favoured. Consider
diagonal matrix configurations for the scalars $X^i$; we ignore the
gauge fields on the world-sheet, even though these have interesting
roles to play in the dynamics and the
spectrum~\cite{WITTENBOUND,GURALRAM,GANORRAMTAYLOR,SFIRST}. 
Gauge transformations
that permute the diagonal elements form a symmetry
of this setup. In general, we need to consider boundary conditions on
the scalar matrices allowing for such discrete
gauge transformations
\bb
X^i(y+\Sigma)= g X^i(y) g^{-1}\ ,
\ee
where $g$ is an $N\times N$ matrix permuting the diagonal elements. 
All such matrices $g$ can be decomposed into transformations involving 
the operation of cycling 
acting on subsets of the diagonal elements. Without loss of generality,
we consider $g$ as a cyclic transformation in $Z_N$; the boundary
condition above then sews the various diagonal elements together
into one `long string' with world-sheet length $N \Sigma$, instead of the
periodicity $\Sigma$ that one obtains from the simpler condition
$X^i(y+\Sigma)= X^i(y)$. This is a basic technique known, for example, from
string theory on orbifolds. The picture depicts a IIA string encoded in
the IR of the $1+1$d SYM as a D string wrapping, like a `slinky', $N$
times a circle which is seen as the M cycle from which the IIA string
theory descends. Note that the $SO(8)$ symmetry of the scalar fields 
is precisely the corresponding symmetry of IIA string theory in the Light-Cone
frame. It is then easy to show that the quantized
Hamiltonian of~\pref{SYMaction} leads
to the free Light-Cone string spectrum~\pref{ELCstring} with $p_+=N/R_+$.

The appearance of a Matrix string with $S\gg 1$ in our phase diagrams
implies that there exists rich dynamics that seeds a
certain ordering, a $Z_N$ holonomy, in the
SYM excitations as we navigate past certain critical curves. 
The closest analogy may be a gas-solid phase transition,
where the symmetries of the emerging lattice encode
characteristics of the underlying microscopic dynamics.
The picture of the Matrix string we outlined here will be an integral
part of our discussion in Chapter~\ref{bhformchap}.

\chapter{More Thermodynamics}\label{morethermchap}

\section{Strategy}

In this chapter, we investigate 
phase diagrams of $p+1$d SYM theories on tori for $0<p<7$; we also
study the phase diagram for
a system of intersecting D1 and D5 branes. Unlike the
$p=0$ case, we will deal with finite size effects
resulting from the torus on which the thermodynamics lives.
In the D1D5 system, 
the role of angular momentum in criticality
will be one of the new ingredients.

The basic idea is
a systematic analysis of various 
black supergravity vacua; the underlying
strategy goes as follows:
A $10$D or lower-dimensional near-extremal supergravity 
solution must satisfy the following restrictions:
\begin{itemize}
\item The dilaton at the horizon must be small. Otherwise, in a IIA
theory, we need to lift to an $11$D M-theory; in a IIB theory, we need
to go to the S-dual geometry. 
This amounts to a change of description --
a reshuffling of the dominant degrees of freedom -- 
without any change in the equation of state. 
\item The curvature at the horizon must be smaller than the string scale.
Otherwise, the dynamics of massive string modes becomes relevant. 
By the Horowitz-Polchinski correspondence principle~\cite{CORR1},
as in equation~\pref{curvaturep}, a string theory
description emerges -- an excited string, 
or a perturbative SYM gas reflecting
weakly coupled D-brane dynamics. This is generally associated with a change
of the equation of state; in the thermodynamic limit, we may expect 
critical behavior associated with a phase transition.
This criterion can easily be estimated for various cases
when one realizes that the 
curvature scale is set by the horizon area divided by cycle sizes 
measured at the horizon; \ie\ the localized horizon area should be
greater than order one in string units.\footnote{In general,
the horizon will be localized in some dimensions and delocalized (stretched) 
in others.  The area of the `localized part of the horizon'
means the area along the dimensions in which the horizon is localized.}
\item Cycles of tori on which the geometry may be wrapped, as measured at
the horizon, must be greater than the string scale~\cite{RABIN}. 
Otherwise, light winding
modes become relevant and the T-dual vacuum describes the proper
physics~\cite{GIVEON}.  
We expect no critical behavior in the thermodynamic limit,
since the duality is merely a change of description.
\item The horizon size of the geometry must be smaller than the torus 
cycles as measured at the horizon~\cite{LAFLAMME1,LAFLAMME2}. 
Otherwise, the vacuum smeared
on the cycles is entropically favored. We expect this to be associated
with a phase transition, one due to finite size effects, and it is
associated generally with a change of the equation of state. However, it
is also possible that there is no such entropically favored transition
by virtue of the symmetry structure of a particular
smeared geometry, so we expect no change of phase.  Intuitively,
a system would only localize itself in a more symmetrical solution 
to minimize free energy.  We saw a manifestation of this
transition mechanism already in the D0 phase diagram, where the
black wave W11 smeared along the longitudinal direction localized
into a boosted black hole (see Figure~\ref{locfig}). In addition
to this longitudinal localization transition, this phenomena is now to
appear in the context of localization along the transverse compactified
dimensions.
\end{itemize}

\noindent
On the other hand,
given an $11$D supergravity vacuum, 
a somewhat different set of restrictions applies:
\begin{itemize}
\item The curvature near the horizon must be smaller than the Planck scale.
By the criterion outlined above, we see that, for unsmeared geometries,
this is simply the statement that $S> 1$; \ie\ quantum gravity effects are
relevant for low entropies. For large enough longitudinal momentum $N$, 
this region of the phase diagram is well away from the region
of interest.
This is one of the reasons for taking throughout this thesis 
$N\gg 1$.
\item The size of cycles of the torus as measured at the horizon must be greater
than the Planck scale. Otherwise, we need to go to the IIA solution descending
from dimensional reduction on a small cycle. We expect no change of 
equation of state or critical behavior.
\item The size of the M-theory cycles, including the 
Light-Cone longitudinal box,
as measured at the horizon, must be bigger than the horizon size. Otherwise,
the geometry gets smeared along the small cycles. This is expected to be
a phase transition due to finite size physics.
\end{itemize}

\noindent
Applying these criteria, we then select
the near extremal geometry dual to a phase in SYM theory on the
torus, and navigate the phase diagram via duality transformations
suggested by the various restrictions.
We will then be charting the phase
diagram of SYM quantum field theories or little string theories.
Alternatively, we are tracing through the 
phases visited by a DLCQ M theory black hole, the latter being a phase that
arises on all our diagrams for low entropies; 
the Schwarzchild black hole being a generic state in M theory,
we may propose that we are
mapping the thermodynamic phase diagram of DLCQ M theory.

We will avoid presenting the calculations that lead to
the construction of the upcoming phase diagrams. 
The details can be found in~\cite{MSSYM123,MSFIVE}.
The basic idea was
illustrated in the previous chapter. Further background material
needed to decode the physics of the
phase diagrams can be found in Appendix~\ref{dualapp}.
The details of the scaling of the various transition curves and the
equations of state of the bulk phases 
for all our phase diagrams are tabulated in Appendix~\ref{thedetailsapp}.
Note also that
the structure of all the diagrams can be checked by
minimizing the Gibbs energies between the various phases.

\section{Phase diagrams of Super Yang-Mills on tori}

\begin{figure}
\epsfxsize=9cm \centerline{\leavevmode \epsfbox{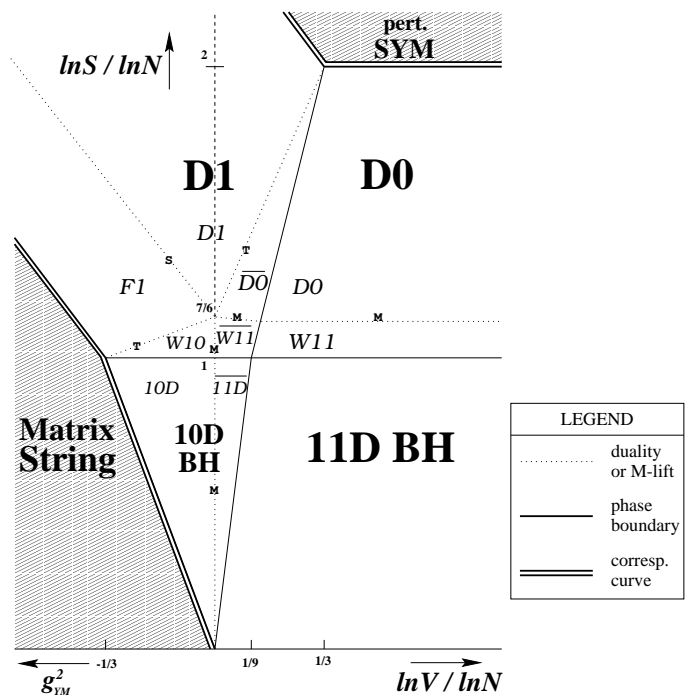}}
\caption{\sl Phase diagram of SYM theory on $T^1$;
$S$ is entropy, $V$ is the radius of the circle in Planck units, $N$ is
the longitudinal momentum. The geometry label dictionary is as follows:
D0: black D0; 
$\overline{D0}$: black D0 smeared on $V$;
D1: black D1; 
F1: black IIB string; 
W10: black IIA wave; 
W11: $11$D black wave; 
$\overline{W11}$: $11$D black wave smeared on $V$; 
$10$D BH: IIA Light-Cone black hole;
$11$D BH: Light-Cone M-theory black hole; 
$\overline{11D}$ BH: Light-Cone M-theory BH smeared on $V$. 
$M$, $T$ and $S$ stand for respectively an M-duality
(such as reduction, lift or M flip on $T^3$), a T-duality 
curve, and an S duality transition.}
\label{phase1}
\end{figure}

The phase diagrams for Dp branes on tori
have a number of common features.  
We focus on these aspects first; the reader is referred to, for example,
Figure~\ref{phase1} throughout the first part of this discussion.
The vertical axis of the diagrams will be entropy; for the
horizontal axis we take the size $V$ of cycles on the torus $T^p$,
in eleven dimensional Planck
units, as measured in the Light-Cone M-theory phase appearing in the
lower right corner (the phase of boosted 11d black holes).
$N$ is the charge carried by the system: brane number in
the high entropy regimes and longitudinal
momentum in the low-energy, Light-Cone M-theory phase.
The SYM torus radii $\Sigma$ and the effective SYM coupling for large $N$
can be written in terms of the parameters of the boosted black hole phase
\bb\label{torusgeff}
\Sigma=\frac{\lp^2}{\r11 V}\quad,\qquad g_{\rm eff}^2\sim 
\gym^2 N T^{p-3} \sim V^{-p} N \lk(\frac{T \lp^2}{\r11}\re)^{p-3}
\ee
($T\sim E/S$ is the temperature).
The unshaded areas
are described by various supergravity solutions, while the shaded
regions do not have dual geometrical descriptions. 
Throughout the various phases, the corresponding gravitational
couplings vanish in the Maldacena limit (except for $p=6$,
where the limit keeps the Planck scale of the high-entropy
phase held fixed), implying the decoupling
of gravity for the dual dynamics.
Solid lines on the diagrams denote thermodynamic transitions separating
distinct phases, while dotted lines represent
symmetry transformations which change the appropriate 
low-energy description. 
We do not expect sharp phase transitions along these dotted
curves since the scaling of 
the equations of state is unchanged across them.
This, does not in principle exclude 
the possibility of smoother (\ie\ higher order) transitions.

The structure of the phase diagram for $V>1$ is identical in all the cases
we encounter. 
At high entropies and large M theory torus, 
we have a perturbative p+1d SYM gas phase.  Its Yang-Mills
coupling $\gym$ increases toward the left.
The effective dimensionless coupling is of order one 
on the double lines bounding this phase, 
which are Horowitz-Polchinski correspondence curves.
As the entropy decreases at large $V$,
there is a D0 brane phase arising at
$S\sim N^2$ on the right and middle of the diagrams.
From the perturbative SYM side, 
this is where the thermal wavelength becomes of order
the size of the box dual to $V$,
$T_c\sim \Sigma^{-1}$; 
from the D0 phase side, it is a Horowitz-Polchinski correspondence curve. 
This transition may be associated with rich microscopic physics.
From the thermodynamic perspective, as the
transition is crossed, dynamics is transferred from
local excitations in $p+1$d SYM to that of its zero modes; 
and Dp brane charge of the perturbative SYM is
traded for longitudinal momentum charge of Light-Cone M-theory.
This process is one of several paths on the phase diagram
relating the Maldacena and Matrix conjectures.

The description of the D0 phase within strongly coupled
SYM theory would be highly interesting.
We see that this phase localizes into a Light-Cone 11d
black hole phase for entropies $S<N$.
This region of the phase diagram is then
a reproduction of Figure~\ref{D0phasefig};
\ie\ for $V\rightarrow \infty\Rightarrow \Sigma \rightarrow 0$, we connect
to the $0+1$d SYM physics.
The line $S\sim N$ separates the 11d phases that
are localized on the M-theory circle (whose coordinate size is $\r11$)
from those that are delocalized, uniformly across the diagram
~\cite{BFKS1,BFKS3,HORMART,BFKS2}.
The 11d black hole phase at small entropy
becomes smeared across the $T^p$ when
the horizon size becomes smaller than the torus scale $V$;
we denote generally such smeared phases by an overline
(in this case $\overline{11D}$).  This (de)localization
transition of the horizon on the compact space extends
above the $S\sim N$ transition, separating the black
Dp brane phase from the black $D0$ brane phase.
Initially, the $D0$ brane phase becomes smeared 
to $\overline{D0}$; as the entropy increases, the effective
geometry of the latter patch
becomes substringy at the horizon, and one should
T-dualize into the black Dp brane patch.  Both the
$\overline{D0}$ and $Dp$ patches have the same equation
of state, since they are related by a symmetry transformation
of the theory; they are different pieces of the same phase.
The localization transition line runs into the correspondence
curve separating the SYM gas phase from the geometrical
phases at $S\sim N^2$.  Thus
as we move to the left (decreasing $V$, \ie\ increasing bare
SYM coupling) at high entropy $S> N^2$, the 
SYM gas phase reaches a correspondence point;
on the other side of the transition is the phase of
black Dp-branes.

A further common feature of the diagrams is a `self-duality'
point at 
\bb
V\sim 1\ \ \mbox{and}\ \ S\sim N^{\frac{8-p}{7-p}}\ ,
\ee
where a number of U-duality curves meet.
That the various patches do not overlap
is a self-consistency check
on the logical structure of the picture.
Basically, there is always only one set of degrees of freedom that dictate
the low energy dynamics in a given window of the thermodynamic
parameter space.{\em
The duality transformation that patch the various string theories together
now mold various thermodynamical vacua into one phase diagram
for the SYM theory}.
We also note the following connections to previous work.
For high entropies, localization effects are circumvented and
the phases are the ones studied in~\cite{MALDA2};
the triple point on the upper right corner was the one studied 
in~\cite{RABIN}.

On the diagram, the behavior of the effective SYM coupling depends on the
equation of state governing a given region under consideration.
We find the
equipotentials of the effective coupling in the Dp phase
\bb
g_{\rm eff}^2\sim \lk( N^{6-p} S^{p-3} V^{-3p}\re)^{\frac{5-p}{9-p}}\ .
\ee
For $p<4$ and in the Dp phase domain, 
the effective coupling increases diagonally on the diagrams
as we move toward lower entropies and smaller volumes $V^p$.
Using the equation of state of the localized D0 phase, we obtain
the equipotentials in the D0 phase for all diagrams
\bb
g_{\rm eff}^2\sim \lk( \frac{N^2}{S}\re)^{5/3}\ ,
\ee
\ie\ equation~\pref{D0geff}.
The coupling increases from one at $S\sim N^2$ as we lower the entropy toward
the $11$D black hole phase.
From SYM physics, both correspondence curves are where the effective 
coupling is of order one; the localization effect at $S\sim N^2$ changes
this effective coupling appropriately. 

In contrast, the structure of the phase diagrams for $V<1$
depends very much on the specific case at hand.
The $p=1,2,3$ diagrams will be displayed and discussed next.
The $p=4,5,6$ cases will be tackled in the 
next section.

Figure~\ref{phase1} depicts the thermodynamic phase diagram of 
SYM theory
on the circle.
In total, we have six different thermodynamic phases.
The salient feature of this diagram is
the Matrix string phase, characterized by $Z_N$ order,
appearing in the IR of the SYM
on the left and at strong coupling.
This phase is an interesting platform to explore some
of the phase transitions through a statistical mechanical setting.
Note that the dynamics is such that the Matrix string will
emerge from the adjacent black geometrical phases as we move toward the
left of the diagram; this path is however {\em not} reversible
as discussed earlier. The physics of
this phenomena will be explored 
in Chapter~\ref{bhformchap} in detail.

\begin{figure}
\epsfxsize=9cm \centerline{\leavevmode \epsfbox{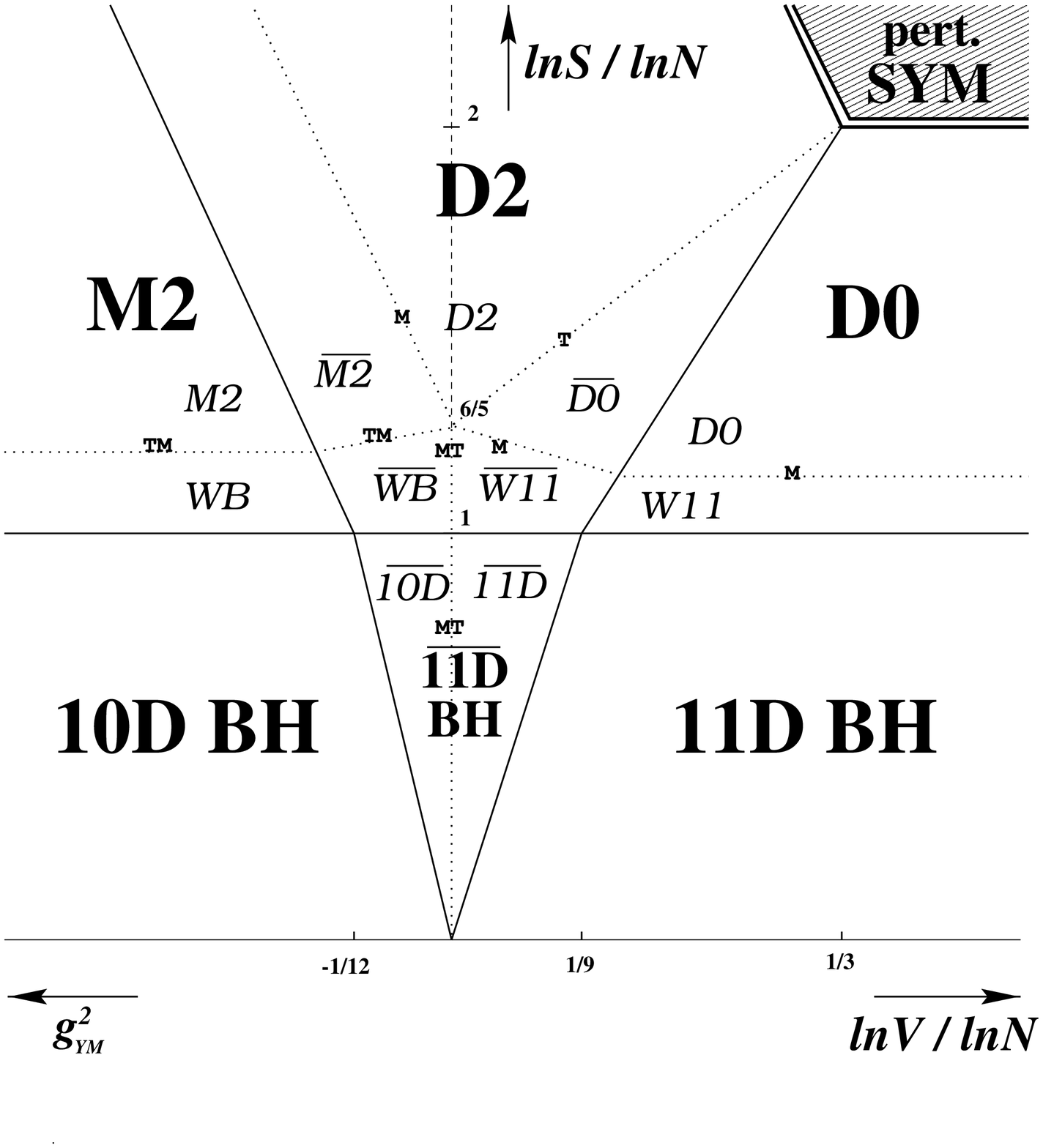}}
\caption{\sl Phase diagram of SYM theory on $T^2$.
The geometry label dictionary is as follows:
D0: black D0; 
$\overline{D0}$: black D0 smeared on $V$;
D2: black D2; 
M2: black membrane; 
$\overline{M2}$: black membrane smeared on a dual circle;
WB: black IIB wave; 
$\overline{WB}$: black IIB wave smeared on a dual circle;
W11: $11$D black wave; 
$\overline{W11}$: $11$D black wave smeared on $V$; 
$11$D BH: Light-Cone M-theory black hole; 
$\overline{11D}$ BH: Light-Cone M-theory black hole smeared on $V$;
$10$D BH: IIB Light-Cone black hole; 
$\overline{10D}$ BH: IIB Light-Cone black hole smeared on a dual circle.
}
\label{phase2}
\end{figure}
\begin{figure}[p]
\epsfxsize=9cm \centerline{\leavevmode \epsfbox{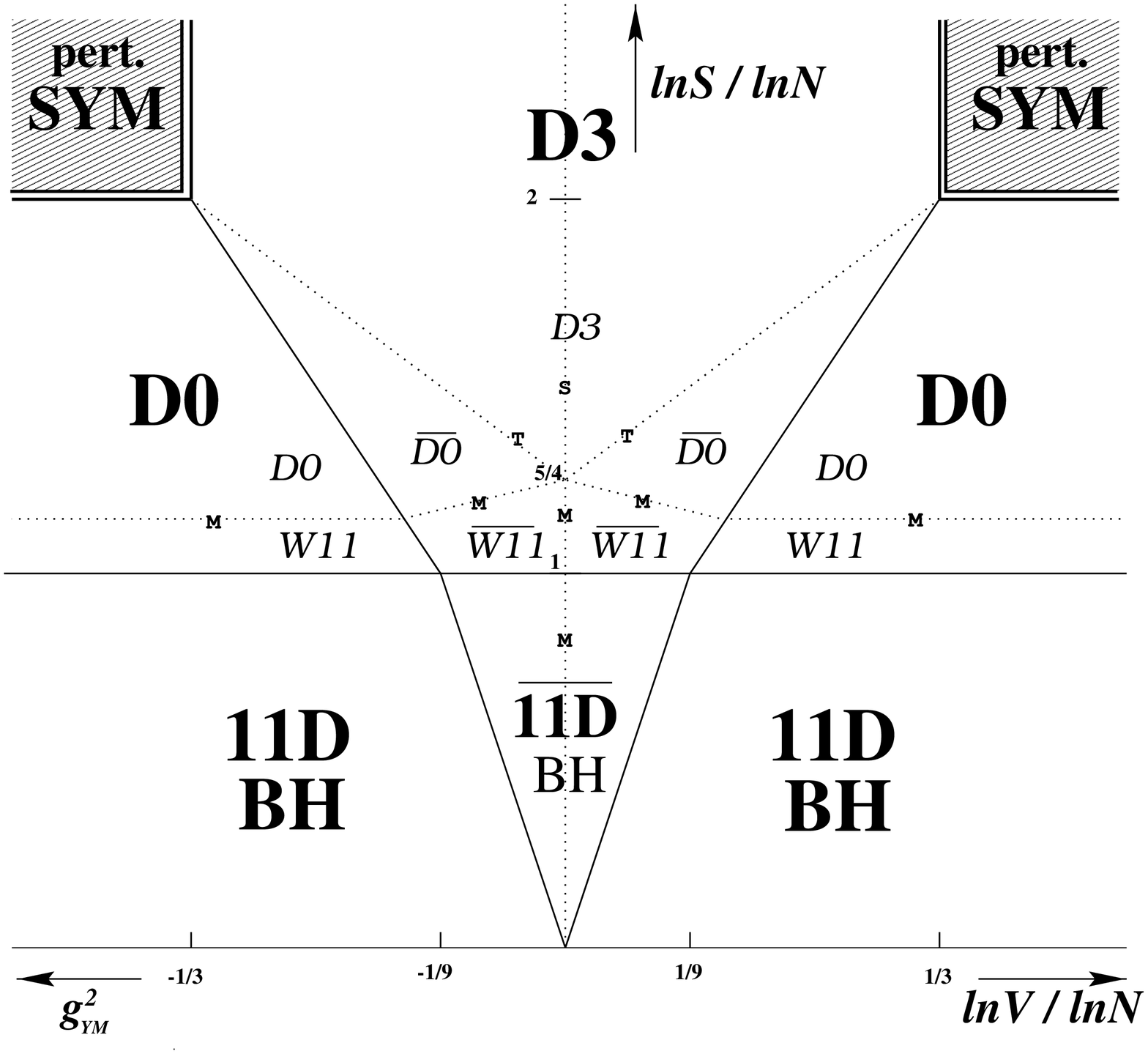}}
\caption{\sl Phase diagram of SYM theory on $T^3$.
D0: black D0; 
$\overline{D0}$: black D0 smeared on $V$;
D3: black D3;
W11: $11$D black wave; 
$\overline{W11}$: $11$D black wave smeared on $V$;
$11$D BH: Light-Cone M-theory black hole; 
$\overline{11D}$ BH: Light-Cone M-theory black hole smeared on $V$.
}
\label{phase3}
\end{figure}

\newpage
Figures~\pref{phase2} and~\pref{phase3} depict the phase diagrams
for SYM theory on $T^2$ and $T^3$; $V$ here is 
again the radius of the cycles 
(which are chosen to be equal) measured in Planck units. 
In the strong coupling region of SYM on $T^2$, 
the SYM dynamics approaches the
infrared fixed point governing the dynamics of M2 branes --
the conformal field theory dual to M-theory on $AdS_4\times S^7$ 
(in `Poincare' coordinates).
The proper size of the $T^2$ shrinks toward the origin;
at high entropy, the black M2 geometry accurately describes
the low-energy physics, while at low entropy 
the near-horizon geometry is best described
in terms of the IIB theory dual to M-theory on $T^2$~\cite{INCREDIBLE}.
In the $T^3$ case, the diagram reflects the self-duality 
of the D3 branes and M-theory on $T^3$
as reflection symmetry about $V\sim 1$.
The 't Hooft scaling limit, $\gym^2 N$ fixed with $N\rightarrow \infty$,
focusses in on the neighborhood of
the vertical line at $\ln\,V/\ln\,N\rightarrow\pm\frac13$.

Finally, we conclude by restating a previous observation.
Starting with a thermodynamic phase in Light-Cone M-theory, 
say for example the lower right corner
phase of the $11$D boosted black hole, using geometrical considerations,
the duality symmetries of M-theory, and the Horowitz-Polchinski
correspondence with the perturbative SYM phase, 
we would be led to conclude that 
Light-Cone M-theory thermodynamics is encoded in the
thermodynamics of SYM QFT. We saw that the DLCQ limit is dual to the
Maldacena energy regime; this mapping is depicted on the
phase diagrams by the two dotted curves on the right half of the
figures converging to the self-duality point.
Maldacena's conjecture asserts that underlying all these phases
is super Yang-Mills theory in various regimes of its parameter space.
Having not known the Matrix conjecture, we would
then have been led to it from Maldacena's proposal. The
Matrix conjecture is a special realization of the
more general statement of Maldacena.
Correspondingly, our ability to discover the low-energy 
theories that yield Matrix theory on some background depends
on our ability to understand duality structures with less supersymmetry
in sufficient detail to construct the phase diagram analogous
to figures~\ref{phase1}-\ref{phase3}.

\section{Phase diagrams of Five Brane Theories}

In this section, we extend our analysis to $p+1$d SYM theories
for $p=4,5$, where the relevant theories involve the dynamics
of five-branes~\cite{BRS,SEIBLITTLE,WHYSEIB,ASHOKE}; 
and $p=6$, where the decoupling issue
is problematic~\cite{WHYSEIB,ASHOKE,
BRUNNERKARCH,HANANYLIFSCHYTZ,LOSMOORES}.
In the process of generating
the phase diagrams, we will rediscover the known prescriptions
for generating Matrix theory compactifications on $T^4$, $T^5$, and 
$T^4/Z_2$; we will also
comment on the difficulties encountered for $p=6$.

In addition, we will analyze the phase diagram of the D1D5 system,
which arises in diverse contexts:
\begin{itemize}

\item 
It has played a central role in our understanding of black
hole thermodynamics~\cite{STROMVAFA}; 

\item
It is a prime example of Maldacena's conjecture,
due to the rich algebraic structure of 
1+1d superconformal theories which are proposed duals to string theory on 
$AdS_3\times S^3\times \MM_4$~\cite{MALDA1,MALDASTROMADS3,MARTINECADSMM,GKS}; 

\item
It describes the little string theory of fivebranes
~\cite{SEIBLITTLE,DVVM5}, 
where the little strings carry both winding and momentum charges.

\item
It is related to the DLCQ description of fivebrane 
dynamics~\cite{ABKSS,SETHISEIBERG,SETHI,GANORSETHI}.
\end{itemize}

\noindent
The analysis will clarify the relation of the D-brane description
of the system to one in terms of NS fivebranes and fundamental
strings~\cite{GKS}, as low-energy descriptions of
different regions of the phase diagram
(for earlier work, see \cite{JOHNSON}).

\subsection{Phase diagrams for $T^4$, $T^5$, and $T^6$}

\begin{figure}
\epsfxsize=9cm \centerline{\leavevmode \epsfbox{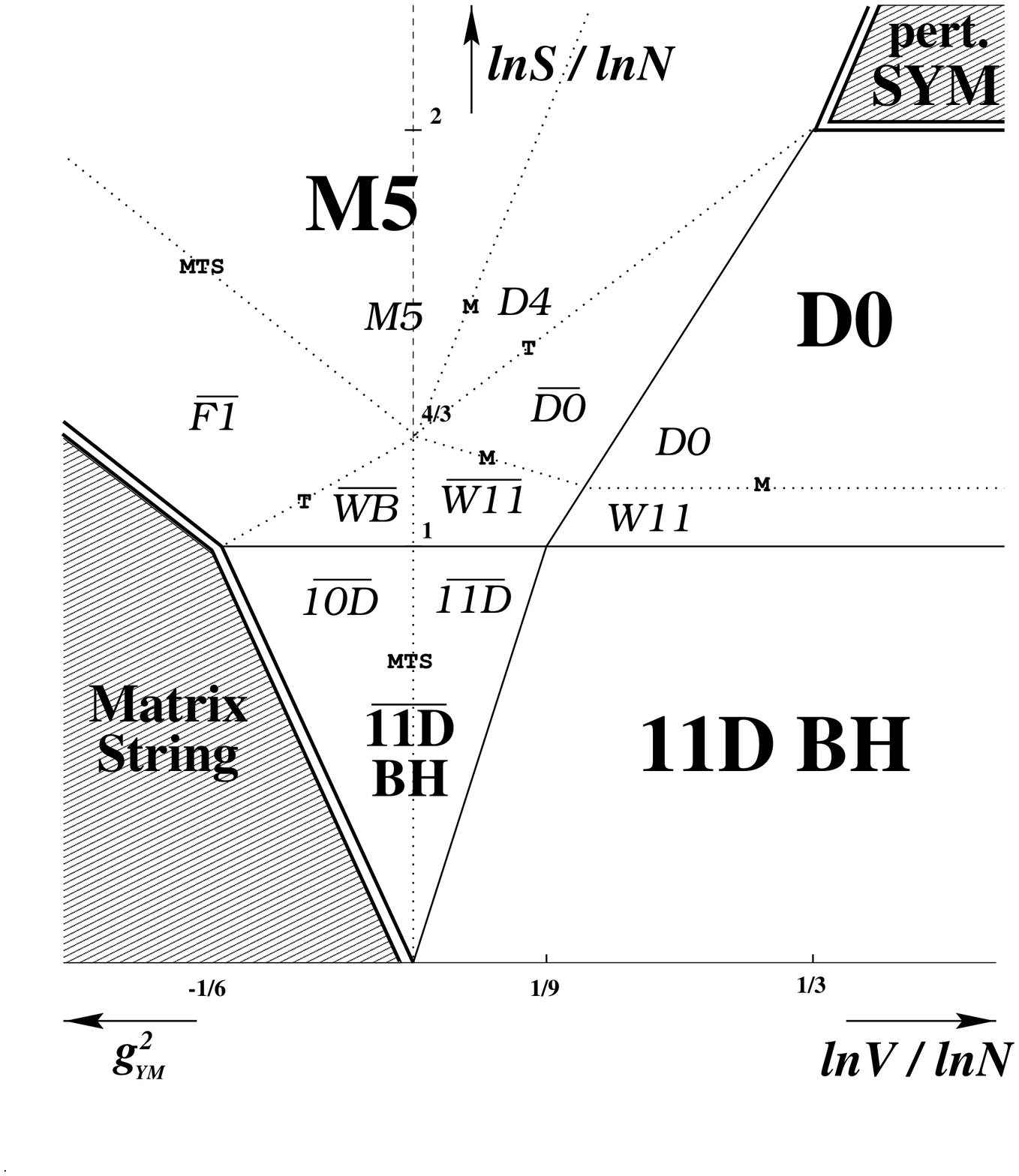}}
\caption{{\small \sl Phase diagram of the six-dimensional
$(2,0)$ theory on $T^4\times S^1$.
$V=R/\lp$ is the size of a cycle on the $T^4$
of light-cone M theory.
The dashed line is the extension of the axis $V=1$, and
is merely included to help guide the eye.
The label dictionary is as follows:
$D0$: black D0 geometry;
$W11$: black 11D wave geometry;
$11DBH$: 11D Light-Cone black hole;
$\overline{D0}$: black smeared D0 geometry;
$\overline{W11}$: black smeared 11D wave geometry;
$\overline{11D}BH$: 11D smeared Light-Cone black hole;
$D4$: black D4 geometry;
$M5$: black M5 geometry;
$\overline{F1}$: black smeared fundamental string geometry;
$\overline{WB}$: black smeared IIB wave geometry;
$\overline{10D}BH$: IIB boosted black hole.
The phase diagram can also be considered that of the 
$(2,0)$ theory on $T^4/Z_2\times S^1$
by reinterpreting the 
$\overline{F1}$, $\overline{WB}$, $\overline{10D}$ phases,
and the Matrix string phase as those of a Heterotic theory.
.}}
\label{SYM4fig}
\end{figure}

Figure~\ref{SYM4fig} is the phase diagram of $T^4$ compactification.
There are six different phases, several of which -- the 11d 
and $\overline{11d}$ black
hole, black $D0$ and $Dp$ brane, and SYM gas phases -- 
were discussed above.  In a slight shift of emphasis,
we have relabeled the black $D4$ brane phase as the black $M5$ brane
phase, since its description in terms of the latter
object extends to the region $V<1$ (in fact, even for a patch of $V>1$
the D4 brane becomes strongly coupled and must be lifted to
M-theory).  The appropriate dual non-gravitational description
involves the six-dimensional $(2,0)$ theory on $T^4\times S^1$,
where the last factor is the M-theory circle;
the scale of Kaluza-Klein excitations given by the size of this circle 
(times the number of branes) sets the
transition point between the $(2,0)$ and SYM descriptions.
This $M5$ phase consists of six patches
that we cycle through via duality transformations
required to maintain a valid low-energy description. 
The energy per entropy increases toward the left and toward higher entropies;
this is to be contrasted with the cases analyzed above where the
IR limit appears toward the left of the diagrams.
This behavior is a consequence of the reversal of the direction
of RG flow between $p<3$ and $p>3$.
As we continue to the left and/or down on the figure
at small volume $V<1$, the $T^4$ is small while the M-theory circle
remains large; eventually one reduces to string theory along
the cycles of the $T^4$, and the M5-brane dualizes into
a string.  Somewhat further in this direction, we encounter
a Horowitz-Polchinski correspondence curve, and a transition to a phase 
consisting of a Matrix String~\cite{MOTL,DVV,BSMAT} 
with effective string tension set by
the adjacent geometries. Using Maldacena's conjecture, we thus
validate earlier suggestions to describe Matrix strings using the
$(2,0)$ theory~\cite{BRS,SEIBLITTLE,WHYSEIB}. 
This Matrix string phase has a correspondence curve also for low entropies,
now with respect to a phase of smeared Light-Cone M theory black holes 
(or equivalently boosted IIB holes). 

Figure~\ref{SYM4fig} is trivially modified to give the phase
diagram of the $(2,0)$ theory on $T^4/Z_2\times S^1$. 
The additional structure does not affect
the critical behavior of the diagram. The change appears in the chain of
dualities we perform on the dotted lines of the diagram. 
Appendix~\ref{orbapp} contains the details.
The orbifold quotient metamorphoses into world-sheet parity, 
and the fundamental string patch (labeled $\overline{F1}$)
becomes that of the Heterotic string. The emerging Matrix
string phase at the correspondence point is then that of a Heterotic
theory. We thus confirm the suggestion~\cite{BR,DIACGOMISMS}
to describe Heterotic Matrix
strings via the $(2,0)$ theory on $T^4/Z_2\times S^1$. One can also
propose to extend the dual theory of an intermediate state obtained in the
chain of dualities between the $M5$ and the $\overline{F1}$ patches
into the Matrix string regime; we then have Heterotic Matrix strings
encoded in the $O(N)$ theory of type I D strings,
as suggested in~\cite{BANKSMOTL,LOWE,HORAVA}.
Similar statements can be made about Matrix theory orbifolds/orientifolds
in other dimensions. 

\begin{figure}
\epsfxsize=9cm \centerline{\leavevmode \epsfbox{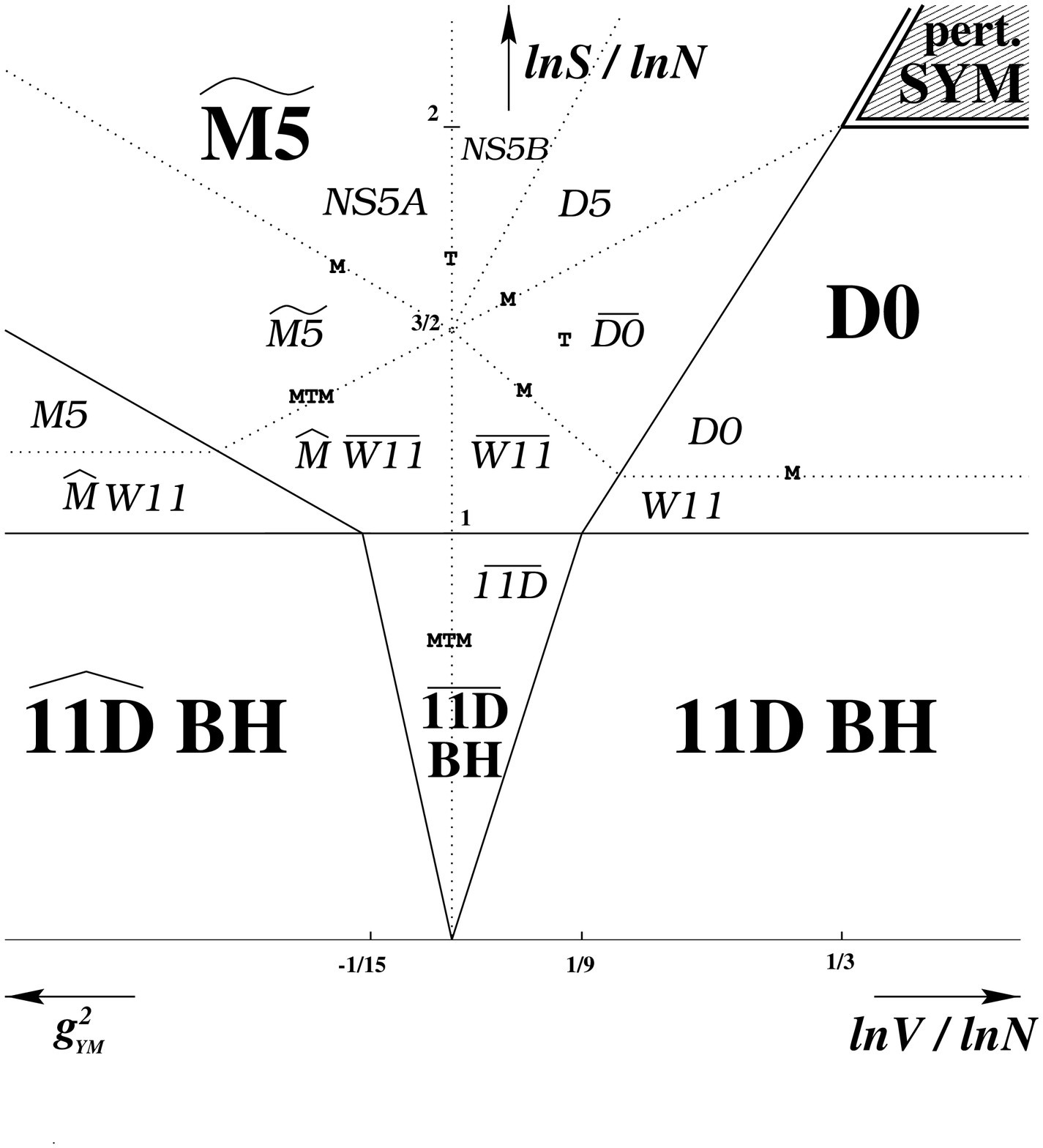}}
\caption{{\small\sl Phase diagram of little string theory on $T^5$.
$D0$: black D0 geometry;
$W11$: black 11D wave geometry;
$11DBH$: 11D Light-Cone black hole;
$\overline{D0}$: black smeared D0 geometry;
$\overline{W11}$: black smeared 11D wave geometry;
$\overline{11D}BH$: 11D smeared Light-Cone black hole;
$D5$: black D5 geometry;
$NS5B$: black five branes in IIB theory;
$NS5A$: black five branes in IIA theory;
$M5$: black M5 brane geometry;
$\wtilde{M5}$: black smeared M5 brane geometry;
$\what{M}\overline{W11}$: black smeared wave geometry in $\what{M}$ theory;
$\what{M}W11$: black smeared wave geometry in the $\what{M}$ theory;
$\what{11D}BH$: smeared boosted black holes in the $\what{M}$ theory.
}}
\label{SYM5fig}
\end{figure}

The thermodynamic phase diagram of fivebranes 
(sometimes called the theory of little strings
~\cite{DVV5D,SEIBLITTLE,DVVM5})
on $T^5$ is shown in Figure~\ref{SYM5fig}. We have a total of seven 
distinct phases.  We again shift the notation somewhat,
relabeling the black $D5$ phase as a black $\wtilde{M5}$ phase,
since the latter extends the validity of the description 
to $V<1$~\footnote{The tilde is 
meant to distinguish this eleven-dimensional phase
(where the M-circle is transverse to the five-branes) from the
eleven dimensional Light-Cone phase on the lower right, 
whose M-circle has a different origin.\vspace{8pt}}.
The equation of state of this high-entropy regime is 
\bb
  S\sim E N^{\frac12}\left(\frac{\lp^2}{\r11}\right) V^{-\frac52}\ ,
\ee
characteristic of a string in its Hagedorn phase.
We have a patch of black NS5 branes 
in the middle of the diagram. They appear near the $V\sim 1$
line, at which point a T duality transformation
exchanges five branes in IIA and IIB theories. 
The IIB $NS5$ patch connects to a $D5$ brane patch via S-duality.
The IIA $NS5$ patch lifts to a patch of $M5$ branes on $T^5\times S^1$
at strong coupling on the left.  The extra circle is the M-circle
transverse to the wrapped $M5$-branes; the horizon undergoes
a localization transition on this
circle at lower entropy and/or smaller $V$ to a phase
whose equation of state is that of a 5+1d gas.
It is interesting that the Hagedorn transition is seen here
as a localization/delocalization transition in the black geometry.
Yet further in this direction,
the system localizes at $N\sim S$ to a 
dual Light-Cone $\what M$ theory on a $T^4\times S^1\times S^1$;
here the horizon is smeared along the square $T^4$,
localized along both $S^1$ factors, and carrying momentum
along the last $S^1$. 
This $\what{M}$ phase on the lower left is U-dual to
the Light-Cone M-theory on the lower right.

\begin{figure}
\epsfxsize=9cm \centerline{\leavevmode \epsfbox{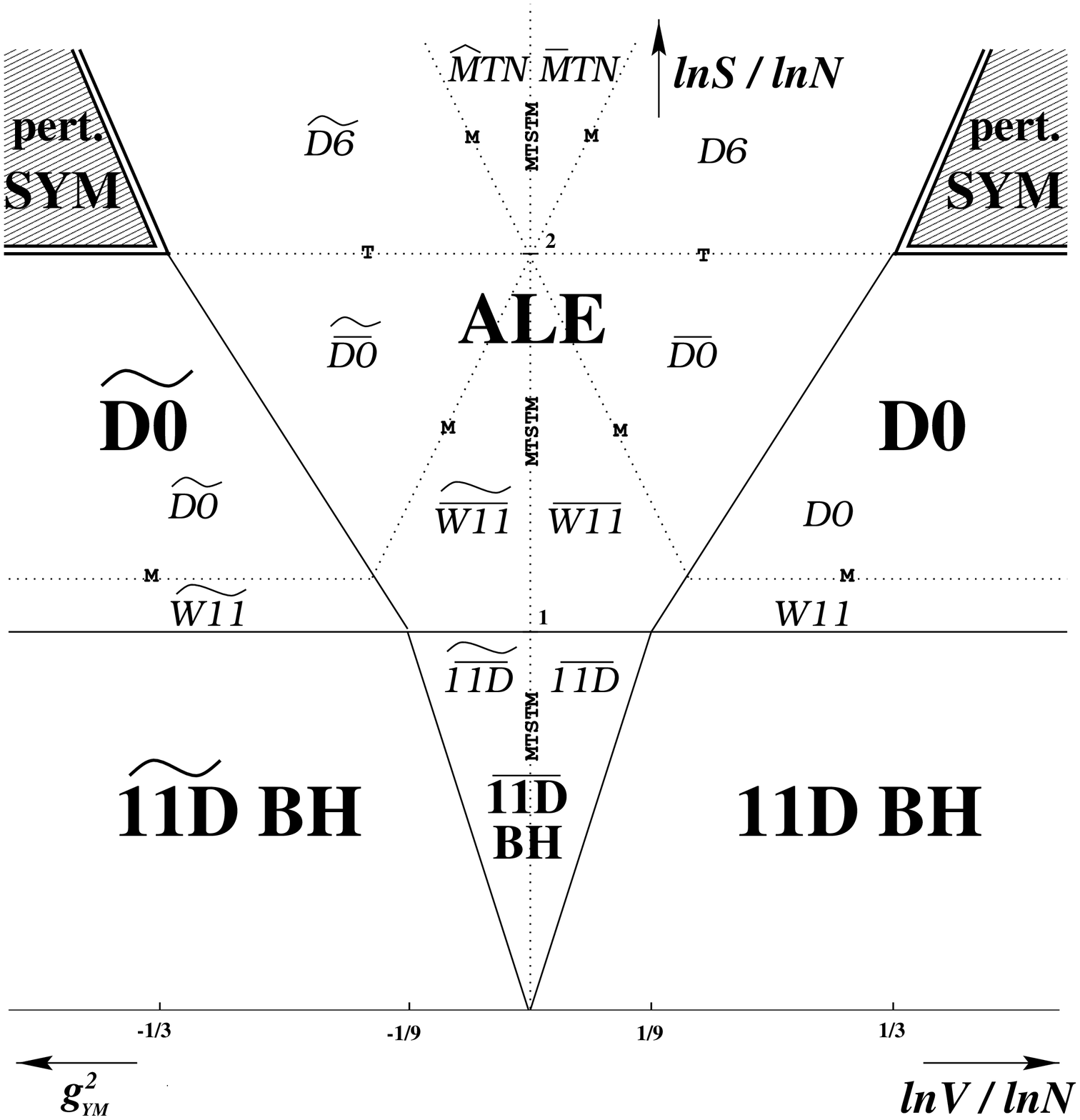}}
\caption{{\small\sl Phase diagram of the $D6$ system.
The label dictionary is as follows:
$\overline{M}TN$,$\what{M}TN$: black Taub-NUT geometry;
$D6$,$\wtilde{D6}$: black D6 geometry;
$D0$,$\wtilde{D0}$: black D0 geometry;
$W11$,$\wtilde{W11}$: black 11D wave geometry;
$11DBH$,$\wtilde{11D}BH$: 11D Light-Cone black hole;
$\overline{D0}$,$\wtilde{\overline{D0}}$: black smeared D0 geometry;
$\overline{W11}$,$\wtilde{\overline{W11}}$: black smeared 11D wave geometry;
$\overline{11D}BH$,$\wtilde{\overline{11D}}BH$: 11D smeared 
Light-Cone black hole.
}}
\label{SYM6fig}
\end{figure}

The $D6$ phase diagram has two important features (see Figure~\ref{SYM6fig}).
First of all, the Maldacena limit keeps fixed the Planck scale
$\tlp\sim\frac{\lp^2}{\r11}V^{-2}$ of the high-entropy
black Taub-NUT geometry\footnote{In the Maldacena limit, the near horizon
geometry is that of an ALE space with $A_{N-1}$ singularity.}~\cite{MALDA2}.
Thus, gravity does not decouple,
and the limit does not
lead to a non-gravitational dual system that would
serve as the definition of M-theory in such a spacetime.
A symptom of this lack of decoupling of gravity 
is the negative specific heat $S\propto E^{3/2}$ 
of the high-entropy equation of state.  This property
is related to the breakdown of the usual UV-IR
correspondence of Maldacena duality~\cite{SUSSWIT,PEETPOLCH}.
The energy-radius relation of~\cite{PEETPOLCH}
determined by an analysis of the scalar wave equation in the
relevant supergravity background, is in fact the
relation between the horizon radius and the Hawking temperature
of the associated black geometry; thus, 
for $p=6$ {\it decreasing energy}
of the Hawking quanta is correlated to {\it increasing radius}
of the horizon, as a consequence of the negative specific heat.
This is to be contrasted with the situation for $p<5$, where
the positive specific heat means increasing horizon radius
correlates to increasing temperature; and $p=5$, where the
Hawking temperature is independent of the horizon radius
in the high-entropy regime.
Now, temperature in any dual description must be the same
as in the supergravity description.  For $p<5$, the dual is
a field theory; high temperature means UV physics dominates
the typical interactions, leading to the UV-IR correspondence.
For $p=5$, the dual is a little string theory; the temperature
is unrelated to the horizon radius (and thus the total energy)
on the gravity side, and unrelated to short-distance physics
in the dual little string theory (since high-energy collisions
of strings do not probe short distances).  Hence the UV-IR
correspondence already breaks down at this point.  For $p=6$,
there is nothing to say -- large radius (large total energy)
corresponds to low temperature of probes (Hawking quanta);
and any dual description could not have high energy/temperature
related to short distance physics, since it is a theory
that contains gravity (so high energy makes big black holes).

A second key feature is the duality symmetry 
(\cf~\cite{SENTP}) $V\rightarrow V^{-1}$
of the diagram relating the $V<1$ structure to that discussed
above for $V>1$.  Note that this duality symmetry inverts
the $T^6$ volume as measured in {\it Planck} units
rather than string units.  The duality interchanges momentum
modes with fivebrane wrapping modes, while leaving membrane
wrapping modes fixed; in other words, the dual space is
that seen by the $M5$ brane.  

\subsection{The D1D5 system}\label{d1d5sec}

As a further example of our methods, we have examined the D1D5 system
on $T^4\times S^1$, which as we mentioned above can be considered
as the little string theory of $Q_5$ fivebranes, with $Q_1$ units of
string winding along the $S^1$. 
\begin{figure}
\epsfxsize=10cm \centerline{\leavevmode \epsfbox{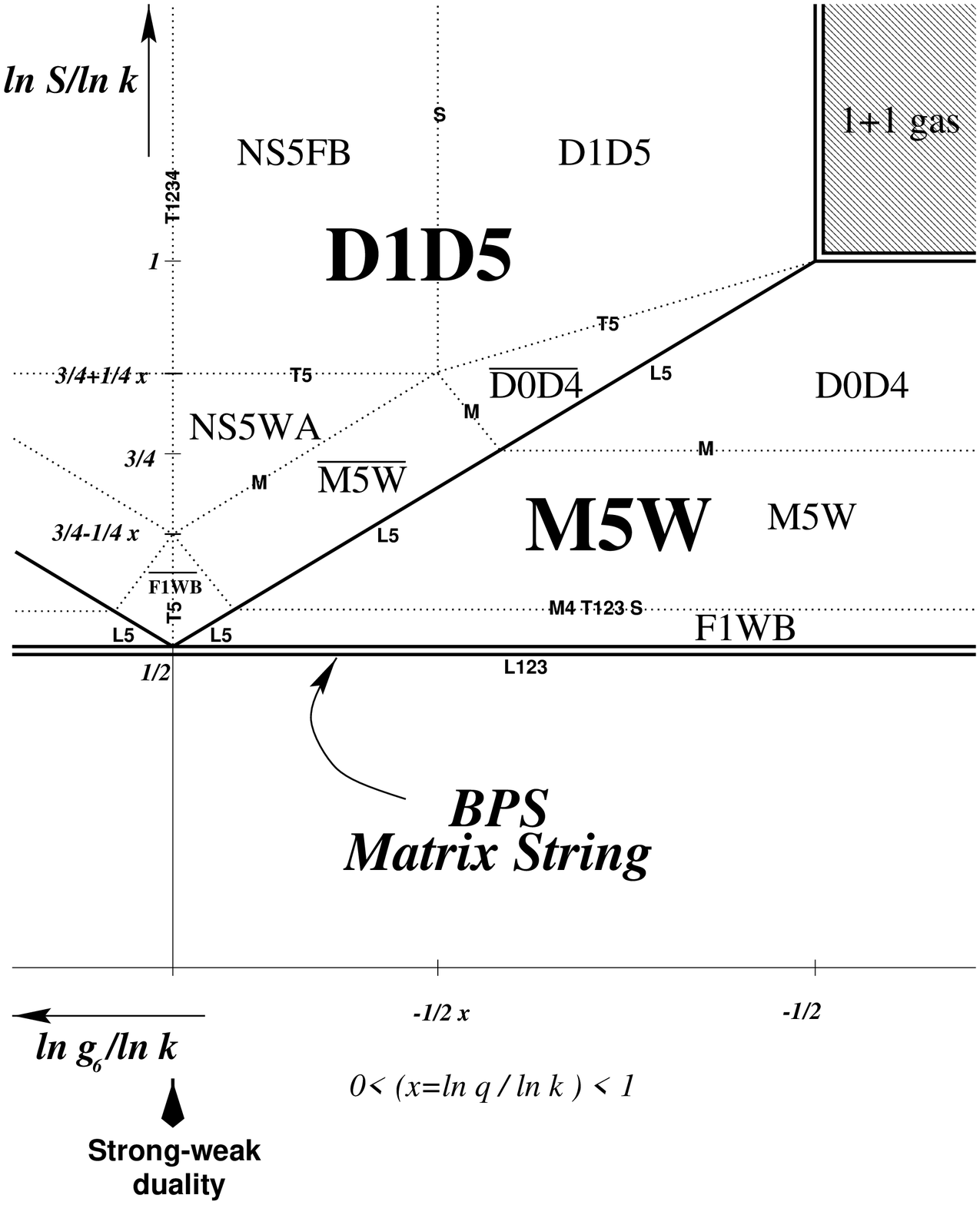}}
\caption{{\small\sl Thermodynamic phase diagram of little strings
wound on the $S^1$ of $T^4\times S^1$,
with $Q_1$ units of winding and $Q_5$ five branes. $k\equiv Q_1 Q_5$
and $1<q\equiv Q_1/Q_5<k$. $g_6$ is the six dimensional string coupling
of the $D1D5$ phase.
The label dictionary is as follows:
$D1D5$: black $D1D5$ geometry;
$NS5FB$: black NS5 geometry with fundamental strings in IIB theory;
$D0D4$: black D0D4 geometry;
$\overline{D0D4}$: black smeared D0D4 geometry;
$M5W$: black boosted M5 brane geometry;
$\overline{M5W}$: black smeared boosted M5 brane geometry;
$NS5WA$: black boosted NS5 branes in IIA theory;
$F1WB$: black boosted fundamental strings in IIB theory;
$\overline{F1WB}$: 
black smeared and boosted fundamental strings in IIB theory;
L: localization transitions.
}}
\label{D1D5fig}
\end{figure}
Figure~\ref{D1D5fig} shows the thermodynamic phase diagram.
In the Maldacena limit\footnote{
This regime corresponds to taking $\alp\rightarrow 0$, with $R$, $g_6$,
and $v$ held fixed. This sends the little string tension to infinity
and drives the equation of state of the smeared phase to Light-Cone scaling.
More general limits of decoupling can be considered and yield somewhat
richer structures for the phase diagram~\cite{MSD1D5}.},
this theory is a representation of the algebra of $\NN=(4,4)$
superconformal transformations in 
1+1d~\cite{MALDA1,STROMBHMICRO,BROWNHEN,MALDASTROMADS3,MARTINECADSMM,GKS}.
We have defined $k\equiv Q_1 Q_5$ and $q\equiv Q_1/Q_5$.  
We keep $k$ fixed, but $q$ may be viewed as a variable
ranging between $1<q<k$, thus moving some of the dotted curves
of duality transformation, but not altering phase transition curves.
For $q\sim 1$, we can exchange
the roles of $Q_1$ and $Q_5$ via duality transformations across the
diagram; the structure is unchanged. 
The other limit, $q=k$, is the $Q_5=1$ bound.
The vertical axis on the diagram is again the entropy, 
while the horizontal axis is the six-dimensional string
coupling $g_6\equiv g_s/\sqrt{v}$ of the $D1D5$ patch,
where $v=V_4/\alp^2$ is the volume of the $T^4$ in string units
(equivalently $g_6^{-2}$ is the 
volume of the $T^4$ in appropriate string units of the NS5FB phase). 
The phase diagram has a symmetry $g_6\rightarrow 1/g_6$
(inversion of the torus in the NS5FB phase); 
this is the T-duality symmetry of the little string theory.
From the perspective of the $D1D5$ patch,
we can consider the entire phase diagram as that of the
1+1d conformal theory that arises in the IR of this gauge
theory, which is conjectured to be dual to the near-horizon 
geometry $AdS_3\times S^3\times T^4$ of the $D1D5$ system.
In this patch, the D strings are wrapped
on a cycle of size $R_5$.  This parameter is absent from the scaling 
relations of all curves because of conformal symmetry.
Analogous to the singly-charged brane systems we have been
discussing, at high entropies there is a `1+1d gas' phase at
small $g_6$ (large effective $V_4$), which passes across a correspondence curve
to the black brane phase as the coupling increases.
Being determined by conformal symmetry and quantization of the
central charge, the equation of state does not change across
this `phase transition'.
Starting in the `1+1d gas' phase and decreasing the entropy,
$S\sim k$ corresponds to the point 
where the thermal wavelength in the 1+1d conformal theory
becomes of order the size of the box $R_5$. 
This is again a Horowitz-Polchinski correspondence
curve from the side of lower entropies, 
analogous to the SYM theories at $S\sim N^2$
\footnote{There is similarly a hidden phases of zero specific
heat between the gas phase and the lower, localized phase,
as can be seen by the discontinuity in temperatures
that occurs between $S>k$ and $S<k$.}.
There is a localization transition on the $R_5$ cycle
cutting obliquely across the diagram. The localized phase
can be interpreted as that of $M5$ branes with
a large boost, thus connecting with 
the proposal of~\cite{ABKSS}
for a Matrix theory of this system.
The lower boundary of this phase occurs at
entropies of order $S\sim \sqrt{k}$,
where a BPS Matrix string phase emerges and the diagram is
truncated at finite entropy.   
We find agreement with Vafa's argument~\cite{VAFAHAGE}
that the BPS spectrum in the R sector of the D1D5 system 
is that of fundamental IIB strings carrying
winding and momentum 
(sometimes called Dabholkar-Harvey states~\cite{DABHOLHARV}).
Similarly, chasing through the sequence of dualities for
the D1D5 system on $K3\times S^1$, one finds the BPS spectrum
consists of Dabholkar-Harvey states of the heterotic string.

For simplicity, we have restricted the set of parameters 
we have considered in the phase diagram to the entropy and
the coupling $g_6$.  It is straightforward to see what
will happen as other moduli of the near-horizon geometry
are varied.  Consider for instance decreasing one of the $T^4$
radii keeping the total volume fixed.  At some point, the
appropriate low energy description will require T-duality
on this circle, shifting from 
$D1$-branes dissolved into $D5$-branes, to $D2$-branes
ending on $D4$-branes.  One can then chase this duality
around the diagram: The $NS5FB$ phase becomes $M2$-branes
ending on $M5$-branes; the $NS5WA$, $\overline{D0D4}$, and
$D0D4$ phases become $D1$-branes ending on $D3$-branes;
and the $\overline{M5W}$ and $M5W$ phases become those of
fundamental strings ending on $D3$ branes.  The near-extremal
$F1WB$ phase is unaffected.
One can also imagine replacing the $T^4$ by K3.  Moving around
the K3 moduli space, when a two-cycle becomes small,
a $D3$-brane wrapping the vanishing cycle becomes light;
one should consider making a duality transformation 
that turns $Q_1$ or $Q_5$ into the wrapping number on this cycle.

Thus the D1D5 system appears to have a remarkably varied life. 
On the one hand, it can describe low-energy supergravity
on a 6d space, namely $AdS_3\times S^3$; the common
coordinate of the branes is the angle coordinate on $AdS_3$.
This space parametrizes physics of the Coulomb branch of the
gauge theory.  On the other hand, the same system describes 
the `decoupled' dynamics of the five-brane, another 6d system
\footnote{Seven-dimensional, if we include the circle
transverse to the $M5$-brane.} --
except that the spatial coordinates are now $T^4\times S^1$,
with the $T^4$ apparently related to the physics of the Higgs
branch of the gauge theory, and the $S^1$ the dimension common 
to the branes.  
In the Maldacena limit, the theory is
a representation of the 1+1d superconformal group.

\chapter{Black hole formation from~Super~Yang-Mills}\label{bhformchap}

\section{Summary}

A characteristic of the SYM phase diagrams we encountered in the 
previous chapters was the emergence, at strong Yang-Mills coupling
and low entropies,
of a Light-Cone black hole phase. We also encountered
in the analysis for the phase diagram for the D1 system a phase
describing a highly excited Matrix string.
Both Matrix black hole and Matrix string have explicit realizations
as excitations in the SYM fields as discussed 
in Sections~\ref{mbsec} and~\ref{matstsec}.
It would then be interesting to study 
the microscopic dynamics of the formation of
the Light-Cone black hole phase as the {string coupling} of the
Matrix string is tuned up. Previous analysis~\cite{CORR2} of the collapse of a
gravitationally interacting string into a black hole 
has yielded the conclusion that such a collapse accurs at {\em weak}
string coupling provided that the system is compactified to low
enough dimensions. This is because gravitational forces fall 
Coulombically as a power of the distance between the interacting
objects and this power is smaller when the system is embedded in lower
dimensions. Hence, the gravitational forces become stronger when we
compactify to lower dimensions. The purpose of this chapter is to arrange
a setting in the SYM theory where a Matrix string-Matrix black hole
transition occurs readily; we then try to analyze the
statistical mechanics of the formation of the black hole from SYM
physics.

We would like $1+1$d SYM dynamics {\em and} a space-time compactified
to low enough dimensions. We can achieve this dynamical setting by
considering $p+1$d SYM theory compactified on a skewed torus $T^p$; \ie\ $p-1$
radii of this torus will be accorded a different size
from the $p$th radius. We consider Light-Cone M theory 
compactified on this torus,
and we descend from it along the $p$th torus cycle to Light-Cone
IIA theory; this cycle then gives us the string coupling $\gs$.
The longitudinal boost direction is accorded radius
size $R_+$. The remaining $p-1$ cycles 
fo the torus are tuned to the string scale. Our SYM theory then 
describes DLCQ IIA theory parametrized by $\gs$, $\ls$, $N$ and
$R_+$. We pick an object in this IIA theory, a black hole
or a highly excited string, and trace its history on a
thermodynamic phase diagram.

\begin{figure}
\epsfxsize=9cm \centerline{\leavevmode \epsfbox{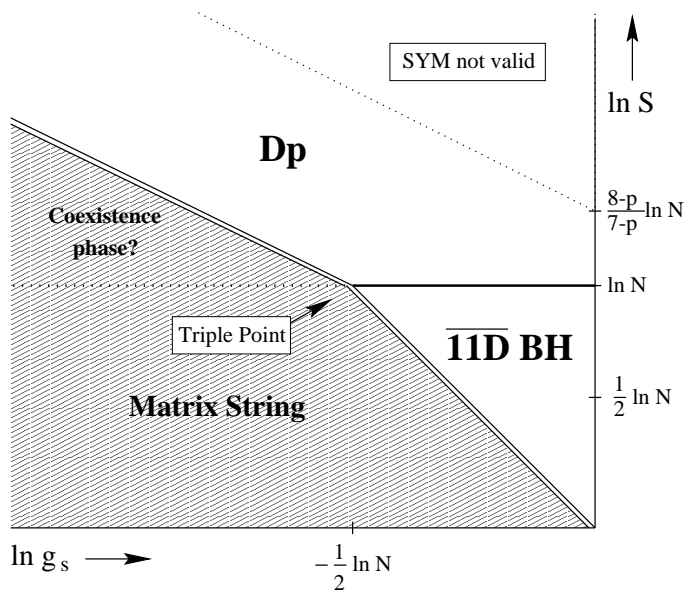}}
\caption{\sl The proposed thermodynamic phase diagram for the $p+1$d SYM
on the skewed torus for $p=4,5$. Equivalently, this is the phase
diagram for DLCQ IIA theory on $T^{p-1}$.
On the horizontal axis is the IIA string coupling,
which is the aspect ratio of the SYM torus.  The vertical
axis is entropy.
}
\label{fig3}
\end{figure}

Figure~\ref{fig3} is the phase diagram for the system of interest.
We choose to work on
a two dimensional cross section in the entropy-string coupling $S$-$g_s$ plane,
with fixed longitundinal momentum $N\gg 1$. 
In principle, one is to take the thermodynamic limit $N\rightarrow \infty$
with $N/S$ fixed, to see criticality; transition between phases 
at finite but large $N$ are smooth crossovers. 
It is expected that, in the infinite $N$ limit, 
the physics tends to the appropriate critical behavior.
The figure is a correct representation of physics for compactifications on 
$T^4$ or $T^5$, \ie\ for $p=4,5$; these are the scenarios where
the transverse non-compact space is small enough in dimensions for a black 
to form, while the torus is of low enough in dimensions for 
gravity to decouple from the SYM theory.
The limit of validity of the SYM description for the DLCQ string theory 
is determined by the upper right curve. Above this line,
the physics is not accurately described by super Yang-Mills theory;
rather, one must pass to the six-dimensional little string theory
as was done in the previous chapter.
We will see that the dynamics of interest
to us occurs outside this region.
We identify several phases in this setting;
a smeared black hole phase ($\overline{10D}$ BH), a Matrix string phase,
a phase of $p+1$ dimensional strongly interacting SYM (\ie\ the phase
described by black Dp branes), 
and perhaps a `coexistence phase'
of a Matrix string with SYM vapor.  
There is a `triple point', a thermodynamic
critical point of the DLCQ string theory where three 
transition manifolds coincide.

A brief description of the physics of the diagram is as follows: 
In type IIA DLCQ string theory on $T^{p-1}$, with $p=4,5$, 
there exists a (longitudinally wrapped) D$p$-brane phase; 
it is unstable at $S\sim N$ to the formation
of a black hole because of longitudinal localization effects.
Along another critical curve (the diagonal line above $S\sim N$), 
the D$p$ brane freezes its strongly
coupled excitations onto a single direction of the torus, 
making a transition to a perturbative string through the Horowitz-Polchinski
correspondence principle. 
In this regime, the thermodynamics is that of a 
near-extremal fundamental (IIB) string supergravity solution,
with curvature at the horizon becoming of order the string scale. 
The correspondence mechanism also applies on the other side
of the $S\sim N$ transition; in this case, a Matrix black hole
makes a transition to a Matrix string when it acquires
string scale curvature at the horizon.
A coexistence phase, where both Matrix string and SYM gas
excitations contribute strongly to the thermodynamics,
may exist in the region indicated on the diagram; 
this depends on the extent to which the object persists
long enough to treat it using the methods of 
equilibrium thermodynamics.

\begin{figure}
\epsfxsize=7cm \centerline{\leavevmode \epsfbox{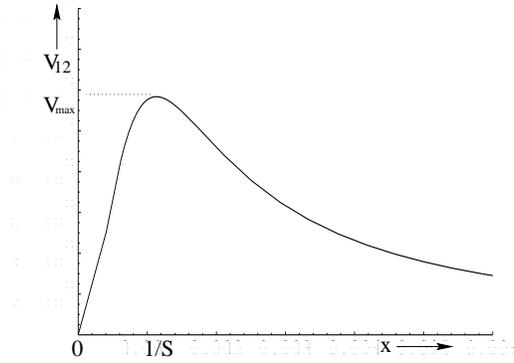}}
\caption{\sl The string self-interaction potential as a function of 
relative separation $x$ along the string, for $p=4,5$.}
\label{potl}
\end{figure}

Our second set of results concerns the dynamics that
leads to the correspondence transitions,
and is summarized by Figure~\ref{potl}.
The plot depicts the mutual gravitational interaction energy
between a pair of points on a typical (thermally excited)
macroscopic Matrix string, as a function of 
the world-sheet distance $x$ along the string separating the two points.
This potential governs the
dynamics of the Matrix string near
a black hole or black brane transition,
as it is approached from the weak string coupling side. 
A bump in the potential occurs at the thermal wavelength $N/S$ 
for $p=4,5$ (five or four noncompact spatial directions);
in these dimensions, the correspondence transition
to a black hole is indeed caused by the
string's self-interaction, as discussed in~\cite{CORR2}. 
For smaller $p$ (more noncompact spatial directions),
there is no bump; similarly, in \cite{CORR2}
the self-interactions could not cause a spontaneous
collapse to a black object.
We will see that the height of the bump is proportional to the
gravitational coupling, such that it `confines'
excitations of the string on the strong-coupling
side of the correspondence transition.

This result supports the suggestion~\cite{LIMART} 
to describe the black hole phase as clustered Matrix SYM 
excitations of size $N/S$.  These correlated
clusters were invoked in order that the
object with $N>S$ can be localized in the longitudinal direction. 
Such a localization necessarily 
involves the longitudinal momentum physics 
of Matrix theory.  We find that a plausible argument for the dynamics 
with this potential gives the two correspondence curves
as the boundaries of validity of the Matrix string phase:
for $N<S$, one finds the transition to the Dp brane 
phase shown in Figure~\pref{fig3};
while for $N>S$, one finds the transition to the Matrix
black hole phase.  Accounting for the latter transition requires taking into 
consideration longitudinal momentum transfer effects as in~\cite{LIMART};
we justify this by a string theory amplitude calculation 
involving winding number exchange in a dual picture
given in Appendix~\ref{stramplapp}.
We thus conclude that we have identified the characteristics of the
microscopic mechanism of black hole formation 
from the SYM point of view.

In the next section, we map the phase diagram in the region of
the thermodynamic parameter space
of interest to us. In Section~\ref{dynmsec}, we analyze the statistical
mechanics of the SYM theory near the triple point. 

\section{The phase structure}

\subsection{Preliminaries}

A DLCQ IIA theory descends from the DLCQ $M$
theory described above; we choose string scale compactification
\bb \label{IIAsetup}
R_i \sim \ls\ \ \
\mbox{for } i=1\ldots p-1\ ,
\ee
with
\bb \label{IIAsetup2}
R_p=g_s \ls\ \ \ \lp^3=g_s \ls^3\ ,
\ee
and a perturbative IIA regime
\bb \label{IIAsetup3}
g_s\ll 1\ .
\ee
We can in principle relax~\pref{IIAsetup} at the expense of introducing new
state variables, and a more complicated (and richer) phase diagram; 
for simplicity, we will stick to this `IIA regime'. 
We write the dictionary between our IIA
theory and the Matrix SYM
\bbb 
g_Y^2&=&~(2 \pi)^{p-2} (a g_s)^{p-3}\ ,\nonumber\\
\Sigma_i&=&~g_s a\ \ \ \mbox{for } i=1\ldots p-1\ ,\nonumber\\
\Sigma_p&=&~a\ ,\label{maindict}\\
V&\equiv &~\Sigma_i^{p-1} \Sigma_p=g_s^{p-1} a^p\ ,\nonumber
\eee
with
\bb
a\equiv \frac{\alp}{R_+}\ .
\ee
We chose~\pref{IIAsetup3} so that we have
$\Sigma_i\ll \Sigma_p$, simplifying our analysis later. Note that $V$
here stands for the {\em volume} of the SYM torus.

We study finite temperature physics of
this IIA theory with the finite temperature vacuum of the corresponding SYM.

\subsection{The phase diagram}

Given that we are working with Matrix theory on $T^4$ and $T^5$, the first
question that must be addressed concerns the validity of the
description, given that SYM 4+1d and 5+1d are non-renormalizable.
New degrees of freedom are required to make sense of the SYM
dynamics as we probe it in the UV,
\ie\ as we navigate outward 
in the corresponding supergravity solution.
These new degrees of freedom for SYM 4+1d and 5+1d are associated with 
the onset of strong coupling dynamics; 
the validity of the theories at different energy scales
is then determined by looking at the 
size of the dilaton vev at different locations
in the supergravity solution.
For finite temperatures, 
physics at the thermal wavelength of the SYM is identified with 
physics at the horizon of the near extremal solution~\pref{Dpmetric}.
The finite temperature vacuum of the $4+1$d and $5+1$d
SYM is a valid thermodynamic description of the DLCQ IIA theory when
\bb \label{dillim}
\lk.e^\phi \re|_{r_o}\ll 1 \Rightarrow S \ll N^{\frac{8-p}{7-p}} g_s^{-1}\ .
\ee
This is a purely geometric statement, in terms of 
the horizon area and string coupling. For entropies satisfying this
bound, the SYM statistical mechanics obeys the equation of state
\bb \label{eosint}
\lk( a E\re)^{p-9} \sim \lk(\frac{S^2}{N}\re)^{p-7} \gs^4\ .
\ee
From our discussion in the earlier chapters, we know that this
phase will localize into a black hole for $S<N$. This is a $11$D
Schwarzchild hole smeared over $T^p$; it appears as a black hole
living in DLCQ IIA theory smeared on $T^{p-1}$.
Its equation of state is
\bb \label{eosbh}
E^{p-9}_{bh}\sim E_{int}^{p-9} \lk(\frac{N}{S}\re)^2\ .
\ee

The Schwarzschild black hole geometry will become stringy when
its curvature near the
horizon becomes of the order of the string scale; the emerging state
is a Matrix string in the Matrix conjecture language, \ie\ a $1+1$d 
dynamics with $Z_N$ holonomy on $\Sigma_p$.
Minimizing the Gibbs energy between the Matrix string 
\bb\label{mseosf}
E\sim a^{-1} \frac{S^2}{N}
\ee
and the Matrix black
hole phases leads to the Horowitz-Polchinski 
correspondence curve~\cite{CORR1,CORR2}
\bb \label{corr}
S\sim g_s^{-2}\ .
\ee
This is a statement independent of $N$ and $p$. 
At $g_c\sim N^{-1/2}\sim N_\osc^{-1/4}$, 
where $N_\osc$ is the string oscillator level,
there exists an interesting critical point.

We next deal with finite size effects in the Dp brane phase
which are due to the other radii $\Sigma_i$.
Because the torus is skewed, the thermal wavelength probes
a $p-1$ dimensional torus. We expect a localization
transition to a phase consisting of the geometry of D1 branes.
Equating the energies of~\pref{eosint} for $p=1$ and for general
$p$, we get the transition point
\bb \label{spec1}
S\sim \sqrt{N} g_s^{-1}\ .
\ee
We find however that, near this transition point,
the D1 vacuum is strongly coupled
at the horizon; the S-dual geometry is that of black fundamental 
strings in IIB theory. No change of equation of state occurs 
through this duality transformation. 
The curvature at the horizon of the black IIB string solution is found to
become of order the string scale at precisely~\pref{spec1},
beyond which a Matrix string description emerges. 
Our analysis leading to~\pref{spec1} is then valid.
We can further check the correctness of this conclusion by
matching the equation of state~\pref{eosint}
with that of the Matrix string~\pref{mseosf},
the latter being the
dominant phase on the other side of this correspondence curve.
The result is again~\pref{spec1}.
We conclude that the Dp brane phase makes
a transition to a Matrix String at~\pref{spec1}. 

Finally, we note that we assumed above that there exists a well
defined Matrix string
description for $N<S$. In this
regime, the thermal wavelength on the Matrix string is smaller than the 
UV cutoff imposed by the discretized nature of the matrices.
Our procedure may be equivalent to an analytical continuation of
the Matrix string phase into a regime where the description may not be fully
justified; this is in the same spirit as the extension of the Van der Waals
equation of state into the gas-liquid coexistence region,
which one uses to identify the
emergence of the liquid phase\cite{RIEF}. Furthermore,
the regime $N<S$ is similar 
to the Hagedorn regime~\cite{HAG,ATTWIT},
in that the temperature remains constant as the system absorbs heat.
We speculate that the $N<S$ regime of the Matrix string near the triple point
is characterized by a coexistent phase of a string with SYM vapor.

As a unifying probe for all the transitions, we observe that the `mass
per unit charge' $q$ defined by
\bb
q\equiv \frac{M}{N}
\ee
scales on the various transition curves as

\vspace{12pt}
\begin{tabular}{lcl}
Matrix String-Dp brane transition         
	& $\rightarrow$ & $q^{-1}\sim h_{\rm eff} \ls$ \\
Matrix String-Coexistence Phase Transition 
	& $\rightarrow$ & $q^{-1}\sim \ls$ \\
Matrix String-Black Hole Transition        
	& $\rightarrow$ & $q^{-1}\sim h_{\rm eff}^2 \ls$ \\
Black Hole-Dp brane transition            
	& $\rightarrow$ & $q^{-1}\sim h_{\rm eff}^{2/(9-p)} \ls$ 
\end{tabular}
\vspace{12pt}

\noindent
with the effective coupling
\bb
h_{\rm eff}^2\equiv g_s^2 N\ .
\ee

From the point of view of the DLCQ string theory 
characterized by the parameters
$g_s$, $\ls$ and $N$, this scaling on the transition curves
is a non-trivial signature of a unifying framework
underlying the physics of criticality of the theory. 
Note also that the $g_s^2 N$ combination 
is {\em not} the 't Hooft coupling of the
SYM description;
recall that $g_s$ is a modulus
of the torus compactification.
From the point of view of field theory, the various transitions
that we have identified are predictions about
the thermodynamics of 4+1d, 5+1d SYM on the torus well into
non-perturbative field theory regimes.

\section{The interacting Matrix string}\label{dynmsec}

In this section, we study
the dynamics near the triple point in greater detail
from the side of the Matrix string. In Section~\ref{mbsec}, it was argued that
the Matrix black hole phase, as a configuration of the SYM fields,
can be thought of as a gas of $S$ distinguishable clusters of
D0 branes, each cluster consisting of $N/S$ partons. The system
is self-interacting through the $v^4/r^7$ interaction, or its smeared
form on the torus.
This phase of clustered D0 branes may be an effective description, \ie\
thermodynamically strongly correlated regions of a metastable state; 
or more optimistically, it might 
be a microscopic description associated with formation of bound
states like in BCS theory. We will try here to investigate the Matrix
string dynamics so as to reveal the signature of the clusters as we
approach the correspondence curve. The aim is to identify a possible
dynamical mechanism for black hole formation, 
and determine the correspondence curve from such
a microscopic consideration.

In Section~\ref{potsec}, we derive 
the potential between two points on the Matrix string;
in view of the Matrix conjecture, 
we can do this by expanding the Dirac-Born-Infeld (DBI) action
of a D-string in the background of a D-string. We then evaluate the
expectation value of this potential in the free string ensemble at 
fixed temperature. In Sections~\ref{bumpsec1} and~\ref{bumpsec2}, 
we analyze the characteristic features of
the potential, particularly noting the bump for $p=4,5$ that we alluded to
in the Introduction. In Section~\ref{dynsec}, we comment on the dynamics implied
by the potential, particularly in regard to the phase diagram derived 
earlier.

\subsection{The potential}\label{potsec}

In this section, we derive the potential between two points on the
Matrix string by expanding 
the DBI action for a D-string probe in the background
geometry of a D-string.
We then check the validity of the DBI expansion, and
evaluate the expectation value of the potential in the thermal ensemble 
of highly excited Matrix strings. The details of the finite temperature
field theory calculations are collected in Appendix~\ref{potapp}.

A IIA string in Matrix string theory is constructed in a sector of
field configurations described by diagonal matrices, 
with a holonomy in $Z_N$;
a nonlocal gauge transformation converts this
into 't Hooft-like twisted boundary conditions on the
transverse excitations of the eigenvalues of the matrices, as described
in detail in \cite{DVV,BSMAT,WYNMAT,MOTL}.
The conclusion is that the eigenvalues are sewn
together into a long string, and a IIA string emerges as an object looking
much like a coil or `slinky' 
wrapped on $\Sigma_p$. The self-interactions of this
string are described by integrating out off-diagonal modes between the
well-separated strands. Alternatively, making use of the Matrix
conjecture, this effective action can be obtained from supergavity, by
expanding the Born-Infeld action of 
a D-string in the background of a D-string
\footnote{Note that, for D-string strands closer to each other than
the Plank scale, the W bosons cannot be integrated out in the problem;
the physics is described by the full non-abelian degrees of freedom. 
We are assuming here that this `UV' physics does not effect the analysis 
done at a larger length scale.}.
We will follow this prescription to calculate 
the gravitational self-interaction
potential between two points on a highly excited Matrix string.

The Born-Infeld action for N D-strings is given by~\cite{POLCHTASI}
\bb \label{DBI}
S=-\frac{1}{2 \bar{\alpha}' \bar{g}_s}
 \lk[ \int d^2 \sigma e^{-\phi} \mbox{Tr} \mbox{Det}^{1/2}
\lk( G_{ab} + B_{ab}+2 \pi \alp F_{ab} \re) - N \int C_{RR}^{(2)}
\re] ,
\ee
where we have assumed commuting matrices so that there is no ambiguity in
matrix orderings in the expansion, and $\bar{g}_s$ is the dilaton vev 
at infinity. We choose $\sigma^1$ to have radius $\Sigma_p$,
turn off gauge and NS-NS fluxes,
\bb
B_{ab}=F_{ab}=0\ ,
\ee
and choose the static gauge
\bbb
X^0&=&~\sigma^0 \mbox{1}\ ,\\
X^1&=&~\sigma^1 \mbox{1}\ .\nonumber
\eee
A single D-string background in the string frame is given by~\cite{SL2Z}
\bbb
ds_{10}^2&=&~h^{-1/2} (-dt^2+dx^2)+h^{1/2} (d\vec{x})^2 \nonumber\\
\medskip
e^\phi&=&~h^{1/2}\hfill \\
C_{01}&=&~h^{-1}\hfill\nonumber\\
h&=&~\lk( \frac{r_0}{r} \re)^{7-p} \ ,\nonumber
\eee
(recall $p$ is the torus dimension)
and we define $D\equiv 7-p$. 
The string is taken to have no polarizations
on the torus, nor any Kaluza-Klein charges.
Here, we have followed the prescription in \cite{BECKERS,KRAUS}, 
where we T-dualized the D-string 
solution to a D0 brane, lifted to 11 dimensions, 
compactified on a lightlike direction,
and T-dualized the solution to the one above.
The only change is in replacing $1+(r_0/r)^{7-p}\rightarrow
(r_0/r)^{7-p}$. By Gauss' law, 
\bbb
r_0^{D}&=&~c_D \frac{g_s^2}{R_+^2} \ls^{9-p}\ ,\\
c_D&\equiv&~ \frac{(2\pi)^7}{2\pi^{D/2}} \Gamma(D/2)\ ,\nonumber
\eee
where we have made use of the needed dualities to express things in our
IIA description.
Putting in the background, we have
\bbb\label{bistr}
S&=&~-\frac{1}{\alp}\int^{\Sigma_p N}\ 
h^{-1}\lk(1+hK+h^2V\re)^{1/2}-h^{-1}\\
K&\equiv&~ {X'}^2-{\dot{X}}^2=4\del_+ X.\del_- X\nonumber\\
\medskip
V&\equiv&~ 4((\del_+X.\del_-X)^2-(\del_+X)^2 (\del_-X)^2)\ .\nonumber
\eee
We note that, as the limit of the action indicates, we have made use of
the $Z_N$ holonomy that sews the rings of the slinky together.
Expanding the square root yields the Hamiltonian
\bbb \label{hamiltonian}
H&=&\int^{\Sigma_p N} \frac{1}{2 \alp} ({\dot{X}}^2+{X'}^2)
+c_D'\frac{g_s^2 \ls^{7-p}}{R_+^2 r^{7-p}} \times \nonumber \\
& & \lk\{(\del_+X)^2 (\del_-X)^2-\lk[ (\del_+X)^2+(\del_-X)^2\re] 
(\del_+X.\del_-X) \re\}\ .
\eee
Let us check the validity of the DBI expansion we have performed. 
We would like to study dynamics of the string squeezed 
at most up to the string scale, the
correspondence point; setting $r\sim \ls$ in $h$, we get
\bb \label{hh}
h\sim \lk(\frac{g_s \ls}{R_+}\re)^2\ .
\ee
From elementary string dynamics, we have
\bb\label{kkk}
\vev{K}\sim \lk(\frac{R_+ S}{\ls N}\re)^2\ ,
\ee
where brackets indicate thermal averaging at fixed entropy $S$. It can be
shown from the results of the next section that
\bb\label{vvv}
\vev{V}_{max}\sim \vev{K}^2\ ,
\ee
and that $\vev{V}$ will have a definite maximum for all $p$. 
We now see from~\pref{bistr}, \pref{hh}, \pref{kkk}, and~\pref{vvv},
that our DBI expansion is a perturbative expansion in
\bb
\varepsilon= g_s \frac{S}{N}\ .
\ee
We then need
\bb
S \ll N g_s^{-1}\ .
\ee
A glance at Figure~\ref{fig3} reveals that we are well within the region
of interest.

The potential in this expression is the interaction energy 
between a D-string probe and a D-string source.  Using the residual
Galilean symmetry in the DLCQ, and assuming string thermal wavelengths
$> \Sigma_p$ (the `slinky regime'), we deduce that the potential between
two points on the Matrix strings denoted by the labels $1$ and $2$ is
\bb \label{slinky}
V_{12} = {K}_D \frac{g_s^2 \ls^{5-p}}{R_+} \frac{
\lk\{(\del_+X_r)^2 (\del_-X_r)^2-\lk[ (\del_+X_r)^2+(\del_-X_r)^2\re] 
(\del_+X_r.\del_-X_r)
\re\}}{\lk(X_r^2\re)^{D/2}}\ ,
\ee
where
\bb
X_r \equiv X_2-X_1 .
\ee
and $K_D$ is a horrific numerical coefficient we are
not interested in.

Ideally, one should self-consistently determine the shape distribution
of the string in the presence of this self-interaction;
however, this is rather too complicated to actually carry out.
To first order in small $g_s$, 
the effect of the potential is to weigh different regions of the
energy shell in phase space~\cite{GOLDENFELD} by a factor derived from 
its expectation value in the free string ensemble. 
We will discuss the dynamics in the presence of the potential
in somewhat more detail below.
For now, in light of this weak-coupling approximation scheme,
we would like to calculate the expectation value of the potential
in a thermodynamic ensemble consisting of a highly excited 
{\it free} string with fixed entropy $S$.
From the Matrix string theory point of view, this is essentially a 
problem in finite
temperature field theory, where we will deal with a two dimensional 
Bose gas (ignoring supersymmetry; the fermion contribution is similar) 
on a torus with sides $\Sigma_p N$ and $\beta=1/T$,
$\beta$ being the period of the Euclidean time. Using Wick contractions,
we can then express the potential in terms of the free Green's functions;
we defer the details to Appendix~\ref{potapp}.
We get
\bb \label{thepotential}
V_{12} = \alpha_D g_s^2 \frac{\ls^{5-p}}{R_+}
\frac{K^{zz}_{12} K^{\bar{z}\bar{z}}_{12}}{\lk( -K_{12} \re)^{D/2}} ,
\ee
where $\alpha_D$ is a dimension dependent numerical coefficient,
\bb
K_{12}\equiv K_\Delta\equiv -\alp \lk<X_1 X_2\re>
\equiv -\alp G_{12}\ ,
\ee
is the Green's function of the two dimensional Laplacian
on the torus, and $K^{zz}_{12}$ is its double derivative with respect
to the $z$ complex coordinate of the Riemann surface representing
the Euclideanized world-sheet.

\subsection{The thermal free string}\label{bumpsec1}

The thermodynamic properties of the Matrix string at
inverse temperature $\beta$ are determined by the Green's function
of the Laplacian on the worldsheet torus of sides
$(\Sigma,\beta)$, where $\Sigma\equiv \Sigma_p N$.
It is known from CFT on the torus 
that this is given by\cite{DIFRANCESCO,KIRITSIS}
\bb
G_{12}=-\frac{1}{2 \pi} \ln \lk| 
\frac{\theta_1\lk(\frac{z}{\Sigma} | \tau \re)}
{\theta'_1\lk( 0| \tau \re)} \re| +\frac{1}{2 \tau_2} 
\lk( \mbox{Im} \frac{z}{\Sigma}
\re)^2 ,
\ee
where
\bb
\tau\equiv i\frac{\beta}{\Sigma}\equiv i \tau_2= \frac{i}{S} .
\ee
Here $\beta$ can be obtained from 
free string thermodynamics.
All correlators and their derivatives must eventually 
be evaluated on a time slice
corresponding to the real axis in the $z$ plane.

Divergences will be seen
in correlators due to infinite zero point energies.
The conventional approach
is to introduce a normal ordering scheme giving the vacuum zero
expectation value in such situations, \ie\ throwing away disconnected
vacuum bubbles. 
In our case, the string has a classical background due
to its thermal excitation.
To renormalize finite $T$ correlators, 
we subtract the zero temperature limit from each propagator. 
This corresponds to
\bb
\ave{f(X)} \sim f(\coeff\delta{\delta J}) \ln Z_T[J]
\rightarrow f(\coeff\delta{\delta J}) \ln Z_T[J] - 
f(\coeff\delta{\delta J}) \ln Z_{T=0}[J]=
f(\coeff\delta{\delta J}) \ln \Bigl(\frac{Z_T[J]}{Z_{T=0}[J]}\Bigr) .
\ee
From the expression for $Z[J]$, we see that this amounts to correcting the
Green's functions as
\bb
K \rightarrow K_T - K_{T=0} .
\ee
This subtraction removes the divergent zero-point fluctuations
of nearby points on the string, while leaving the effects
due to thermal fluctuations.

\begin{figure}
\epsfxsize=7cm \centerline{\leavevmode \epsfbox{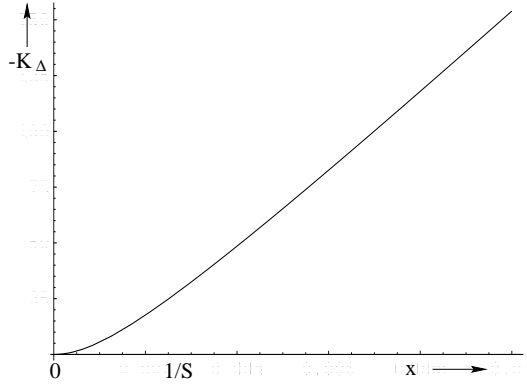}}
\caption{\sl $-K_\Delta$ as a function of 
the string separation parameter $x$;
we see the change of scaling from $x^2$ to $x$.}
\label{f2}
\end{figure}

Defining
\bb
x\equiv \frac{z}{\Sigma}\ ,
\ee
we then have, subtracting the zero temperature part,
\bb
K_\Delta\rightarrow \alp \lk(\frac{1}{4 \pi} 
\ln g \bar{g} +\frac{1}{8 \tau_2} \lk( x-\bar{x} \re)^2 \re) ,
\ee
with
\bb
\ln g=\sum_{n=1} \ln \lk( \frac{1- 2 q^n \cos (2 \pi x) + q^{2 n}}
{\lk( 1-q^n \re)^2} \re) .
\ee
We can now make use of $\tau_2\ll 1$ for $S\gg 1$, to write
this sum as an integral
\bb
\ln g=\frac{1}{2 \pi \tau_2} \int_0^{e^{-2 \pi \tau_2}} \frac{dv}{v} \ \ 
\ln \lk( \frac{1-2v \cos (2 \pi x) + v^2}{\lk( 1-v \re)^2} \re) .
\ee
This integral can be evaluated to yield
\bb\label{Kfinal}
\ln g =\frac{1}{2 \pi \tau_2} \lk( 2 Li_2 e^{-2 \pi \tau_2}
-Li_2 e^{-2 \pi \tau_2 + 2 \pi x i}
-Li_2 e^{-2 \pi \tau_2 - 2 \pi x i} \re) ,
\ee
where $Li_2$ is the PolyLog function of base 2, related to the Lerch $\Phi$
function~\cite{MATH}. We then have
\bb\label{kdelta}
K_\Delta=\frac{\alp}{2\pi} \ln g\ ,
\ee
with $x$ here being real, and
representing the equal-time separation between
two points on the string, $x=x_1-x_2$, 
as a fraction of the total length $\Sigma$ ($0<x<1$).
The asymptotics are
\bb
K_\Delta \simeq \lk[ 
\begin{array}{cl}
\alp S x & \mbox{     for     } \tau_2 \ll 1\\
\frac{\alp}{\pi} S^2 x^2 & \mbox{     for     } 2 \pi x \ll 1
\end{array}
\re . ,
\ee
The first line is a well-known 
result of Mitchell and Turok~\cite{MITCHTUR} calculated 
originally using the microcanonical
ensemble. It shows random walk scaling $\sqrt{\ave{R^2}}\sim N_\osc^{1/4}
x^{1/2}$. The second line is new 
and valid for small separations on the string;
it is the statement that within 
the thermal wavelength $\beta$ of the
the excited string, the string is stretched, scaling as 
$\sqrt{\ave{R^2}}\sim N_\osc^{1/2} x$. 
This is intuitively expected, as regions
on the string within the typical thermal 
wavelength will be strongly correlated
in the thermodynamic sense. This change in the scaling is crucial to
what we will soon see in the behavior of the potential between strands.
$-K_\Delta$ is plotted in Figure~\ref{f2}.

Next, consider the derivatives of the correlators, 
evaluated on the real axis. 
We have
\bb
\del_x K_\Delta = \del_{\bar{x}} K_\Delta =
\frac{-i \alp}{2 (2 \pi)^2 \tau_2} \ln 
\lk( \frac{1-e^{-2\pi \tau_2 -2 \pi i x}}
{1-e^{-2\pi \tau_2 +2 \pi i x}} \re) .
\ee
We also have
\bb
\del_x \del_{\bar{x}} K_\Delta =\frac{-\alp}{4 \tau_2} ,
\ee
or
\bb
K^{z \bar{z}}_\Delta\rightarrow 0 ,
\ee
since we subtract the zero temperature result.
The most relevant term is
\bb
\del_x^2 K_\Delta= \del_{\bar{x}}^2 K_\Delta =
-\frac{\alp}{\tau_2} \frac{1-e^{2 \pi \tau_2} \cos (2 \pi x)}
{e^{4 \pi \tau_2} - 2 e^{2 \pi \tau_2} \cos(2 \pi x) +1}
+\frac{\alp}{4 \tau_2} ,
\ee
or
\bb \label{Kzzfinal}
\Sigma^2 K^{zz}_\Delta=\Sigma^2 K^{\bar{z} \bar{z}}_\Delta \rightarrow
-\frac{\alp}{\tau_2} \frac{1-e^{2 \pi \tau_2} \cos (2 \pi x)}
{e^{4 \pi \tau_2} - 2 e^{2 \pi \tau_2} \cos(2 \pi x) +1}
-\frac{\alp}{\tau_2} \frac{1}{e^{2 \pi \tau_2}-1} 
\ee
(again we subtract the zero temperature part).

\begin{figure}
\epsfxsize=7cm \centerline{\leavevmode \epsfbox{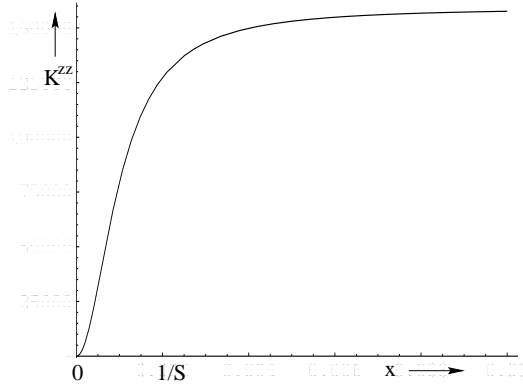}}
\caption{\sl $K^{zz}_\Delta$ as a function 
of the string separation parameter $x$;
we see the flattening of the correlation at large $x$. For small $x$, small
relative stretching or motion is implied; for larger $x$, the flattening
indicates a constant correlation in the relative stretching of the string.}
\label{barkzz}
\end{figure}

This yields the asymptotics
\bb
(N\Sigma_p)^2 K^{zz}_\Delta\simeq \lk[ 
\begin{array}{cl}
\alp S^2 +\frac{\pi}{12} (5+\cos(2\pi x)) (\csc(\pi x))^2 +O(\tau_2^2)
\rightarrow \alp S^2 & \mbox{     for     } \tau_2 \ll 1\\
\alp S^4 x^2 & \mbox{     for     } 2 \pi x \ll 1
\end{array}
\re . .
\ee
$K^{zz}_\Delta$ is plotted in Figure~\ref{barkzz} as a function of $x$.

\subsection{The bump potential}\label{bumpsec2}

We now put together equations~\pref{kdelta} and~\pref{Kzzfinal}
in the potential of~\pref{thepotential} to get
the asymptotics
\bb
V_{12}\simeq g_s^2\frac{R_+^3}{\alp^3 N^4} \lk[
\begin{array}{cl}
S^{(8-D)/2} x^{-D/2} & \mbox{     for     } \tau_2 \ll 1\\
S^{8-D} x^{4-D} & \mbox{     for     } 2 \pi x \ll 1
\end{array}
\re . .
\ee
For $p>3$, $V_{12}\rightarrow 0$ as $x\rightarrow 0$; 
at larger $x$, it decays as $x^{-D/2}$. 
At the thermal wavelength $x\sim 1/S$, both expressions give
\bb
V_{max}\simeq g_s^2 \frac{R_+^3}{\alp^3 N^4} S^4 .
\ee
Note that in this expression the dimension dependence in the power of $S$
conspires to vanish.
For $p=3$, $V_{12}\sim S^4$ for $x\rightarrow 0$, while for $p<3$,
$V_{12}\rightarrow \infty$ for $x\rightarrow 0$;
in both of these latter cases,
the potential decays as $x^{-D/2}$ for larger $x$.

\begin{figure}
\epsfxsize=7cm \centerline{\leavevmode \epsfbox{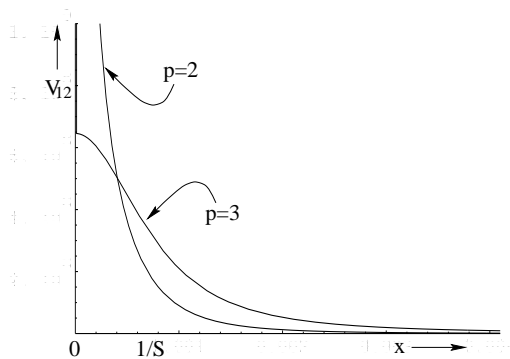}}
\caption{\sl The potential as a function of $x$ 
for dimensions $p=3$ and $p=2$.}
\label{pot2}
\end{figure}

The conclusion can be summarized as follows. 
For $p=4$ and $p=5$, there exists a
bump in the potential of height proportional to $S^4$ at the
thermal wavelength on the string; for $p=3$, the bump smoothes to a flat
configuration where the difference between the potential at the thermal
wavelength separation and at $x=0$ is of order unity. Finally, for $p<3$,
the bump disappears altogether and the potential blows up at the origin
signaling the breakdown of the description.
This potential is plotted in various cases 
in Figure~\ref{potl} of the Introduction and Figure~\ref{pot2}.

The presence or absence of the bump is a result of two competing
effects: First of all, the increasingly singular short-distance
behavior of the Coulomb potential \pref{slinky} 
with increasing dimension $D$;
and secondly, the strong correlation of neighboring points
on the string, which makes $(\del X_1-\del X_2)$
decrease as the separation along the string decreases
(inside a thermal wavelength).

We observe that:
\begin{itemize}
\item
The bump occurs at separations of $1/S$ of a fraction of the 
whole length of the string; in the Matrix theory language, this corresponds to
a bump about matrices of size $N/S$.
\item
The presence or absence of the bump 
as a function of the number of non-compact
space dimensions correlates with the observations of~\cite{CORR2}, given
that in the DLCQ, the light-like direction reduces the number of 
non-compact dimensions by one.
In~\cite{CORR2}, considering the gravitational self-interactions of a highly
excited string, the authors demonstrated that the string collapses
into a black hole at weak string coupling~\pref{corr}
provided that there are less than six non-compact
space dimensions. Correspondingly, we observe the bump in the potential for the
scenarios where we have four or five non-compact space dimensions in
Light-Cone IIA theory. 
\item
As described in~\cite{LIMART}, a Matrix black hole
can be described by SYM excitations clustered within matrices of
size $N/S$, the location of the bump. Furthermore, we will shortly
reproduce, from scaling arguments regarding the dynamics of this
potential, the two correspondence lines determined from thermodynamic
considerations above.
\end{itemize}

\noindent
We then conclude that we have identified the characteristic signature of
black hole formation in the Matrix SYM.

\subsection{Dynamical issues and black hole formation}\label{dynsec}

The dynamics of this potential near 
a phase transition point is certainly complicated. 
Intuitively, we expect that as we approach a critical point, instabilities
develop, an order parameter fluctuates violently, perhaps 
related to some measure of the $Z_N$ symmetry;
it is reasonable to expect the characteristic feature of the potential,
the confining bump, plays a crucial role in the dynamics of the 
emerging phase.
Let us try to extract from these results the scaling of the
correspondence curves.

First let us motivate the use of 
the expectation value of the potential
in the free string ensemble.  We indicated earlier that
this quantity is qualitatively related to the effect of the
interactions, assuming they are weak enough, on the
energy shell in phase space covered by the free string.
The partition function becomes, schematically
\bb
Z \sim \mbox{Tr } e^{H_0+V} \sim e^{\lk< V\re>_0} \mbox{Tr } e^{H_0}\ ,
\ee
so that phase space is weighed by an additional factor related to
the expectation value of the potential in the free ensemble 
$\lk< V\re>_0$. This is also similar to 
the RG procedure applied to the 2d Ising model, where the 
context and interpretation 
is slightly different~\cite{GOLDENFELD}.

Using equation~\pref{thepotential}, 
the potential energy content of the Matrix
string is given by
\bb\label{potwind}
V\sim \int_0^{N\Sigma_p} d\sigma_-\ 
\lk( S^4 g_s^2 \frac{R_+^3}{\alp^3 N^4}\re) \lk( N\Sigma_p \re) v_{12}\ ,
\ee
where we have integrated over one of the two intergals of 
the translationally invariant two-body potential, and scaled $v_{12}$
such that its maximum is of order 1, independent of any state variables;
however, the shape of $v_{12}$ still depends on $N$ and $S$. This
expression represents the interaction energy between two points on the
coiled Matrix string at fixed separation $\sigma_-$. 
From the point of view of Matrix theory physics, the string's fundamental
dynamical degrees of freedom are the windings on the coil; we expect 
a transition in the dynamics of the object 
when there is a competition between forces on an individual winding.
In the present case, the two forces are nearest neighbor elastic
interaction and the gravitational interaction. 
A single string winding being wrapped on $\Sigma_p$ worth of world-sheet,
the maximum potential energy it feels
can be read from equation~\pref{potwind}
\bb\label{vmaxwind}
v_{max}\sim S^4 g_s^2 \frac{R_+^3}{\alp^3 N^4} N\Sigma_p^2 \ ,
\ee
and is due to its interaction with strands a thermal wavelength away. 
Its thermal energy caused by nearest neighbour interactions
is read off equation~\pref{kkk}
\bb\label{kinwind}
\kappa\sim \frac{\lk<K\re>}{\alp} \Sigma_p\ .
\ee
The two forces compete when
\bb \label{corrr1}
S\sim \sqrt{N} g_s^{-1}\ .
\ee
At stronger coupling, the forces due to the gravitational interaction
dominate those of the nearest neighbor stretching and
decohere neighboring strands' velocities.  The free string
evaluation of the interaction, equation~\pref{thepotential},
is no longer valid; one expects a phase transition to occur.
Equation~\pref{corrr1}
is our matching result of~\pref{spec1} between the string and $p+1$d
interacting SYM phase. Here, we are assuming an analytical continuation
of the Matrix string phase to the region $N<S$ in the phase diagram;
our suggestion that this region is associated with a coexistence phase
is consistent with this procedure.

To account for the correspondence curve for $N>S$,
we now recall that in the discussion of clustered
D0 branes of~\cite{LIMART}, the virial treatment of the $v^4/r^7$ 
interaction had to be corrected by a factor in order to reproduce
the black hole equation of state; the origin of this correction was
argued to be interaction processes between the clusters involving
the exchange of longitudinal momentum. Under the assumption that these
effects are of the same order 
as zero momentum transfer processes, a correction
factor of $N/S$ was applied. 
Using a chain of dualities, we can quantify the effect of longitudinal
momentum transfer physics by studying the scattering amplitude
in IIB string theory with winding number exchange. We do this
in Appendix~\ref{stramplapp}, 
where we find that, for exchanges of windings up to order
$N/S$, the winding exchange generates an interaction
identical to that of zero longitudinal momentum exchange; for higher winding
exchanges, the interactions are much weaker. These winding modes,
represent the sections of the Matrix string within the thermal wavelength, 
$N/S$ worth of D-string windings. 
Thus we modify the $v_{12}$ potential above by the factor $N/S$,
which accounts in the scaling analysis for the effect
of longitudinal momentum transfer physics in the Matrix string
self-interaction potential. Applying the virial theorem between 
equation~\pref{kinwind} and $N/S$ times equation~\pref{vmaxwind}
yields the Matrix string-Matrix black hole correspondence
point at
\bb \label{corr2}
S\sim g_s^{-2}\ ,
\ee
as needed.

We can now interpret our results as follows. The bump potential accounts
for the matching of the string phase onto {\it both} 
$N<S$ and $N>S$ phases, one involving partons interacting
without longitudinal momentum exchange 
(the Matrix string-Dp brane curve in
Figure~\ref{fig3}), and the other being the 
Matrix black hole phase of parton clusters of
size $N/S>1$ interacting in addition 
by exchange of longitudinal momentum 
(the Matrix string/Matrix black hole
correspondence curve of Figure~\ref{fig3}). 
In the latter case, the location
of the confining bump correlates with 
matrices of size $N/S$. In the former
case, the correlations are finer than 
the UV matrix cutoff; a better understanding
of this latter issue obviously 
needs a more quantitative analysis of the $N<S$ 
Matrix string regime. This analysis further substantiates the
identification of the bump potential as the signature of black
hole formation from Matrix SYM, as well as justifying the new
Matrix string-Dp brane transition
microscopically.

\appendix

\chapter{A few words about black holes}
\label{grapp}

We review in this appendix some background material relevant to black 
hole physics.
Black holes appear to obey the four laws of thermodynamics with the
following identifications: the area of the horizon $A$ is mapped to
the entropy $S$ of the black hole
\bb\label{Hawking}
S\sim \frac{A}{2\kappa_D^2}\ ,
\ee
where $D$ is the space-time dimension, and $2\kappa_D^2$ is the 
gravitational coupling; the temperature is given by the surface gravity
\footnote{
For example, for a Schwarzschild black hole, the temperature is
\bb
T\sim \lk( 2\kappa_D^2 M\re)^{\frac{1}{3-D}}\ ,
\ee
where $M$ is the mass of the black hole.}; and the mass is identified
with the energy of the given thermodynamic state. 

A geometry in general relativity 
is accorded a mass or energy by studying probe dynamics at
large distances away from the source responsible for the 
curving of the space-time. 
This gives a measure for
the energy content of the system as seen by an observer at infinity.
For a metric of the form~\cite{ADM}
\bb
ds^2=-A(r) dt^2+B(r) dr^2 + r^2 C(r) d\Omega_{D-p-2}^2+D(r) dy^i\ dy^i\ ,
\ee
$i$ running over $p$ space coordinates,
the mass per unit $p$ dimensional volume is given by
\bbb\label{ADMenergy}
M&=&-\frac{\Omega_{D-p-2}}{2\kappa_D^2} \lk[
(D-p-2) r^{D-p-2} \del_r C + p r^{D-p-2} \del_r D \re. \nonumber \\
& - & \lk. (D-p-2) r^{D-p-3} (B-C) \re]_{r\rightarrow \infty}\ ,
\eee
where $\Omega_{D-p-2}$ is the surface area of the unit $D-p-2$-sphere.
Using equations~\pref{Hawking} and~\pref{ADMenergy}, we can
write an equation of state for a black 
geometry. The simplest example is that of a Schwarzchild black hole
in $D$ space-time dimensions. The metric is
\bb
ds^2=-f dt^2+f^{-1} d{{r}}^2 +{{r}}^2 d\Omega_{D-2}^2\ ,
\ee
with $f \equiv 1-\lk(\frac{r_0}{{r}}\re)^{D-3}$.
The equation of state is then given by
\bb\label{bhmass}
M\sim S^{\frac{D-3}{D-2}} \lk(2 \kappa_D^2\re)^{\frac{1}{2-D}}\ .
\ee

We will often need to consider boosted black holes. For this, we change
to isotropic coordinates~\cite{MSSYM123}, 
and apply a large boost to the corresponding geometry.
The Light-Cone energy of a boosted black hole then becomes
$E_{LC}=M^2/p_{11}$, where $M$ is the mass of the black hole given in
equation~\pref{bhmass}.

We say a geometry localizes on a torus parametrized by the
coordinates $y_{(p)}$ when the following transition in the metric
\bb
d{\vec y}_p^2+h^{-1} dr^2+r^2 d\Omega_{D-p-2}^2\rightarrow
h^{-1} dr^2 + r^2 d\Omega_{D-2}
\ee
minimizes the free energy. Here $h$ is a harmonic function whose
scaling changes accordingly
\bb
h=1-\frac{q}{r^{D-p-3}}\rightarrow 1-\frac{q V}{r^{D-3}}\ ,
\ee
where $V$ is the volume of the torus.

When dealing with string theoretical geometries, care must be taken
to convert the string frame metric
to the Einstein frame metric $ds_E^2=e^{-\phi/2} ds_{str}^2$ 
before applying the formulae~\pref{Hawking} and~\pref{ADMenergy}. 
This puts the supergravity action in the canonical form $\int \sqrt{-g}\ R$.
The Dp brane metrics given in the text are written in the string frame.

\chapter{A few comments about dualities}\label{dualapp}

We briefly review in this appendix 
the various duality transformations we make use of in the text.
The reader is referred, for example,to~\cite{GIVEON,SL2Z,NARAIN,NARAINSARMADI,
SCHSENDUAL,SCHSENDUALTWO} for more elaborate expositions to this
subject.

\section{T duality}

T duality relates a string theory compactified
on a radius of size $R$ to another
string theory compactified on the dual radius $\alp/R$. The origin
of this symmetry has to do with the fact that string theories
describe extended objects that
can wrap compact cycles of the background
geometry. In a local field theory, excitations carrying $n$ integer
units of momentum along a cycle
of compactification of size $R$ appear
in the transverse non-compact space as quanta with mass $n/R$. 
As $R\rightarrow 0$, only field excitations constant along the cycle, \ie\ 
having $n=0$, survive the dynamics. In the opposite decompactification limit 
$R\rightarrow \infty$, 
an infinite number of flavors of particles with masses given by
$n/R$ enter the spectrum of the compactified dynamics. Generically, the
emergence,
as a function of a modulus in the theory,
of such an infinite tower of states 
is the signature of a new dimension opening up. In string
theory, as we compactify on a vanishingly small circle of size $R$, 
while quanta carrying momenta along this cycle are scaled out
of the dynamics, strings winding the cycle $w$ times get their masses
$w (2\pi R)/\alp$ progressively smaller; as $R\rightarrow 0$, a tower of
winding string states comes into focus, indicating that the 
theory being compactified may be equivalent to one decompactifying 
with a cycle of size $\alp/R$. Indeed, 
the equivalence between such ``T dual'' theories has been
established even at non-zero string coupling. It is now believed
that T duality is an exact symmetry of string theories.

T duality transformations
relate the various known string theories to each other.
IIA and IIB string theories are interchanged under 
T-duality on an odd number of circles.
Similarly, the two Heterotic theories, $SO(32)$ and $E_8\times E_8$,
are transformed into each other under T duality.
A T duality is also involved in a relation between
the IIA theory on the orbifold $T^4/Z_2$ (or on K3 more generally)
and the Heterotic theory on $T^4$.

The string coupling transforms under T duality as
\bb
\gs \rightarrow \gs \frac{\ls}{R}\ ,
\ee
while the string tension 
is unchanged; these imply that the gravitational coupling is invariant.
As argued above, the cycle of interest gets inverted $R\rightarrow \alp/R$.
We will also need the 
transformation that the low energy supergravity fields undergo. 
We quote here only the forms relevant to our discussion:
\bbb
& g_{aa} & \rightarrow  \frac{1}{g_{aa}}\ ,\ \ 
g_{ax} \rightarrow  \frac{B_{ax}}{g_{aa}}\ ,\ \ 
g_{xy} \rightarrow  g_{xy}-\frac{g_{xa} g_{ay}+B_{xa} B_{ay}}{g_{aa}}\ ,
\nonumber \\
& B_{ax} & \rightarrow  \frac{g_{ax}}{g_{aa}}\ ,\ \ 
B_{xy} \rightarrow  B_{xy} -\frac{g_{xa} B_{ay} + B_{xa} g_{ay}}{g_{aa}}\ ,\ \ 
\phi   \rightarrow  \phi-\frac{1}{2}g_{aa}\ .
\eee
Here $a$ is the direction along which we apply the T duality, $x$ and $y$
are directions transverse to this, $g_{\mu \nu}$ is the metric,
$B_{\mu\nu}$ is the gauge field of the field strength 
$H_{(3)}=dB_{(2)}$ that appears in equation~\pref{Sgrav}, and $\phi$ is
the dilaton. Note the mixing of the metric with the gauge field.

As the different string theories are transformed into each other by the
action of T duality, the degrees of freedom of the theories 
are shuffled amongst each other. Let us denote by $D^p_{xy\ldots}$ a Dp
brane streched in the $x,y,\ldots$ directions; similarly, for an
NS5 brane in the IIA or the IIB theory,
we write $NS5(A/B)_{xy\ldots}$, for the F1 string we write $F1_x$,
and for a wave we write $W_x$.
We denote a T duality transformation along cycles $a,b,\ldots$ by
$T_{ab\ldots}$. We chart some of the flavor mixings under T duality;
note that the condition of nilpotency $T_{ab\ldots}\times T_{ab\ldots}=1$ 
complements the transformations listed below.

\begin{center}
\begin{tabular}{ll}
$T_x \lk(D^p_{xy\ldots} \re)$ & $\rightarrow D^{p-1}_y\ldots$ \\
$T_a \lk(F1_a\re)$ & $\rightarrow W_a$ \\
$T_x \lk(F1_a\re)$ & $\rightarrow F1_a$ \\
$T_x \lk(W_a\re)$ & $\rightarrow W_a$ \\
$T_x \lk(NS5A_{xy\ldots}\re)$ & $\rightarrow NS5B_{xy\ldots}$ \\
$T_a \lk(NS5A_{xy\ldots}\re)$ & $\rightarrow NS5B_{xy\ldots}$
\end{tabular}
\end{center}

Typically, the full group
structure of T dualities for a $D$ dimensional IIA or IIB string theory
is given by the orthogonal group over the integers $O(10-D,10-D;{\bf Z})$.
This group for the Heterotic theories is $O(26-D,10-D;{\bf Z})$.

\section{S duality}

S duality relates a theory at weak coupling to a theory at strong coupling.
This can be a symmetry within a single theory; this is the case for
type IIB string theory, which is said to be self-dual. The Heterotic
and type I string theories on the other hand transform into each other
under S duality. The action on the string coupling and tension is given by
\bb
\gs\rightarrow 1/\gs\ ,
\ee
\bb
\alp\rightarrow \alp \gs\ ,
\ee
which implies that the gravitational coupling is invariant.
The action on the low energy supergravity fields relevant to our discussion
is given by
\bb
\phi\rightarrow -\phi\ ,\ \  
H_{(3)} \rightarrow F_{(3)}\ ,\ \ 
F_{(3)} \rightarrow -H_{(3)}\ .
\ .
\ee
$F_{(3)}$ is the field strength coupling to the D1 or D5 branes (see
equation~\pref{Sgrav}).
Charting the mapping of some of the
states as was done in the previous section, we 
have (note that $S^2=1$ as in the case of T duality):

\begin{center}
\begin{tabular}{ll}
In IIB theory & \\
$S\lk(D^3_{xyz}\re)$ & $\rightarrow D^3_{xyz}$ \\
$S\lk(D^5_{xy\ldots}\re)$ & $\rightarrow NS5B_{xy\ldots}$ \\
In the IIB, and type I-Het. cases &\\
$S\lk(D^1_x\re)$ & $\rightarrow F1_x$ \\
$S\lk(W_x\re)$ & $\rightarrow W_x$
\end{tabular}
\end{center}

Strictly speaking, S duality is still a conjectured symmetry. Compelling
evidence exists in favor of it being an exact symmetry
of string theories. The full group structure is typically $SL(2,Z)$.

\section{M duality}

M duality is the relation between the strong coupling regime of IIA
theory and M theory which we discussed in the Introduction at some length.
The map between the parameters of the two theories 
was given in equation~\pref{MIIA}.
The low energy fields are related to each other via the
standard Kaluza-Klein dimensional reduction prescription; the metric
in eleven dimensions is
\bb
ds_{11}^2=e^{-2\phi/3} ds_{10}^2+
e^{4\phi/3} \lk( dx_{11} + A_\mu dx^\mu\re)^2\ ,
\ee
where $ds_{10}^2$ is the ten dimensional metric in the string frame,
and $A_{\mu}$ is the gauge field of the field strength $F_{(2)}$
coupling to D0 branes (see equation~\pref{Sgrav}).
We remind the reader the relation between gravitational couplings
that arises in this procedure
\bb
2\kappa_D^2=2\kappa_{D-1}^2 (2\pi R)\ ,
\ee
where $R$ is the cycle size, \ie\ $R_{11}$ in the case of M duality.

Finally, we chart the table of state mapping for this duality:

\begin{center}
\begin{tabular}{ll}
$M \lk( D^0\re)$ & $\rightarrow W11$ \\
$M \lk( D^2_{xy}\re)$ & $\rightarrow M2_{xy}$ \\
$M \lk( D^4_{xy\ldots}\re)$ & $\rightarrow M5_{xy11\ldots}$ \\
$M \lk( D^6_{xy\ldots}\re)$ & $\rightarrow\ $KK Monopole (ALE space) \\
$M \lk( F1_x\re)$ & $\rightarrow M2_{x11}$ \\
$M \lk( W_x\re)$ & $\rightarrow W_x$ \\
$M \lk( NS5A_{xy\ldots}\re)$ & $\rightarrow M5_{xy\ldots}$
\end{tabular}
\end{center}

\begin{figure}
\epsfxsize=10cm \centerline{\leavevmode \epsfbox{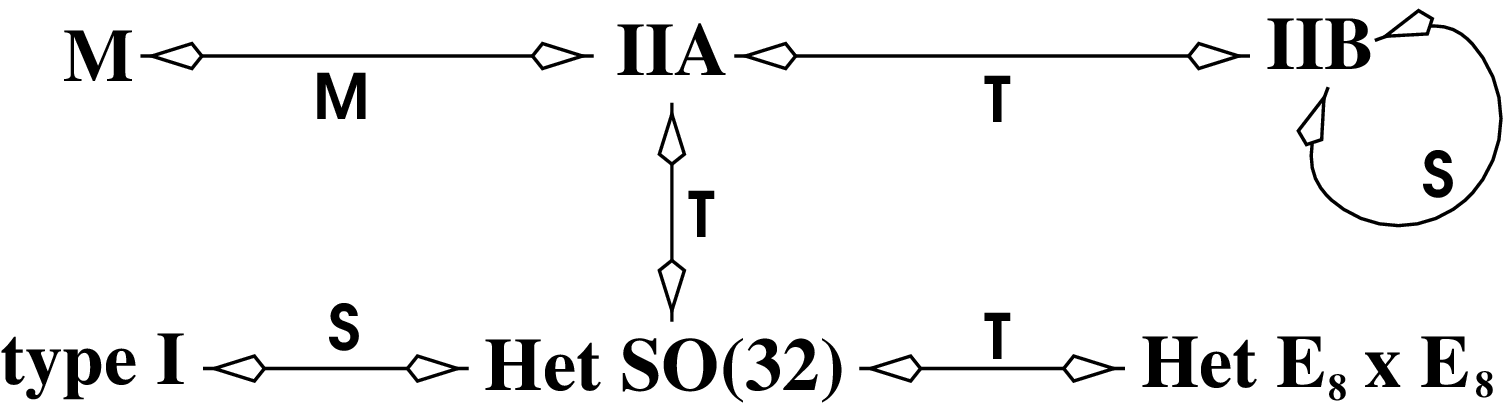}}
\caption{\sl Some of the dualities among the various string theories and
M theory.}
\label{dualities}
\end{figure}

Figure~\ref{dualities} summarizes pictorially some of the various duality
connections we make use of in the text.

\chapter{Scaling of transition curves and equations of states}
\label{thedetailsapp}

\section{Summary}

In the subsequent sections, we will tabulate the scaling of
the transition curves and equations of states for the phase
diagrams studied in the text. The systematic derivation of these
formulae can be found in~\cite{LMS,MSSYM123,MSFIVE}. Note that we use
a uniform labeling scheme across the SYM diagrams.

\section{SYM 1+1}\label{pplusone}

\begin{figure}[p]
\epsfxsize=9cm \centerline{\leavevmode \epsfbox{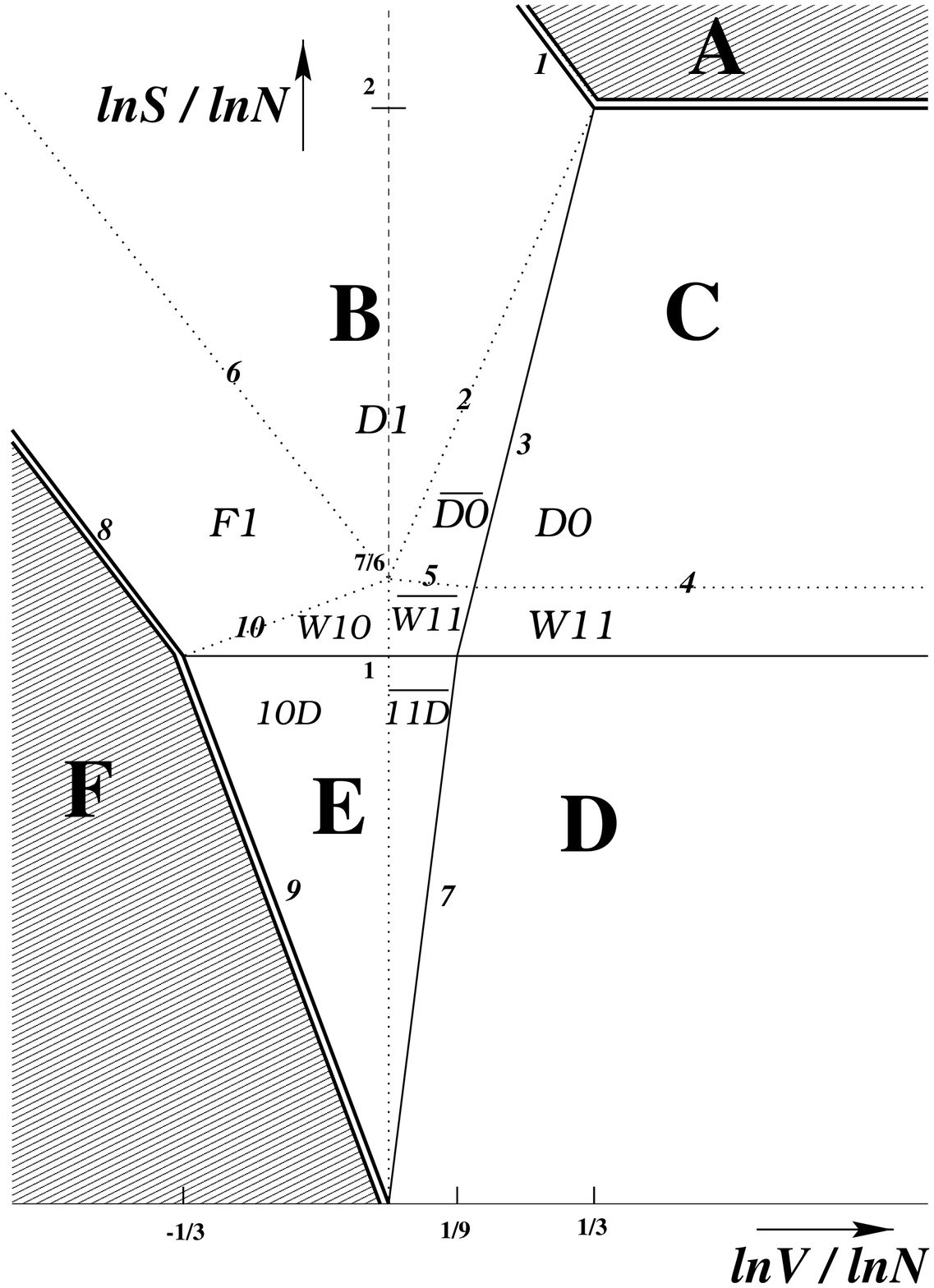}}
\caption{\sl The thermodynamic phase diagram of $1+1$d SYM. 
.}
\label{SYM1sfig}
\end{figure}

\begin{center}
{\em Equations of state of bulk phases for $1+1$d ($p=1$) 
SYM (Figure~\ref{SYM1sfig})}

\vspace{0.5cm}
\begin{tabular}{ll}
Phase A: &
$ E\sim \lk(\frac{\r11}{N} \frac{1}{\lp^2}\re) V N^{(p-2)/p}
S^{(p+1)/p}\ .  $ \\
Phase B: &
$ E\sim \frac{\r11}{\lp^2} \lk( \frac{S^2}{N} \re)^{(7-p)/(9-p)}
V^{2p/(9-p)}\ . $ \\
Phase C: &
$ E\sim \lk(\frac{\r11}{N} \frac{1}{\lp^2}\re) S^{14/9} N^{2/9}\ . $ \\
Phase D: &
$ E\sim \lk(\frac{\r11}{N} \frac{1}{\lp^2}\re) S^{16/9}\ .  $ \\
Phase E: &
$ E\sim \lk( \frac{\r11}{N} \frac{1}{\lp^2} \re) V^{2p/(9-p)} 
S^{2(8-p)/(9-p)}\ .  $ \\
Phase F: &
$ E\sim \lk(\frac{\r11}{N} \frac{1}{\lp^2}\re) V S^2\ .  $
\end{tabular}
\end{center}

\vspace{1cm}
\begin{center}
{\em Scaling of transition curves for $1+1$d ($p=1$) SYM}

\vspace{0.5cm}
\begin{tabular}{ll}
Curve 1:& $ S\sim N^{(6-p)/(3-p)} V^{-3p/(3-p)}\ .  $ \\
Curve 2:& $ S\sim N^{(8-p)/(7-p)} V^{3(6-p)/(7-p)}\ .  $ \\
Curve 3:& $ S\sim V^{9/2} N^{1/2}\ . $ \\
Curve 4:& $ S\sim N^{8/7}\ .  $ \\
Curve 5:& $ S\sim  N^{(8-p)/(7-p)} V^{p/(p-7)}\ .  $ \\
Curve 6:& $ S \sim N^{(8-p)/(7-p)} V^{3p/(p-3)}\ .  $ \\
Curve 7:& $ S\sim V^9\ .  $ \\
Curve 8:& $ S\sim V^{-3/2} N^{1/2}\ .  $ \\
Curve 9:& $ S\sim V^{-3}\ .  $ \\
Curve 10:& $ S\sim V^{1/2} N^{7/6}\ .  $
\end{tabular}
\end{center}

\newpage
\section{SYM 2+1}

\begin{figure}
\epsfxsize=7cm \centerline{\leavevmode \epsfbox{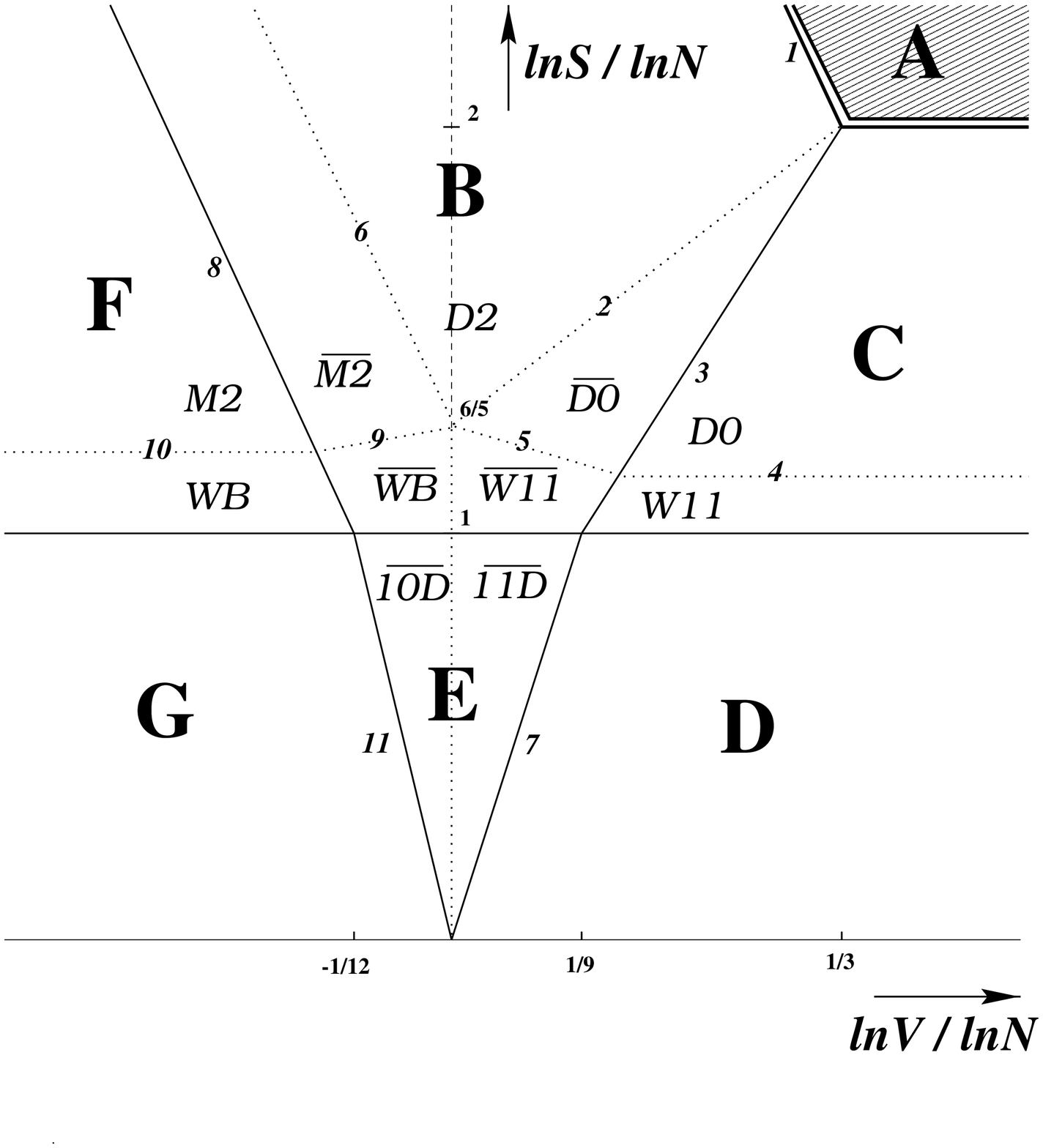}}
\caption{\sl The thermodynamic phase diagram of $2+1$d SYM. 
.}
\label{SYM2sfig}
\end{figure}

\begin{center}
{\em Equations of state of bulk phases for $2+1$d SYM (Figure~\ref{SYM2sfig})}

\vspace{0.5cm}
\begin{tabular}{ll}
Phases A-E:& As in Section~\ref{pplusone} with $p=2$\\
Phase F:&
$ E\sim \lk(\frac{\r11}{N}\frac{1}{\lp^2} \re) V S^{3/2} N^{1/4}\ .  $ \\
Phase G:& $ E\sim \lk(\frac{\r11}{N}\frac{1}{\lp^2} \re) V S^{7/4}\ .  $
\end{tabular}
\end{center}

\vspace{1cm}
\begin{center}
{\em Scaling of transition curves for $2+1$d SYM}

\vspace{0.5cm}
\begin{tabular}{ll}
Curves 1-7:&  As in Section~\ref{pplusone} with $p=2$\\
Curve 8:& $ S\sim V^{-6} N^{1/2}\ .  $ \\
Curve 9:& $ S\sim V^{3/10} N^{6/5}\ .  $ \\
Curve 10:& $ S\sim N^{7/6}\ .  $ \\
Curve 11:& $ S\sim V^{-12}\ .  $
\end{tabular}
\end{center}

\newpage
\section{SYM 3+1}

\begin{figure}
\epsfxsize=10cm \centerline{\leavevmode \epsfbox{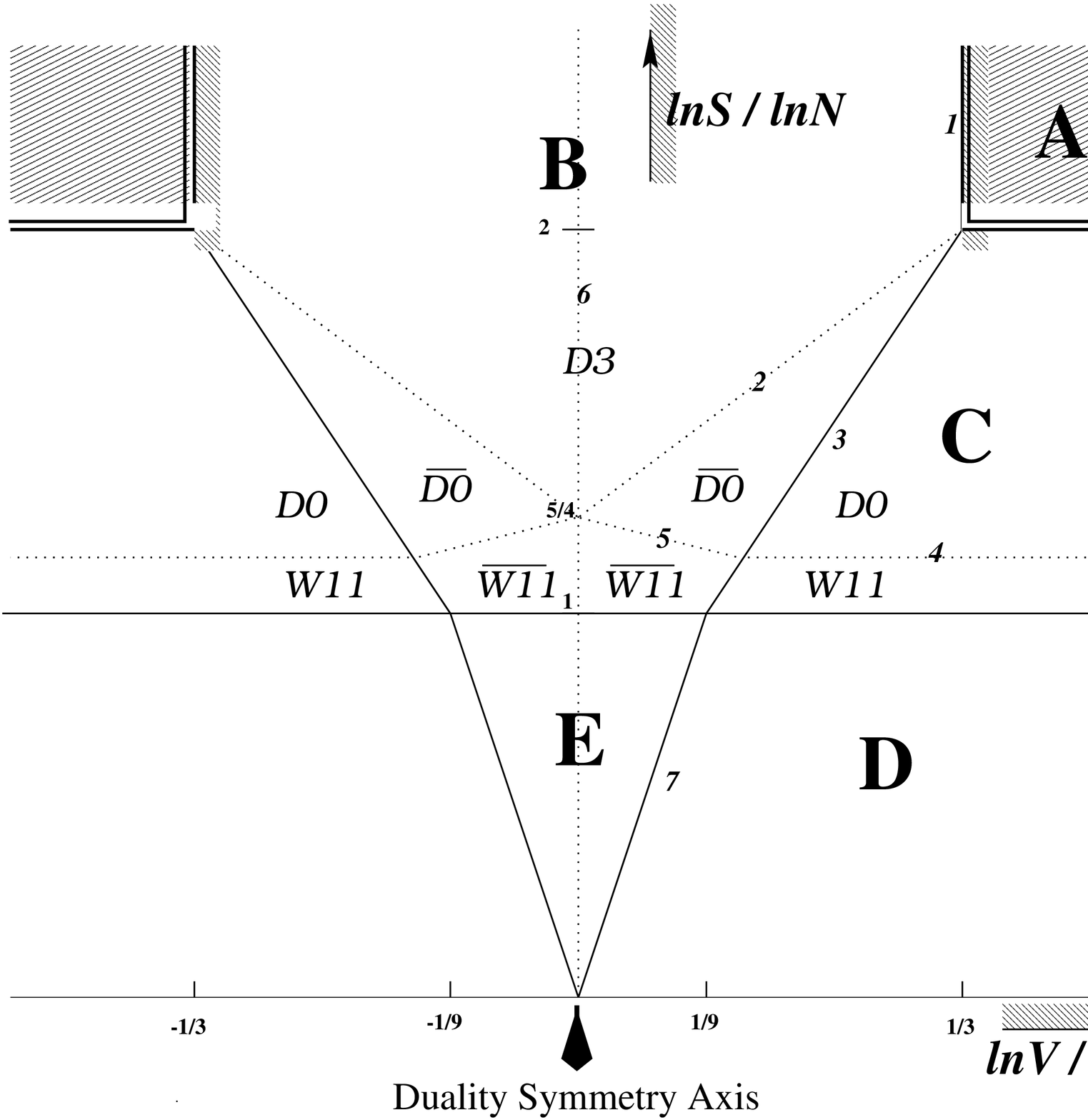}}
\caption{\sl The thermodynamic phase diagram of $3+1$d SYM. 
.}
\label{SYM3sfig}
\end{figure}

\begin{center}
Equations of states and scaling of transition curves for $3+1$d
SYM (see Figure~\ref{SYM3sfig}) are tabulated in Section~\ref{pplusone}
with $p=3$.
\end{center}

\newpage
\section{$(2,0)$ on $T^4\times S^1$}

\begin{figure}
\epsfxsize=7cm \centerline{\leavevmode \epsfbox{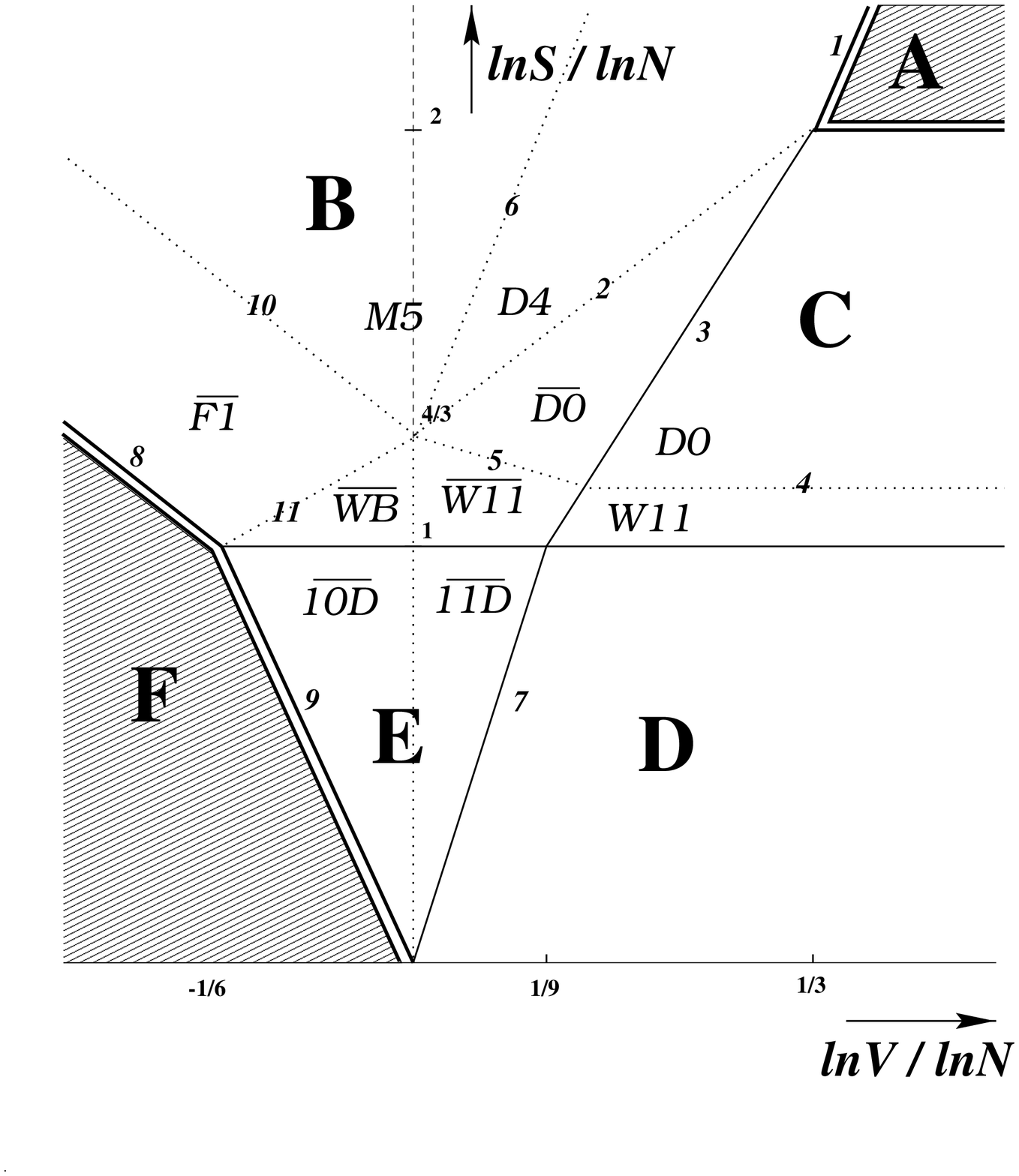}}
\caption{\sl The thermodynamic phase diagram of the $(2,0)$ on $T^4\times S^1$. 
.}
\label{SYM4sfig}
\end{figure}

\begin{center}
{\em Equations of states of bulk phases for $4+1$d SYM (Figure~\ref{SYM4sfig})}

\vspace{0.5cm}
\begin{tabular}{ll}
Phases A-E:& As in Section~\ref{pplusone} with $p=4$\\
Phase F:& $ E\sim \frac{\r11}{\lp^2} N^{-1} S^2 V^4\ .$
\end{tabular}
\end{center}

\vspace{1cm}
\begin{center}
{\em Scaling of transition curves for $4+1$d SYM}

\vspace{0.5cm}
\begin{tabular}{ll}
Curves 1-7:&  As in Section~\ref{pplusone} with $p=4$\\
Curve 8:& $ S\sim V^{-3} N^{1/2}\ .  $ \\
Curve 9:& $ S\sim V^{-6}\ .  $ \\
Curve 10:& $ S\sim V^{-3} N^{4/3}\ .  $ \\
Curve 11:& $ S\sim V^2 N^{4/3} \ .  $
\end{tabular}
\end{center}

\newpage
\section{$(2,0)$ on $T^5$}

\begin{figure}
\epsfxsize=7cm \centerline{\leavevmode \epsfbox{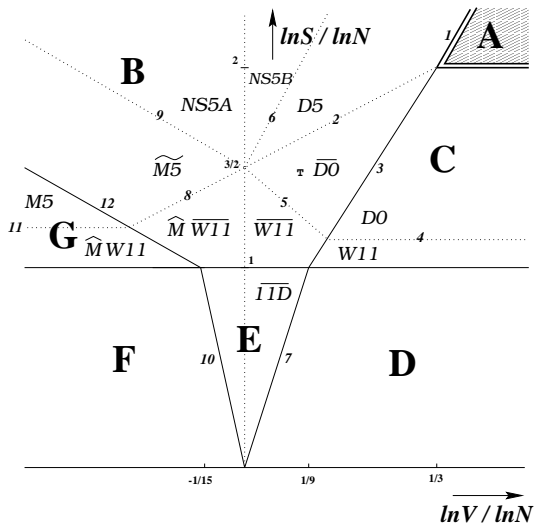}}
\caption{\sl The thermodynamic phase diagram of the $(2,0)$ on $T^5$. 
.}
\label{SYM5sfig}
\end{figure}

\begin{center}
{\em Equations of states of bulk phases for $5+1$d SYM (Figure~\ref{SYM5sfig})}

\vspace{0.5cm}
\begin{tabular}{ll}
Phases A-E:& As in Section~\ref{pplusone} with $p=5$\\
Phase F:& $ E\sim \frac{\r11}{\lp^2} N^{-1} V^4 S^{8/5}\ .  $ \\
Phase G:& $ E\sim \frac{\r11}{\lp^2} S^{6/5} V^4 N^{-3/5}\ . $
\end{tabular}
\end{center}

\vspace{1cm}
\begin{center}
{\em Scaling of transition curves for $5+1$d SYM}

\vspace{0.5cm}
\begin{tabular}{ll}
Curves 1-7:& As in Section~\ref{pplusone} with $p=5$\\
Curve 8:& $ S\sim V^{3/2} N^{3/2}\ . $ \\
Curve 9:& $ S\sim V^{-15/2} N^{3/2}\ . $ \\
Curve 10:& $ S\sim V^{-15}\ . $ \\
Curve 11:& $ S\sim N^{4/3}\ . $ \\
Curve 12:& $ S\sim V^{-15/2} N^{1/2}\ . $ 
\end{tabular}
\end{center}

\newpage
\section{SYM 6+1}

\begin{figure}
\epsfxsize=7cm \centerline{\leavevmode \epsfbox{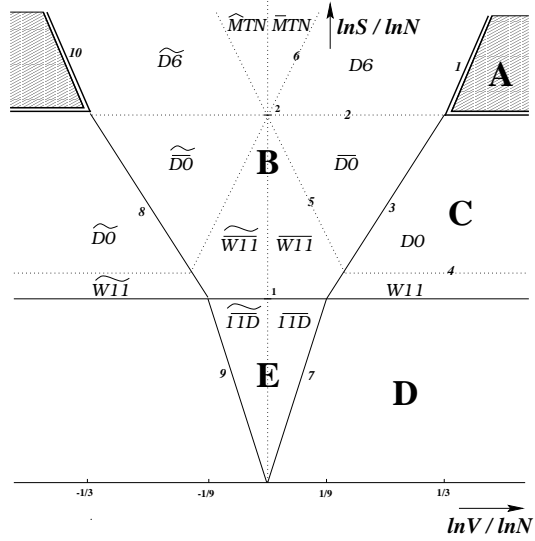}}
\caption{\sl The thermodynamic phase diagram of the D6 system. 
.}
\label{SYM6sfig}
\end{figure}

\begin{center}
{\em Equations of states of bulk phases for $6+1$d SYM (Figure~\ref{SYM4sfig})
are as in Section~\ref{pplusone}} with $p=6$.
\end{center}

\begin{center}
{\em Scaling of transition curves for $6+1$d SYM}

\vspace{0.5cm}
\begin{tabular}{ll}
Curves 1-7:& As in Section~\ref{pplusone} with $p=6$\\
Curve 8:& $ S\sim V^{-9/2} N^{1/2}\ . $ \\
Curve 9:& $ S\sim V^{-9} \ . $ \\
Curve 10:& $ S\sim V^{-6}\ . $
\end{tabular}
\end{center}

\newpage
\section{Little strings with winding charge}\label{winding}

\begin{figure}
\epsfxsize=7cm \centerline{\leavevmode \epsfbox{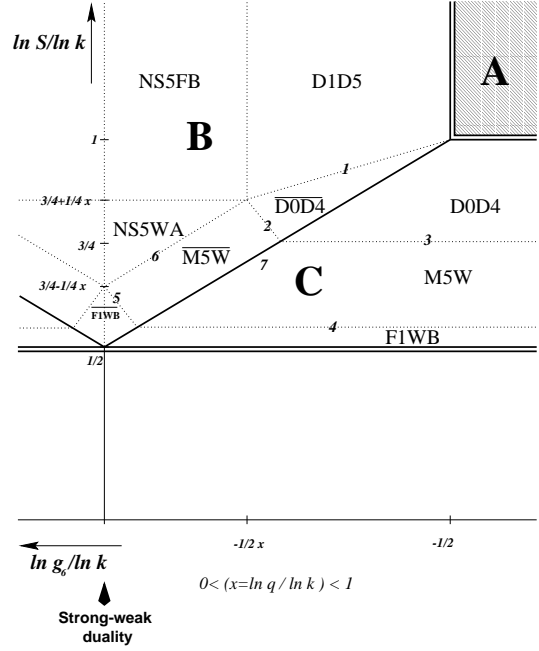}}
\caption{\sl The thermodynamic phase diagram of D1D5 system. 
.}
\label{D1D5sfig}
\end{figure}

\begin{center}
{\em Equations of states of bulk phases for D1D5 system (Figure~\ref{D1D5sfig})}

\vspace{0.5cm}
\begin{tabular}{ll}
Phases A and B:& $ E=\frac{S^2}{8\pi^2 k R_5}\ . $ \\
Phase C:& $ E\sim \frac{g_6}{R_5} \lk(\frac{S}{k^{1/2}}\re)^3\ . $
\end{tabular}
\end{center}

\vspace{1cm}
\begin{center}
{\em Scaling of transition curves for the D1D5 system}

\vspace{0.5cm}
\begin{tabular}{ll}
Curve 1:& $ S\sim g_6^{-1/2} k^{3/4}\ . $ \\
Curve 2:& $ S\sim g_6^{1/2} k^{3/4} q^{1/2}\ . $ \\
Curve 3:& $ S\sim k^{2/3} q^{1/3}\ . $ \\
Curve 4:& $ S\sim k^{2/3} q^{-1/6}\ . $ \\
Curve 5:& $ S\sim g_6^{1/2} k^{3/4} q^{-1/4}\ . $ \\
Curve 6:& $ S\sim g_6^{-1} k^{3/4} q^{-1/4}\ . $ \\
Curve 7:& $ S\sim g_6^{-1} k^{1/2}\ . $
\end{tabular}
\end{center}

\chapter{The $(2,0)$ theory on $T_4/Z_2 \times S^1$}\label{orbapp}

Inspired by the discussion regarding the phase structure of the $(2,0)$
on $T^4\times S^1$, we further consider the phase structure of this theory
on $T^4/Z_2\times S^1$. This corresponds to a corner in the moduli space of 
$K3\times S^1$; particularly, in addition to considering
a square $T^4$, we will be ignoring phase dynamics associated
with the $16\times 4$ moduli that blow up the fixed points~\cite{ASPINWALLREV}.
Our parameter space is again two dimensional, entropy $S$ and the volume of
the $T^4$. There are only two novelties that arise,
both leaving the global structure of the phase
diagram unchanged, modifying only the interpretation of the various patches
of geometry. 

The first change arises from the effect of the orbifold on the duality 
transformations; we will obviously be driven into the other branch of
the web of dualities that converge onto M theory (\cf~\cite{POLCHV2}). 
We proceed from the
$\overline{11D}$ phase of the previous discussion, upward and counter-
clockwise on the phase diagram. We have M theory on a light-cone
circle times $T^4/Z_2$.
We reduce on $\r11$ to D0 branes in IIA living on the
$T^4/Z_2$. 
Under this orbifold, the massless spectrum has
positive parity eigenvalue. We T dualize on $T^4$, 
getting to the
patch of $D4$ branes in IIA wrapped
on $T^4/Z_2$. We remind the reader of the
transformation
\bb
T_{(4)}\beta_{(4)}T_{(4)}^{-1}=\beta_{(4)}\ ,
\ee
where we have used the properties of the reflection operator on the
spinors
\bb
\beta_i=\Gamma \Gamma^i\quad , \qquad
\beta_i^2=(-1)^{F_L}\quad , \qquad
\lk\{ \beta_i,\beta_j\re\}=0\ ,
\ee
with the T duality operation
reflecting the left moving spinors only. Here, $(-1)^{F_L}$
is the left moving fermion operator. 
We then M lift to M5 branes
in $\wtilde{M}$ theory on $T^4/Z_2\times S^1$. 
Next, we have to apply
the chain of dualities $M,T_{(3)},S$. 
From the M reduction we obtain
$\wtilde{D4}$ branes on $T^3/(-1)^{F_L}\Omega$. This is because the
M reduction along an orbifold direction yields the twist eigenvalues for
the massless spectrum
\bb
g_{\mu\nu}\ +;\ \ \phi\ +;\ \ B_{\mu\nu}\ -;\ \ C^{(1)}\ -;\ \ C^{(3)}\ +\ ,
\ee
while the world-sheet parity operator $\Omega$ acts on this spectrum as
\bb
g_{\mu \nu}\ +;\ \phi +;\ B_{\mu\nu}\ -;\ C^{(1),(2),(5),(6)}\ +;\ 
C^{(0),(3),(4),(7),(8)}\ -\ ,
\ee
and the action of $(-1)^{F_L}$ yields
\bb
\rm NSNS\ +\quad;\qquad RR -\ .
\ee
The T duality on $T^3$ brings us to $D1$ branes in IIB theory on
$S^1\times T^3/\Omega$, which is type I theory on $S^1\times T^3$.
This is because
\bb
T_{(3)} \beta_{(3)} (-1)^{F_L} \Omega T_{(3)}^{-1}=(-1)^{F_L}\Omega\ .
\ee
Finally, the S duality culminates in the geometry of $N$ black Heterotic 
strings smeared on the $T^3$. 
The Horowitz-Polchinski correspondence 
curve patches this phase onto that of the
Heterotic Matrix string phase.
We thus verify the following previous
suggestions~\cite{BR,DIACGOMISMS,BANKSMOTL,MOTLSUSSKIND} 
from the perspective of Maldacena's conjecture:

\begin{itemize}
\item Heterotic Matrix string theory emerges in the UV of the $(2,0)$ theory.

\item Heterotic Matrix strings can be described via the $O(N)$ SYM of type I
D strings
\end{itemize}

\noindent
The structure of the phase diagram has not changed, but the labeling of
some of the phases has.
The additional symmetry structure of the orbifold background entered our
discussion trivially; the critical behaviors are unaffected.

To complete the discussion, we need to address a second change 
to the $T^4$ compactification.  The
localization transitions
are of a somewhat different nature than the ones encountered
earlier.  Localized black geometries on orbifold
backgrounds are unstable toward collapse toward the nearest fixed point;
by virtue of being above extremality, there are static forces, and by
virtue of the symmetry structure of the orbifold, there is no balance of
forces as in the toroidal case. 
It is then most probable that the localized D0 branes sit at the
orbifold points, with their black horizons surrounding the
singularity. The most natural geometry is the one corresponding to 16 black
D0 geometries distributed among the singularities, yielding a non-singular
geometry outside the horizons.

\chapter{Calculation of the Potential}\label{potapp}

We need to evaluate
\bb \label{VV}
\mathcal{V}\equiv \lk< \frac{
\lk\{(\del_+X_r)^2 (\del_-X_r)^2-\lk[ (\del_+X_r)^2+(\del_-X_r)^2\re] 
(\del_+X_r.\del_-X_r)
\re\}}{\lk(X_r^2\re)^{D/2}}\re>\ ,
\ee
in the finite temperature vacuum of the SYM.
Let subscripts $(123456)$ denote the argument of $X$, \eg\
$X_1\equiv X(\sigma_1)$.
Writing $X_r\equiv X_5-X_6$, we will encounter in the numerator
only factors of the form
\bb \label{ugly1}
\del_\alpha X^i_1 \del_\alpha X^i_2 \del_\beta X^j_3 \del_\gamma X^j_4\ ,
\ee
with the target indices $i$, $j$ summed over; 
$\alpha$, $\beta$, $\gamma$ are worldsheet indices $\pm$;
and the labels $(1234)$ are set equal to $5$ and $6$ in various ways.
By expanding the numerator of equation~\pref{VV}, 
we get $3\times 16$ terms of the form claimed.
We can write our desired `monomial'~\pref{ugly1} as
\bb \label{monom}
\del_\alpha^1 \del_\alpha^2 \del_\beta^3 \del_\gamma^4
\lk< \frac{X_1^{i} X_2^{i} X_3^{j} X_4^{j}}{((X_5-X_6)^2)^{D/2}} \re>\ .
\ee

Consider
\bbb 
\lk< \frac{X_1^{i} X_2^{i} X_3^{j} X_4^{j}}{((X_5-X_6)^2)^{D/2}} \re>
&=& \frac{\pi^{-d/2}}{\Gamma(D/2)} 
\int_0^\infty ds\ \int d^dp\ s^{(D/2)-1} e^{-p^2}
\delta_1^i\delta_2^i \delta_3^j 
\delta_4^j \lk <e^{\int \tilde{J}.X}\re > \nonumber\\
&=& \frac{\pi^{-d/2}}{\Gamma(D/2)} 
\int_0^\infty ds\ \int d^dp\ s^{(D/2)-1} e^{-p^2}
\delta_1^i \delta_2^i \delta_3^j \delta_4^j e^{\Delta}
\eee
where
\bb
\tilde{J^i}\equiv J^i+2 i \sqrt{s} 
(\delta(\sigma-\sigma_5)-\delta(\sigma-\sigma_6))p^i
\ee
and
\bb
\Delta\equiv \frac{1}{4} \int \tilde{J} K \tilde{J}=
\frac{1}{4} \int J K J+i\sqrt{s} \int J.p K_x-2sp^2 f^2
\ee
We have defined
\bb
K_x\equiv K_{x5}-K_{x6}
\ee
\bb
f^2\equiv K-K_{56}
\ee
Here $K_{ab}$ means $K(a-b)$, the Green's function of the two dimensional
Laplacian
\bb
K_{ab}\equiv -\alp \vev{X_a X_b}\ ,
\ee
and $K\equiv K_{aa}$. The rest is an exercise in combinatorics,
making use of
\bb
\delta_a^i e^\Delta= \lk[ \frac{1}{2} 
\int G_{ax} J^{i}+i\sqrt{s} p^i K_a\re] e^\Delta
\ee
where the $x$ subscript is integrated over 
and it is implied to be the argument of the $J$ as well.
Denoting the number of polarizations in the Lorentz indices by $d$, we get
\bbb
\lk< \frac{X_1^{i} X_2^{i} X_3^{j} X_4^{j}}{((X_5-X_6)^2)^{D/2}} \re>
&=&\frac{\pi^{-d/2}}{\Gamma(D/2)} \int ds \int d^dp\ e^{-p^2} e^{-2sp^2f^2} 
\nonumber \\
\hskip 1cm&\times &
\lk[ T_1 s^{(D/2)-1}-\frac{1}{2}p^2 s^{D/2} 
T_2+(p^2)^2 s^{(D/2)+1} T_4 \re]
\eee
where we have defined
\bb
T_1\equiv \frac{d^2}{4} K_{12} K_{34}+\frac{d}{4} 
K_{13}K_{24} +\frac{d}{4} K_{23} K_{14}
\ee
\bb
T_2\equiv d K_1 K_2 K_{34}+d K_3 K_4 K_{12}+
K_1K_3 K_{24}+K_1K_4K_{23}+K_2K_3K_{14}+K_2K_4K_{13}
\ee
\bb
T_4\equiv K_1 K_2 K_3 K_4
\ee
Evaluating the $s$ integral, we get
\bb
\lk< \frac{X_1^{i} X_2^{i} X_3^{j} X_4^{j}}{((X_5-X_6)^2)^{D/2}} \re>
=\frac{\pi^{-d/2}}{2^{D/2} (f^2)^{D/2}}\int d^dp\ e^{-p^2}(p^2)^{-D/2}
\lk[ T_1-\frac{D}{8f^2} T_2+\frac{(D+2)D}{16 (f^2)^2} T_4 \re]
\ee
Evaluating the $p$ integrals, we get
\bb
\lk< \frac{X_1^{i} X_2^{i} X_3^{j} X_4^{j}}{((X_5-X_6)^2)^{D/2}} \re>
=\frac{\pi^{-d/2}}{2^{(D/2)+1} (f^2)^{D/2}}\Omega_{d-1}
\Gamma\lk(\frac{d-D}{2}\re) \lk[
T_1-\frac{D}{8f^2} T_2+\frac{(D+2)D}{16 (f^2)^2} T_4 \re]
\ee
where $\Omega_{d-1}$ is the volume of the $d-1$ unit sphere.

Going back to~\pref{monom}, we need to differentiate $T_1$, $T_2$ and $T_4$,
according to the map $1234\rightarrow \alpha \alpha \beta \gamma$. Let us
denote the derivatives by superscripts on the $K$'s. We then have
\bb\label{T1a}
T_1^{\alpha \alpha \beta \gamma}=\frac{d^2}{4} K_{12}^{\alpha \alpha}
K_{34}^{\beta \gamma}+\frac{d}{4} 
K_{13}^{\alpha \beta} K_{24}^{\alpha \gamma}
+\frac{d}{4} K_{23}^{\alpha \beta} K_{14}^{\alpha \gamma}\ ,
\ee
\bbb
T_2^{\alpha \alpha \beta \gamma}&=&
d K_1^\alpha K_2^\alpha K_{34}^{\beta \gamma}
+d K_3^\beta K_4^\gamma K_{12}^{\alpha \alpha}+
K_1^\alpha K_3^\beta K_{24}^{\alpha \gamma} \nonumber \\
&+&K_1^\alpha K_4^\gamma K_{23}^{\alpha \beta}
+K_2^\alpha K_3^\beta K_{14}^{\alpha \gamma}+
K_2^\alpha K_4^\gamma K_{13}^{\alpha \beta}\ ,
\eee
\bb\label{T4a}
T_4^{\alpha \alpha \beta \gamma}=
K_1^\alpha K_2^\alpha K_3^\beta K_4^\gamma\ .
\ee
We have used here the translational invariance and 
evenness of the Green's function to interpret the
derivatives as differentiations with respect to the argument $i-j$ of the 
Green's functions (and therefore note some flip of signs); furthermore,
we assume that $K$, $K^\alpha$ 
and $K^{\alpha \beta}$ are zero, \ie\ because of
subtraction of the zero temperature limits, or throwing away bubble 
diagrams. This, it turns out, is not necessary 
for the potential we calculate,
since all expressions would have come out 
as differences, say $K^{\alpha \beta}_{12}
-K^{\alpha \beta}$; it is just convenient for notational purposes to
throw them out from the start. We also note the identities
$K^{\alpha \beta}_{12}=K^{\alpha \beta}_{21}$ 
and $K^\alpha_5=-K_{56}^\alpha= K^\alpha_6$. 
For each term in equations~\pref{T1a}-\pref{T4a}, we have
16 terms associated with taking a map from 
$(1234)$ to a sequence of $5$'s and
$6$'s. This combinatorics yields
\bb
\lk< \frac{
\del_\alpha X_r^i \del_\alpha X_r^i \del_\beta X_r^j \del_\gamma X_r^j
}{\lk(X_r^2\re)^{D/2}}\re> \simeq
\frac{K_{56}^{\alpha \beta} K_{56}^{\alpha \gamma}+
\frac{d}{2} K_{56}^{\alpha \alpha} 
K_{56}^{\beta \gamma}}{(-K_{56})^{D/2}}\ .
\ee
Note that $T_2$ and $T_4$ cancelled; 
we have also dropped numerical coefficients.
There are three terms in~\pref{VV} of this type; this yields 
\bb
\mathcal{V}\simeq \frac{K_{56}^{+-} K_{56}^{+-}
+\frac{d}{2} K_{56}^{++} K_{56}^{--}
-\lk( \frac{d}{2}+1 \re) \lk( K_{56}^{++} K_{56}^{+-}
+K_{56}^{--} K_{56}^{+-} \re)
}{(-K_{56})^{D/2}}\ .
\ee
Using the equation of motion 
(delta singularity subtracted) $K_{56}^{+-}=0$,
we get
\bb
\mathcal{V}\simeq \frac{K_{56}^{++} K_{56}^{--}}{(-K_{56})^{D/2}}\ .
\ee
Using Euclidean time $i \tau=t$, 
we have $\sigma^\pm=\sigma\pm t=z,\bar{z}$;
finally, we get for equation~\pref{slinky}
\bb
V_{12} = \alpha_D g_s^2 \frac{\ls^{5-p}}{R_+}
\frac{K^{zz}_{12} K^{\bar{z}\bar{z}}_{12}}{\lk( -K_{12} \re)^{D/2}} .
\ee

\chapter{Longitudinal momentum transfer effects}\label{stramplapp}

Consider the scattering of two wound strings in IIB theory with winding
number exchange.  We will find that, in the regime of small 
momentum transfer, the interaction
is Coulombic for resonances involving
low enough winding number exchange,
and much weaker otherwise; furthermore, the Coulombic
interaction is winding number independent, and the cumulative
strength of this potential suggests modifying
the Matrix string potential by a factor of $N/S$ for $N>S$.

For simplicity, consider the polarizations of the external states
to be that of the dilaton, and T-dualize the momentum in the compact
direction to winding number.
The resulting four string amplitude is given by~\cite{GSW}
\bb
A_m\sim K_{\alpha \beta \gamma \delta} K^{\alpha \beta \gamma \delta}
\frac{\Gamma(-S\alp/4)\Gamma(-T\alp/4)\Gamma(-U\alp/4)}
{\Gamma(1+S\alp/4) \Gamma(1+T\alp/4)\Gamma(1+U\alp/4)} \ ,
\ee
where $S\equiv -(k_1+k_2)^2$, $T\equiv -(k_2+k_3)^2$, $U\equiv -(k_1+k_3)^2$,
with $S+T+U=0$, and 
\bbb
K^{\alpha \beta \gamma \delta}&=&-\frac{1}{2}\lk(
S T \eta^{\alpha \gamma} \eta^{\beta \delta} + S U \eta^{\beta \gamma}
\eta^{\alpha \delta}+ T U \eta^{\alpha \beta} \eta^{\gamma \delta} \re)
\nonumber \\
&+&S \lk( k_4^\alpha k_2^\gamma \eta^{\beta \delta}+
k_3^\beta k_1^\delta \eta^{\alpha \gamma} + k_3^\alpha k_2^\delta
\eta^{\beta \gamma} + k_4^\beta k_1^\gamma \eta^{\alpha \delta} \re)
\nonumber \\
&+&T \lk( k_4^\gamma k_2^\alpha \eta^{\beta \delta}+
k_3^\delta k_1^\beta \eta^{\alpha \gamma} + k_4^\beta k_3^\alpha
\eta^{\gamma \delta} + k_1^\gamma k_2^\delta \eta^{\alpha \beta} \re)
\nonumber \\
&+&U \lk( k_2^\alpha k_3^\delta \eta^{\beta \gamma}+
k_4^\gamma k_1^\beta \eta^{\alpha \delta} + k_4^\alpha k_3^\beta
\eta^{\gamma \delta} + k_2^\gamma k_1^\delta \eta^{\alpha \beta} \re)\ .
\eee
This gives the amplitude
\bb
A_m\sim \lk[ (S+T)^4+S^4+T^4\re] 
\frac{\Gamma(-S\alp/4)\Gamma(-T\alp/4)\Gamma(-U\alp/4)}
{\Gamma(1+S\alp/4) \Gamma(1+T\alp/4)\Gamma(1+U\alp/4)}\ .
\ee
We want to accord winding 
$n_1$, $n_2$, $n_3$ and $n_4$
to the four strings, on a circle of radius $R$;
without any momenta along this cycle, we can extract easily this process
from the amplitude above by
\bb\label{seq}
s=S+ M^2\ ,
\ee
\bb\label{teq} 
t=T+m^2\equiv-q^2\ ,
\ee
with 
\bb
M^2\equiv \lk( \frac{R (n_1+n_2)}{\alp}\re)^2\ ,
\ee
\bb
m^2\equiv \lk( \frac{R (n_3-n_2)}{\alp}\re)^2\ .
\ee
For large $m_1,m_2$, and small $m$, $q^2$ is the 
spatial momentum transfer between the strings in the
center-of-mass frame.  Thus $M\gg m$, and we are in
the non-relativistic regime $E_{cm}^2\gg q^2$.
From equations~\pref{seq} and \pref{teq}, 
we see that $S\gg T$.  Using this and
the identities $\Gamma(z) \Gamma(1-z) \sin
(\pi z )=\pi$ and $\Gamma(1+z)=z \Gamma(z)$, one obtains the amplitude
\bb\label{ampl}
A_m\sim (s-M^2)^2 
\sin \lk(\pi (q^2+m^2) \alp/4\re) \lk(\Gamma((q^2+m^2)
\alp/4)\re)^2\ .
\ee
In the energetic regime considered, 
\bb
s-M^2\sim m_1 m_2 v_{rel}^2\equiv \sqrt{\TT}\ ,
\ee
where $v_{rel}$ is the relative velocity of strings $1$ and $2$
in the lab frame.

Equation~\pref{ampl} has poles at $q^2+m^2=4 n/\alp$ with $n\le 0$.
We consider scattering processes probing distances $r$ much larger than
the string scale, $q_{max}\sim 1/r\ll 1/\ls$; we also assume that 
it is possible to have
$R\ll \ls$, which we will see is necessary. Given that these poles space the 
masses of the resonances by the string scale, the dominant term to
the amplitude is the one corresponding to the exchange of a wound ground
state, \ie\ the $n=0$ pole. 
Measuring quantities in string units, the amplitude then becomes
\bb
A_m\sim \TT \frac{\sin (q^2+m^2)}{\lk(
q^2+m^2\re)^2}\ .
\ee

The effective potential between the strings is the Fourier
transform of this expression with respect to $q$. 
Let us consider various limits.
Take $m \ll q$; we then have $m\ll 1$. The amplitude becomes
\bb
A_m^{(1)}\sim \frac{\TT}{q^2}\ .
\ee
Next consider $m\gg q$, but $m\ll 1$. The amplitude becomes
\bb
A_m^{(2)}\sim \frac{\TT}{q^2+m^2}\ .
\ee
Finally, for $m\gg q$ and $m\gg 1$, we have a constant
\bb
A_m^{(3)}\sim \TT \frac{\sin m^2}{m^4}\ .
\ee

The effective potentials are then ($d\equiv 9-p$)
\bb
V_{eff}^{(1)}\sim  \int d^d q\ e^{iq.x} A_m^{(1)} 
\sim  \frac{T}{r^{d-2}}\ .
\label{Vone}
\ee
The result is a Coulomb potential, independent of $m$.
The second case gives
\bb
V_{eff}^{(2)} 
\sim  T \lk(2\pi\re)^{d/2} \lk(\frac{m}{r}\re)^{d/2-1}
\sqrt{\frac{\pi}{2 m r}} e^{-mr} \ ,
\label{Vtwo}
\ee
which is weaker than $V_{eff}^{(1)}$ since we have $mr\gg 1$.
Finally, we have
\bb
V_{eff}^{(3)} 
  \sim T \frac{\sin m^2}{m^4}\frac{1}{r^d} \ .
\label{Vthree}
\ee
In addition to a larger power in $r$,
we have $m\gg 1$; this interaction is much weaker than
\pref{Vone},\pref{Vtwo}, especially after averaging
over a range of winding transfers $m$.

We conclude that, for $mr=R r(n_3-n_2)/\alp \ll 1$, 
we have a Coulombic potential independent
of the winding exchange $m$; for $mr\gg 1$, we have
much weaker potentials. 
This implies that in a gas of winding strings
bound in a ball of size at most of order the string scale,
the dominant potential is Coulombic with a multiplicative 
factor given by $w_0\equiv \alp/(R r)$,
provided a mechanism restricts winding exchange processes to 
$n_3-n_2\ll n_1+n_2$. 

The S-dual of this amplitude describes the scattering of wound
D-strings at strong coupling, with winding number exchange.
Under a further T duality, and lifting to M theory,
this amplitude encodes a good measure of the effects of longitudinal
momentum exchange in the problem of a self-interacting Matrix string.
The bound on the winding number
translates in our language to
\bb
w_0=\frac{{\bar{\alpha}}' \bar{g}_s}{\Sigma r}=\frac{R_{11}}{r}\sim
\frac{N}{S}\ ,
\ee
\ie\ the resolution in the longitudinal direction. We also note that,
under this chain of dualities, the string scale used to set
a bound
on the impact parameter $r$ transforms as $\alp\rightarrow \alp$,
where the latter string scale is that of the Matrix string. This justifies
our implied equivalence between the scale of
$r$ and that of the size of the black hole.

In the single Matrix string case we study, 
we saw that regions of size $N/S$ were
strongly correlated and `rigid' in a statistical sense. 
The self-interaction of the large string will then involve processes of
coherent exchange of D-string
winding up to the winding number $N/S\ll N$. 
For larger winding, the D-string is not coherent; one
expects a suppression both from the emission vertex and 
from the highly off-shell propagator.
We saw above that all such processes, up to $N/S$, 
are of equal strength and scale Coulombically.
This implies that the potential between the string strands calculated
from the DBI expansion must be enhanced by a factor of $N/S$ for $N>S$, 
and justifies the scaling arguments used in Section~\ref{dynsec}.


\providecommand{\href}[2]{#2}\begingroup\raggedright

\endgroup

\end{document}